\documentclass[preprint,superscriptaddress,showpacs,nofootinbib,tightenlines,amsmath,lmodern]{revtex4}
\usepackage{graphics}
\usepackage{epsfig}
\usepackage{slashed}
\usepackage{amssymb}
\usepackage{color}
\usepackage{bbm}
\usepackage{gensymb}
\usepackage{bbold}
\def\bra#1{\ensuremath{\langle{#1}\vert}}
\def\ket#1{\ensuremath{\vert{#1}\rangle}}

\newcommand{\no}{\nonumber}

\begin{document}
\title{$\eta$-$\eta'$ Mixing in Large-$N_c$ Chiral Perturbation Theory}
\author{P.~Bickert}
\affiliation{PRISMA Cluster of Excellence, Institut f\"ur Kernphysik, Johannes Gutenberg-Universit\"at Mainz,
D-55099 Mainz, Germany}
\author{P.~Masjuan}
\affiliation{PRISMA Cluster of Excellence, Institut f\"ur Kernphysik, Johannes Gutenberg-Universit\"at Mainz,
D-55099 Mainz, Germany}
\affiliation{Grup de F\'{\i}sica Te\`orica, Departament de F\'{\i}sica,
  Universitat Aut\`onoma de Barcelona,
  and Institut de F\'{\i}sica d'Altes Energies (IFAE),
  The Barcelona Institute of Science and Technology (BIST),
  Campus UAB, E-08193 Bellaterra (Barcelona), Spain}
\author{S.~Scherer}
\affiliation{PRISMA Cluster of Excellence, Institut f\"ur Kernphysik, Johannes Gutenberg-Universit\"at Mainz,
D-55099 Mainz, Germany}

\date{16 December 2016}
\preprint{MITP/16-139}
\begin{abstract}
We present a calculation of the $\eta$-$\eta'$ mixing in the
framework of large-$N_c$ chiral perturbation theory. A general
expression for the $\eta$-$\eta'$ mixing at next-to-next-to-leading order (NNLO)
is derived, including higher-derivative terms up to fourth order in the four
momentum, kinetic and mass terms. In addition, the axial-vector
decay constants of the $\eta$-$\eta'$ system are determined at NNLO.
The numerical analysis of the results is performed successively at
LO, NLO, and NNLO. We investigate the influence of one-loop
corrections, OZI-rule-violating parameters, and
$\mathcal{O}(N_c p^6)$ contact terms.
\end{abstract}

\maketitle

\nopagebreak[4]

\section{Introduction}

   The mixing of states is a feature of quantum mechanics and quantum field theory,
which is intimately related to the symmetries of the underlying dynamics and the
eventual mechanisms leading to their breaking.
   Prominent examples in the realm of subatomic physics include the
$K^0$-$\bar{K}^0$, $D^0$-$\bar{D}^0$, and $B^0$-$\bar{B}^0$ mixing and oscillations,
neutrino mixing, the Cabibbo-Kobayashi-Maskawa quark-mixing matrix,
and the Weinberg angle \cite{Agashe:2014kda}.
   In the low-energy regime of QCD, we observe a fascinating
interplay between the dynamical (spontaneous) breaking of chiral
symmetry, the explicit symmetry breaking by the quark masses, and
the axial $\text{U(1)}_A$ anomaly.
   In this context, the pseudoscalar mesons $\eta$ and $\eta'$ represent
an ideal laboratory for investigating the relevant symmetry-breaking
mechanisms in QCD.
   For example, hadronic decays, such as $\eta^{(\prime)}\to\pi\pi\pi$ and $\eta'\to\eta\pi\pi$, test our knowledge of low-energy effective field
theories (EFTs) and provide information on the light-quark
masses.\footnote{For an overview of the main topics in $\eta$ and
$\eta'$ physics from both theoretical and experimental sides, see
Refs.~\cite{Adlarson:2012bi,Amaryan:2013eja,Adlarson:2014hka} and references therein.}
   On the other hand, electromagnetic decays such as $\eta^{(\prime)}\to\gamma^{(\ast)}\gamma^{(\ast)}$
proceed through the Adler-Bell-Jackiw
anomaly~\cite{Adler:1969gk,Bell:1969ts,Adler:1969er}.
   In the case of virtual photons, the corresponding amplitudes reveal the electromagnetic structure in terms
of the transition form factors.

   For vanishing up-, down-, and strange-quark masses,
the QCD Lagrangian has a global $\text{U(3)}_L\times\text{U(3)}_R$
symmetry at the classical level (see, e.g.,
Ref.~\cite{Scherer:2012zzd} for a discussion).
   The transition to the quantum level results in two main features:
   First, the QCD vacuum is assumed to be invariant only under the subgroup
$\mbox{SU(3)}_V\times\mbox{U(1)}_V$, i.e., the symmetry of the
Lagrangian is dynamically broken in the ground state.
   Second, quantum corrections destroy the
singlet axial-vector current conservation, i.e., the corresponding
four divergence has an anomaly proportional to the square of the
strong coupling constant $g$
\cite{Adler:1969gk,Bell:1969ts,Adler:1969er}.
   As a consequence of the Goldstone theorem \cite{Goldstone:1962es},
one expects an octet of massless, pseudoscalar bosons
$(\pi,K,\eta_8)$.
   Because of the $\text{U(1)}_A$ anomaly, the singlet eta, $\eta_1$, is
massive even in the chiral limit of massless quarks
\cite{'tHooft:1976up,Witten:1979vv,Veneziano:1979ec}.
   However, invoking the large-number-of-colors (L$N_c$) limit of QCD~\cite{'tHooft:1973jz,Witten:1979kh}
(see, e.g., Refs.~\cite{Bhaduri:1988gc,Manohar:1998xv} for an
introduction), i.e., $N_c\to\infty$ with $g^2N_c$ fixed, the
$\mbox{U}(1)_A$ anomaly disappears, and the assumption of an
$\mbox{SU(3)}_V\times\mbox{U}(1)_V$ symmetry of the ground state
implies that the singlet state is also massless.
   In other words, in the combined chiral and L$N_c$ limits,
QCD at low energies is expected to generate the nonet
$(\pi,K,\eta_8,\eta_1)$ as the Goldstone bosons
\cite{DiVecchia:1980yfw,Coleman:1980mx}.

   Massless L$N_c$ QCD is an approximation to the real world.
   In fact, chiral symmetry is explicitly broken by the quark masses, and SU(3) flavor symmetry is broken by the
fact that the strange quark is substantially heavier than the up and
down quarks \cite{Leutwyler:2013wna}. As a result (of the flavor symmetry breaking), the physical $\eta$ and $\eta'$ states are mixed octet and singlet states.
   By means of an orthogonal transformation with mixing angle $\theta$, the physical $\eta$ and $\eta'$ states, i.e., the mass eigenstates,
are usually expressed as linear combinations of the octet and
singlet states $\eta_8$ and $\eta_1$ \cite{Amsler}.
   Such a change of basis entails the diagonalization of the general quadratic mass matrix in the basis of SU(3)-octet and -singlet states, where the diagonal entries are given by
the squares of the octet and the singlet
masses~\cite{Isgur:1976qg,Fritzsch:1976qc}, while the off-diagonal
terms account for the SU(3)-symmetry-breaking
effects~\cite{Gasser:1984gg,Donoghue:1986wv,Gilman:1987ax,Schechter:1992iz,Bramon:1997va}.

   In the chiral limit, the $\mbox{U}(1)_A$ anomaly contributes only to the singlet mass \cite{'tHooft:1976up}.
   As a result of the mixing, the anomaly contribution is transferred to both the $\eta$ and $\eta'$
states, such that the $\eta'$ remains heavier than the
$\eta$.
   A discussion of the $\eta$-$\eta'$ mixing in the framework of
effective field theory (EFT) should consider both states as
dynamical degrees of freedom and, for a perturbative treatment, the
respective masses should be small in comparison with a typical
hadronic energy scale.
   Now, in the chiral limit, the $\eta'$ still remains massive.
   For that reason, in the low-energy expansion of conventional
$\text{SU}(3)_L\times\text{SU}(3)_R$ chiral perturbation theory (ChPT), the $\eta'$ does not play a
special role as compared to other states such as the $\rho$ meson
\cite{Gasser:1984gg}.
   However, the combined chiral and L$N_c$ limits may serve as a starting point for Large-$N_c$ chiral perturbation theory
(L$N_c$ChPT) as the EFT of QCD at low energies including the singlet
field~\cite{Moussallam:1994xp,Leutwyler:1996sa,HerreraSiklody:1996pm,Leutwyler:1997yr,Kaiser:1998ds,HerreraSiklody:1998cr,Kaiser:2000gs,Borasoy:2004ua,Guo:2015xva},
which we will also refer to as $\text{U}(3)_L\times\text{U}(3)_R$ effective theory.\footnote{For the sake of notational brevity, from now on we will use the terminology SU(3) and U(3) ChPT instead of $\text{SU}(3)_L\times\text{SU}(3)_R$ and $\text{U}(3)_L\times\text{U}(3)_R$ ChPT, respectively.}

   In the framework of L$N_c$ChPT, one performs a simultaneous expansion of (renormalized) Feynman diagrams
in terms of momenta $p$, quark masses $m$, and $1/N_c$.\footnote{It
is understood that dimensionful variables need to be small in
comparison with an energy scale.}
   The three expansion variables are counted as small quantities of order
\cite{Leutwyler:1996sa}
\begin{equation}
\label{powerexp}
p=\mathcal{O}(\sqrt{\delta}),\ \ \ m=\mathcal{O}(\delta),\ \ \
1/N_c=\mathcal{O}(\delta).
\end{equation}
   The corresponding power-counting rules will be discussed in
Sect.\ \ref{sec_lagrangian_power_counting}.
   The leading-order chiral Lagrangian
is not able to reproduce the experimental result for the $\eta$ and
$\eta'$ masses~\cite{Georgi:1993jn}, and higher-order terms in the
$1/N_c$ (and quark-mass) expansion must be taken into
account~\cite{Peris:1993np}.
   The inclusion of loop effects in the scheme of
Eq.~(\ref{powerexp}) increases the order by $\delta^2$.
   Thus, any calculation in this framework at the loop level needs then to be performed at least at next-to-next-to-leading order (NNLO).
   This order would demand the knowledge of the low-energy constants (LECs) of $\mathcal{O}(p^4)$ and of those of ${\cal O}(p^6)$ which are leading in $1/N_c$.
   The proliferation of (a priori unknown) LECs poses a challenge for any prediction within this theory and information from other sources, e.g., from a matching to physical observables or lattice simulations, will be required in order to determine the LECs.
   For SU(3) ChPT, the LECs at ${\cal O}(p^4)$ are well known, and information on some of the ${\cal O}(p^6)$ LECs is also provided \cite{Bijnens:2014lea}.
   With a suitable matching, one can translate the SU(3) values into the corresponding ones within the U(3)
effective theory.

    Since we take higher orders of the $1/N_c$ expansion into account, terms violating the Okubo-Zweig-Iizuka (OZI) rule appear perturbatively in our calculations.
   They will be accompanied by a set of LECs which are rather poorly known at $\mathcal{O}(\delta)$ and basically unknown at higher orders.

   If we include higher-order corrections in our effective Lagrangian, the connection between the physical $\eta$ and $\eta'$ states and the singlet and octet states is more complicated than a simple rotation. Furthermore, the description of the $\eta$-$\eta'$ mixing with a single mixing angle $\theta$ is not appropriate to describe the experimental data and also the axial-vector decay constants of the $\eta$-$\eta'$ system (at NLO) cannot be described by a simple rotation with angle $\theta$.
   This problem was solved by invoking a mixing scheme with two different angles,
the so-called two-angle mixing scheme~\cite{Feldmann:1998vh,Feldmann:1998sh}.
   In recent years, the use of the two-angle scheme has been very
popular and resulted in well-established phenomenological
determinations of the mixing~\cite{Feldmann:1998vh,Feldmann:1998sh,Benayoun:1999au,Escribano:2005qq,Escribano:2010wt,Escribano:2013kba,Escribano:2015nra,Escribano:2015yup},
a procedure that can also be extended to include an eventual gluonium content of these pseudoscalars
(see, e.g., Refs.~\cite{Ball:1995zv,Thomas:2007uy,Escribano:2007cd}).

   This work is organized as follows.
   In Sec.~\ref{sec_lagrangian_power_counting} we describe the
effective field theory we will consider for our calculation by
specifying the Lagrangian and the power counting.
   In Sec.~\ref{section_mixing_angle} we present the calculation of
the mixing angles at NNLO.
   Section \ref{section_decay_constants} deals with the $\eta$
and $\eta'$ decay constants.
   In Sec.~\ref{section_numerical_analysis} we elaborate on the
numerical analysis of the mixing, decay constants, and pseudoscalar
masses with different input sets of LECs.
   Finally, in Sec.~\ref{section_summary} we conclude with a few remarks
and an outlook of possible future work.

\section{Lagrangians and power counting}
\label{sec_lagrangian_power_counting}

   The most general Lagrangian of L$N_c$ChPT is organized as an
infinite series in terms of derivatives, quark-mass terms, and,
implicitly, powers of $1/N_c$, with the scaling behavior given in
Eq.~(\ref{powerexp}):
\begin{equation}
\label{Leff}
\mathcal{L}_{\text{eff}}=\mathcal{L}^{(0)}+\mathcal{L}^{(1)}+\mathcal{L}^{(2)}+\dots,
\end{equation}
   where the superscripts $(i)$ denote the order in $\delta$.
   The rules leading to the assignments of these orders will be explained below.
   The properties of the building blocks are defined in Appendix \ref{appendix_building_blocks}.

   The dynamical degrees of freedom are collected in the unitary $3\times 3$ matrix
\begin{equation}
\label{definitionU}
 U(x)=\exp\left(i\frac{\phi(x)}{F}\right),
\end{equation}
where the Hermitian $3\times 3$ matrix
\begin{align}
\label{eq:pseudoscalar_mesons}
\phi=\sum_{a=0}^8\phi_a\lambda_a=
\begin{pmatrix}
\pi^0+\frac{1}{\sqrt{3}}\eta_8+\sqrt{\frac{2}{3}}\eta_1 & \sqrt{2}\pi^+ & \sqrt{2}K^+\\
\sqrt{2}\pi^- &  -\pi^0+\frac{1}{\sqrt{3}}\eta_8+\sqrt{\frac{2}{3}}\eta_1 & \sqrt{2}K^0\\
\sqrt{2}K^- & \sqrt{2}\bar{K}^0 & -\frac{2}{\sqrt{3}}\eta_8+\sqrt{\frac{2}{3}}\eta_1\\
\end{pmatrix}
\end{align}
contains the pseudoscalar octet fields and the pseudoscalar singlet field $\eta_1$, the
$\lambda_a$ ($a=1,\ldots,8$) are the Gell-Mann matrices, and
$\lambda_0=\sqrt{2/3}\, {\mathbbm 1}$.
   In Eq.~(\ref{definitionU}), $F$ denotes the pion-decay constant in the three-flavor chiral limit\footnote{Here,
we deviate from the often-used convention of indicating the {\it three}-flavor chiral limit by a subscript 0.}
and is counted as $F=\mathcal{O}(\sqrt{N_c})=\mathcal{O}(1/\sqrt{\delta})$.\footnote{Consider
a generic quark bilinear of the type $\bar{q}\,\Gamma F q$, with
$\Gamma$ and $F$ standing for matrices in Dirac and flavor space,
respectively, and a summation over color indices implied.
   In the L$N_c$ limit of QCD, the matrix element for any such quark
bilinear to create a meson from the vacuum scales like $\sqrt{N_c}$
\cite{Witten:1979kh}.}
  The pseudoscalar fields $\phi_a$ ($a=0,\ldots,8$) count as $\mathcal{O}(\sqrt{N_c})$
such that the argument of the exponential function is $\mathcal{O}(\delta^0)$
and, thus, $U=\mathcal{O}(\delta^0)$.
   Besides the dynamical degrees of freedom of Eq.~(\ref{eq:pseudoscalar_mesons}),
the effective Lagrangian also contains a set of external fields $(s,p,l_\mu,r_\mu,\theta)$.
   The fields $s$, $p$, $l_\mu$, and $r_\mu$ are Hermitian, color-neutral $3\times 3$ matrices
coupling to the corresponding quark bilinears, and $\theta$ is a real field
coupling to the winding number density \cite{Gasser:1984gg}.
   The external scalar and pseudoscalar fields $s$ and $p$ are combined
in the definition $\chi\equiv 2B(s+ip)$ \cite{Gasser:1984gg}.
   The LEC $B$ is related to the scalar singlet quark condensate
$\langle\bar{q}q\rangle_0$ in the three-flavor chiral limit and is of
${\cal O}(N_c^0)$.

   In general, applying the power counting of Eq.~(\ref{powerexp}) to the construction of the effective
Lagrangian in the L$N_c$ framework involves two ingredients.
   On the one hand, there is the momentum and quark-mass counting which proceeds as in conventional $\text{SU}(3)$ ChPT
\cite{Gasser:1984gg}: (covariant) derivatives count as ${\cal O}(p)$, $\chi$ counts as ${\cal O}(p^2)$, etc.~(see
Table \ref{TabCounting}).
   We denote the corresponding chiral order by $D_p$.
   The discussion of the $\text{U}(3)$ case results in essentially three major modifications in comparison
with $\text{SU}(3)$ \cite{Leutwyler:1996sa,HerreraSiklody:1996pm,Kaiser:2000gs}: First,
the determinant of $U$ is no longer restricted to have the value 1, second, additional external
fields appear, third, the conventional structures of $\text{SU}(3)$ ChPT will be multiplied by coefficients which
are functions of the linear combination $(\psi+\theta)$, where $\psi\equiv\sqrt{6}\eta_1/F$ such that $\mbox{det}(U)=\exp(i\psi)$.
   According to Eqs.~(\ref{transformations_properties}), the sum $(\psi+\theta)$ remains invariant under
chiral $\mbox{U}(3)_L\times\mbox{U}(3)_R$ transformations.
   For example, denoting the SU(3) matrix of ordinary ChPT by $\hat{U}$,
the leading-order Lagrangian reads \cite{Gasser:1984gg}
\begin{displaymath}
{\cal L}_2=\frac{F^2}{4}\langle D_\mu \hat{U} D^\mu \hat{U}^\dagger\rangle
+\frac{F^2}{4}\langle\chi\hat{U}^\dagger+\hat{U}\chi^\dagger\rangle,
\end{displaymath}
   where the symbol $\langle\ \rangle$ denotes the trace over flavor indices and
the covariant derivatives are defined in Eqs.~(\ref{appendix_covariant_derivatives}).
   This expression is replaced by \cite{HerreraSiklody:1996pm}
\begin{equation}
\label{replacement}
W_1\langle D_\mu U D^\mu U^\dagger\rangle +W_2\langle \chi U^\dagger+U\chi^\dagger\rangle,
\end{equation}
where $W_1$ and $W_2$ are functions of $(\psi+\theta)$ and are also referred to as
potentials \cite{Kaiser:2000gs}.
   In the limit $N_c\to\infty$, these functions reduce to constants \cite{Leutwyler:1996sa}.
   However, for $N_c$ finite, the functions may be expanded in $(\psi+\theta)$ with
well-defined assignments for the L$N_c$ scaling behavior of the expansion coefficients.

   In addition to the potentials, also new additional structures show up which do not exist in the $\text{SU}(3)$ case.
   For example, in ordinary ChPT one finds for the trace $\langle D_\mu \hat{U}\hat U^\dagger\rangle=0$
\cite{Scherer:2012zzd}, whereas in the $\text{U}(3)$ case one has
\begin{equation}
\label{dmupsi}
\langle D_\mu U U^\dagger\rangle=iD_\mu\psi,
\end{equation}
giving rise to a new term of the type $-W_4 D_\mu\psi D^\mu\psi$ \cite{HerreraSiklody:1996pm}.

   The L$N_c$ behavior can be determined by using the following rules (see Refs.~\cite{HerreraSiklody:1996pm,Kaiser:2000gs}
for a detailed account).
   In the L$N_c$ counting, the leading contribution to a quark correlation function is given by a single flavor trace and is of order $N_c$ \cite{'tHooft:1973jz,Witten:1979kh,Manohar:1998xv}.
   In general, diagrams with $r$ quark loops and thus $r$ flavor traces are of order $N_c^{2-r}$. Terms
without traces correspond to the purely gluonic theory and count at leading order as $N_c^2$.
   This argument is transferred to the level of the effective Lagrangian, i.e., single-trace terms are
of order $N_c$, double-trace terms of order unity, etc.\footnote{When applying these counting rules,
one has to account for the so-called trace relations connecting single-trace terms with products of traces
(see, e.g., Appendix A of Ref.~\cite{Fearing:1994ga}).}
   In other words, we need to identify the number $N_{ft}$ of flavor traces.
   In particular, because of  Eq.~(\ref{dmupsi}), the expression $D_\mu\psi$ implicitly
involves a flavor trace (see footnote 7 of Ref.~\cite{Kaiser:2000gs}).
   Furthermore, when expanding the potentials,
each power $(\psi+\theta)^n$ is accompanied by a coefficient of order ${\cal O}(N_c^{-n})$.
   The reason for this assignment is the fact that, in QCD, the external field $\theta$ couples
to the winding number density with strength $1/N_c$.
   In a similar fashion, $D_\mu\theta$ (as well as multiple derivatives) are related to expressions
with ${\cal O}(N_c^{-1})$.\footnote{Note that we do not directly book the quantities $(\psi+\theta)$
or $D_\mu\theta$ as ${\cal O}(N_c^{-1})$, but rather attribute this order to the coefficients
coming with the terms.}
   Denoting the number of $(\psi+\theta)$ and $D_\mu\theta$ terms by $N_\theta$,
the L$N_c$ order reads \cite{HerreraSiklody:1996pm,Kaiser:2000gs}
\begin{equation}
\label{DNc}
D_{N^{-1}_c}=-2+N_{tr}+N_\theta.
\end{equation}
   The combined order of an operator is then given by
\begin{equation}
\label{Ddelta}
D_\delta=\frac{1}{2}D_p+D_{N_c^{-1}}.
\end{equation}
   In particular, using Eq.~(\ref{Ddelta}) allows us to identify the L$N_c$ scaling
behavior of the LEC multiplying the corresponding operator.

   The leading-order Lagrangian is given by \cite{Leutwyler:1996sa,Kaiser:2000gs}
\begin{equation}
\label{eq:lolagrangian}
\mathcal{L}^{(0)}=\frac{F^2}{4}\langle D_\mu U D^\mu U^\dagger\rangle+\frac{F^2}{4}\langle\chi U^\dagger+U\chi^\dagger\rangle-\frac{1}{2}\tau(\psi+\theta)^2.
\end{equation}
   Comparing with Eq.~(\ref{replacement}), we identify
\begin{displaymath}
\frac{F^2}{4}=W_1(0)=W_2(0)
\end{displaymath}
as the leading-order term of the expansion of the functions $W_1$ and $W_2$ which, because
of parity, are even functions.
   On the other hand, the last term of Eq.~(\ref{eq:lolagrangian}) originates from the second-order term of the expansion of $W_0$.
   The constant $\tau=\mathcal{O}(N_c^0)$ is the topological
susceptibility of the purely gluonic theory \cite{Leutwyler:1996sa}.
   Counting the quark mass as ${\cal O}(p^2)$, the first two terms of $\mathcal{L}^{(0)}$
are of $\mathcal{O}(N_c p^2)$, while the third term is of $\mathcal{O}(N_c^0)$, i.e.,~all terms are of ${\cal O}(\delta^{0})$.
   The leading-order Lagrangian contains 3 LECs, namely, $F$, $B$, and $\tau$.

   To explain the power counting of the interaction vertices, we set $r_\mu=l_\mu=0$ and $\chi=2BM$, where
$M=\text{diag}(m_u,m_d,m_s)$ denotes the quark-mass matrix.
   For this case, the leading-order Lagrangian contains only even powers of the pseudoscalar
fields.
   Expanding the first two terms of Eq.~(\ref{eq:lolagrangian}) in terms of the pseudoscalar fields,
results in Feynman rules for the {\it interaction} vertices of the order $p^2 N_c^{1-k/2}$, where $k=4,6,\ldots$ is the number
of interacting pseudoscalar fields \cite{Kaiser:2000gs}.
   The dependence on $N_c$ and $p$ originates from the powers of $F$ and the two derivatives, respectively.
   When discussing QCD Green functions of, say, pseudoscalar quark bilinears, there will be a factor $BF={\cal O}(\sqrt{N_c})$
at each external source (see sec.~4.6.2 of Ref.~\cite{Scherer:2002tk}), such that an $n$-point function is of the order $p^2 N_c$.
   Taking $\phi_a={\cal O}(\sqrt{N_c})$, the interaction Lagrangians count as ${\cal O}(p^2 N_c)$, which
is consistent with referring to the Lagrangian as ${\cal O}(\delta^0)$, with the leading-order contributions
of quark loops being ${\cal O}(N_c)$ and the leading chiral order being ${\cal O}(p^2)$.
   On the other hand, it is also consistent with the expectation of the effective meson vertices containing $k$ external lines
being of the order $N_c^{1-k/2}$ \cite{Witten:1979kh}.

   The NLO Lagrangian $\mathcal{L}^{(1)}$ was constructed in Refs.~\cite{Leutwyler:1996sa,HerreraSiklody:1996pm,Kaiser:2000gs}
and receives contributions of $\mathcal{O}(N_c p^4)$, $\mathcal{O}(p^2)$, and $\mathcal{O}(N_c^{-1})$.
   The terms that are of the same structure as those in ${\cal L}^{(0)}$ may be absorbed in the coupling
constants $F$, $B$, and $\tau$ \cite{Kaiser:2000gs}.
   In particular, $\tau$ now has to be distinguished from the topological susceptibility of gluodynamics.
   We only display the terms relevant for our calculation, in particular, we set
$v_\mu\equiv (r_\mu+l_\mu)/2=0$ and keep only $a_\mu\equiv(r_\mu-l_\mu)/2$, which is needed for
the calculation of the axial-vector-current matrix elements:
\begin{align}
\label{L1}
\mathcal{L}^{(1)}&= L_5 \langle D_\mu U D^\mu U^\dagger (\chi U^\dagger + U \chi^\dagger)\rangle
+ L_8 \langle \chi U^\dagger\chi U^\dagger + U \chi^\dagger U \chi^\dagger \rangle  \no \\
&\quad + \frac{F^2}{12} \Lambda_1 D_\mu \psi D^\mu \psi - i \frac{F^2}{12} \Lambda_2 (\psi+\theta)
\langle \chi U^\dagger - U \chi^\dagger \rangle+\dots,
\end{align}
where
\begin{align}
\label{Dmupsi}
D_\mu\psi&=\partial_\mu\psi-2\langle a_\mu\rangle,\\
\label{Dmutheta}
D_\mu\theta&=\partial_\mu\theta+2\langle a_\mu\rangle,
\end{align}
and the ellipsis refers to the suppressed terms.
   The first two terms of $\mathcal{L}^{(1)}$ count as $\mathcal{O}(N_cp^4)$ and are obtained from the standard
SU(3) ChPT Lagrangian of ${\cal O}(p^4)$ \cite{Gasser:1984gg} by retaining solely terms with a single trace
and keeping only the constant terms of the potentials.
   With $D_p=4$ and $D_{N_c^{-1}}=-1$, Eq.~(\ref{Ddelta}) implies that $L_5$ and $L_8$ are of ${\cal O}(N_c)$.
   According to Eq.~(\ref{Dmupsi}), the expression $D_\mu\psi D^\mu\psi$ implicitly involves two flavor traces
(see footnote 7 of Ref.~\cite{Kaiser:2000gs}), with the result that the corresponding term is ${\cal O}(N_c^0)$.
   Since $F={\cal O}(\sqrt{N_c})$, the coupling $\Lambda_1$ scales as ${\cal O}(N_c^{-1})$ and has to vanish in the
L$N_c$ limit.
   Finally, the structure proportional to $\Lambda_2$ is the leading-order term of the expansion of the
potential $W_3$.
   With $D_p=2$ and $D_{N_c^{-1}}=0$ ($N_{tr}=N_\theta=1$), the LEC
$\Lambda_2$ scales as ${\cal O}(N_c^{-1})$.

   The SU(3) Lagrangian of ${\cal O}(p^6)$ was discussed in
Refs.~\cite{Fearing:1994ga,Bijnens:1999sh,Ebertshauser:2001nj,Bijnens:2001bb}, and the generalization
to the U(3) case has recently been obtained in Ref.~\cite{Jiang:2014via}.
   For the present purposes, at NNLO, the relevant pieces of $\mathcal{L}^{(2)}$ can be split into three different contributions
of ${\cal O} (N_c^{-1}p^2)$, ${\cal O}(p^4)$, and ${\cal O}(N_c p^6)$, respectively:
\begin{align}
\label{L2a}
{\cal L}^{(2, N_c^{-1}p^2)}&=-\frac{F^2}{4}v^{(2)}_2(\psi+\theta)^2\langle \chi U^\dagger+U\chi^\dagger \rangle,\\
\label{L2b}
{\cal L}^{(2, p^4)} &=
L_4 \langle D_\mu U  D^\mu U^\dagger \rangle \langle \chi U^\dagger + U \chi^\dagger\rangle
+L_6 \langle \chi U^\dagger+U\chi^\dagger\rangle^2
+L_7 \langle  \chi U^\dagger - U \chi^\dagger\rangle^2 \nonumber\\
&\quad + i L_{18} D_\mu \psi \langle  \chi D^\mu U^\dagger - D^\mu U \chi^\dagger \rangle
+ i L_{25} (\psi+\theta) \langle  \chi U^\dagger \chi U^\dagger- U \chi^\dagger U \chi^\dagger\rangle\no\\
&\quad+i L_{46}D_\mu\theta\langle D^\mu U U^\dagger(\chi U^\dagger+ U\chi^\dagger)\rangle
+i L_{53}\partial_\mu D^\mu\theta\langle \chi U^\dagger-U \chi^\dagger \rangle+\ldots,\\
\label{L2c}
{\cal L}^{(2,N_cp^6)} &=
C_{12} \langle \chi_{+}h_{\mu\nu}h^{\mu\nu} \rangle
+ C_{14} \langle u_{\mu}u^{\mu}\chi^2_{+}\rangle+C_{17} \langle \chi_{+}u_{\mu}\chi_{+}u^{\mu}\rangle
+ C_{19} \langle\chi^3_{+}\rangle\no\\
&\quad+C_{31} \langle \chi^2_{-}\chi_{+}\rangle+\ldots,
\end{align}
where
\begin{align}
\chi_\pm&=u^\dagger\chi u^\dagger\pm u\chi^\dagger u,\nonumber\\
u&=\sqrt{U},\nonumber\\
u_{\mu}&=i\left[u^\dagger(\partial_\mu-ir_\mu)u-u(\partial_\mu-il_\mu)u^\dagger\right]=iu^\dagger D_\mu U u^\dagger,\nonumber\\
h_{\mu\nu}&=\nabla_\mu u_\nu+\nabla_\nu u_\mu,\nonumber\\
\nabla_{\mu}X&=\partial_\mu X+[\Gamma_\mu , X],\nonumber\\
\Gamma_{\mu}&=\frac{1}{2}\left[u^\dagger(\partial_\mu-ir_\mu)u+u(\partial_\mu-il_\mu)u^\dagger\right].
\end{align}
   The coupling $v^{(2)}_2$ of Eq.~(\ref{L2a}) scales like ${\cal O}(N_c^{-2})$ and originates from
the expansion of the potentials of Refs.~\cite{Leutwyler:1996sa,Kaiser:2000gs} up to and including
terms of order $(\psi+\theta)^2$.
   The first three terms of Eq.~(\ref{L2b}) stem from the standard SU(3) ChPT Lagrangian of ${\cal O}(p^4)$
with two traces and are $1/N_c$ suppressed compared to the single-trace terms in Eq.~(\ref{L1}).
   The remaining terms of Eq.~(\ref{L2b}) are genuinely related to the L$N_c$ $\text{U}(3)$ framework, since they
contain interactions involving the singlet field or the singlet axial-vector current.
   Finally, the $C_i$ terms of Eq.~(\ref{L2c}) are obtained from single-trace terms of the SU(3) Lagrangian of
${\cal O}(p^6)$ \cite{Bijnens:1999sh}.
   As there is, at present, no satisfactory unified nomenclature for the coupling constants, for easier reference we choose
the names according to the respective references from which the Lagrangians were taken.
   In our calculation we do not include external vector fields, i.e., $v_\mu=0$.
   The $L_{46},L_{53}$ terms in $\mathcal{L}^{(1)}$ are not needed for the calculation of the mixing.
   They enter, however, in the calculation of the decay constants of the axial-vector-current matrix elements.

\begin{figure}[htbp]
    \centering
        \includegraphics[width=0.7\textwidth]{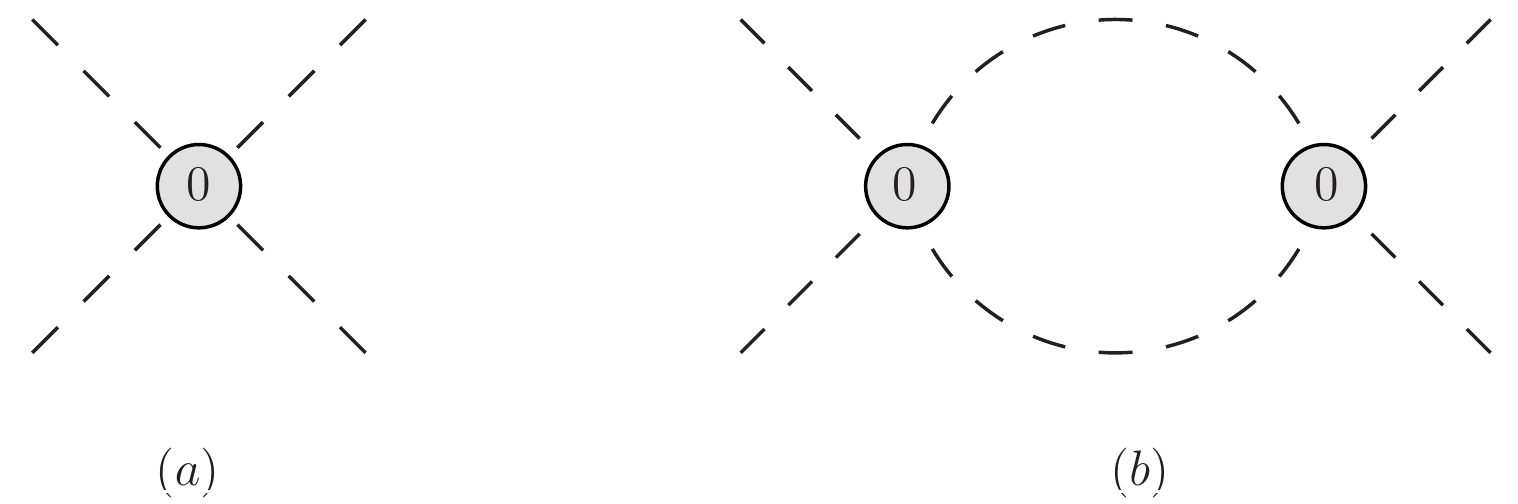}
        \caption{Illustration of the power counting in L$N_c$ChPT. The number 0 in the interaction blobs refers to ${\cal L}^{(0)}$.}
    \label{fig:example_power_counting}
\end{figure}

   Last but not least, we summarize the power-counting rules for a given Feynman diagram, which has been evaluated by using the
interaction vertices derived from the effective Lagrangians of Eq.~(\ref{Leff}).
   Using the $\delta$ counting introduced in Eq.~(\ref{powerexp}), we assign
to any such diagram an order $D$ which is obtained from the following ingredients:
   Meson propagators for both octet and singlet fields count as $\mathcal{O}(\delta^{-1})$.
   Since meson fields are always divided by $F=\mathcal{O}(\sqrt{N_c})=\mathcal{O}(\delta^{-\frac{1}{2}})$, a
vertex with $k$ meson fields derived from $\mathcal{L}^{(i)}$ is $\mathcal{O}(\delta^{i+k/2})$.
   The integration of a loop counts as $\delta^2$.
   The order $D$ is obtained by adding up the contributions of the individual building blocks.
   Figure \ref{fig:example_power_counting} provides two examples of the application of the power-counting rules.
   Since the tree-level diagram of Fig.~\ref{fig:example_power_counting} (a) consists of a single vertex derived
from ${\cal L}^{(0)}$ with four external meson lines, it has $D=2$.
   On the other hand, the one-loop diagram of Fig.~\ref{fig:example_power_counting} (b) has two vertices from ${\cal L}^{(0)}$ with four legs, two
meson propagators, and one loop: $D=2+2-1-1+2=4$.
   As expected, the loop increases the order by two units.
   The power-counting rules are summarized in Table~\ref{TabCounting}.

\begin{table}[htbp]
\tabcolsep2em
\renewcommand{\arraystretch}{1.2}
\begin{tabular}{l c c c}
\hline\hline
Quantity &$N_c$ & $p$ & $\delta$ \\ \hline
Momenta/Derivatives $p$/$\partial_{\mu}$ & $1$ & $p$ & $\delta^\frac{1}{2}$ \\
$1/N_c$ & $N_c^{-1}$ & $1$ & $\delta$ \\
Quark masses $m$ & $1$ & $p^2$ & $\delta$ \\
Dynamical fields $\phi_a$ ($a=1,\ldots, 8$)& $\sqrt{N_c}$ & $1$ & $\delta^{-\frac{1}{2}}$\\
Dynamical field $\psi$ & $1$ & $1$ & $1$\\
External field $\theta$ & $1$ & $1$ & $1$\\
External currents $v_{\mu}$ and $a_{\mu}$ & $1$ & $p$ & $\delta^\frac{1}{2}$ \\
External fields $s$ and $p$ & $1$ & $p^2$ & $\delta$\\
Pion-decay constant $F$ (chiral limit) & $\sqrt{N_c}$ & $1$ & $\delta^{-\frac{1}{2}}$ \\
Topological susceptibility $\tau$  & 1 & 1 &  1\\
$M^2_{\eta'}$ (chiral limit)& $N_c^{-1}$& $1$ & $\delta$\\
Octet-meson propagator & $1$& $p^{-2}$ & $\delta^{-1}$\\
Singlet-$\eta_1$ propagator (chiral limit) & a) & a) & $\delta^{-1}$\\
Loop integration & $1$& $p^4$ & $\delta^2$\\
$k$-meson vertex from ${\cal L}^{(i)}$ & b) & b) & $\delta^{i+k/2}$\\
\hline\hline
\end{tabular}
\caption{Power-counting rules in L$N_c$ChPT.
a) The inverse of the singlet $\eta_1$ propagator is of order $1/N_c$ and $p^2$.
b) The assignment $i$ in ${\cal L}^{(i)}$ receives contributions from both
$1/N_c$ and $p^2$.
   Recall that powers $(\psi+\theta)^n$ come with expansion coefficients of ${\cal O}(N_c^{-n})$ even
though we count $(\psi+\theta)$ as ${\cal O}(1)$.
}
\label{TabCounting}
\end{table}

\section{Calculation of the mixing angle}
\label{section_mixing_angle}
   For $m_u=m_d=\hat{m}\neq m_s$, the physical $\eta$ and $\eta'$ mass eigenstates are linear combinations of the
mathematical octet and singlet states $\eta_8$ and $\eta_1$.
   Our aim is to derive a general expression for the $\eta$-$\eta'$ mixing at the one-loop level
up to and including NNLO in the $\delta$ counting.
   To that end, we start from an effective Lagrangian in terms of the octet and singlet fields
and perform successive transformations, resulting in a diagonal Lagrangian in terms of the
physical fields.
   Because of the effective-field-theory nature of our approach, the starting Lagrangian will
contain higher-derivative terms up to and including fourth order in the four momentum.
   The parameters of the Lagrangian are obtained from a one-loop calculation of the self energies
using the Lagrangians and power counting of Sec.~\ref{sec_lagrangian_power_counting}.
   The Lagrangian after the transformation will have a standard ``free-field'' form.

   Let us collect the fields $\eta_8$ and $\eta_1$ in the doublet
\begin{equation}
\label{eq:eta_{BC}}
\eta_{A}\equiv\begin{pmatrix}\eta_8\\ \eta_1\end{pmatrix}.
\end{equation}
   In terms of $\eta_{A}$, at NNLO the most general effective Lagrangian quadratic in $\eta_{A}$ is of the form
\begin{equation}
\mathcal{L}_{{\rm eff}}={\cal L}_A=
\frac{1}{2}\partial_{\mu}\eta^T_{A}\mathcal{K}_A\partial^{\mu}\eta_{A}-\frac{1}{2}\eta^T_{A}\mathcal{M}^2_A\eta_{A}
+\frac{1}{2}\Box\eta^T_{A}\mathcal{C}_A\Box\eta_{A}.
\label{eq:quadL}
\end{equation}
   Note that the fields $\eta_8$ and $\eta_1$ count as ${\cal O}(\sqrt{N_c})$
and a single derivative as ${\cal O}(p)$.
   The symmetric $2\times 2$ matrices $\mathcal{K}_A$, $\mathcal{M}^2_A$, and $\mathcal{C}_A$ can be written as
\begin{align}
\mathcal{K}_A&=\begin{pmatrix} 1+k_8 &k_{81} \\
                k_{81} &  1+k_1\\
                \end{pmatrix},\\
\mathcal{M}^2_A&=\begin{pmatrix} M^2_8 & M^2_{81} \\
                                  M^2_{81}  &  M^2_{1} \\
                                                       \end{pmatrix},\\
\mathcal{C}_A&=\begin{pmatrix} c_8&c_{81} \\
                                 c_{81} &  c_1\\
                                              \end{pmatrix}.
\end{align}
   Later on, we will provide the one-loop expressions for the matrices $\mathcal{K}_A$ and $\mathcal{M}^2_A$.
   The matrix $\mathcal{C}_A$ is given in Eqs.~(\ref{matrixCAc8})--(\ref{matrixCAc81}) of Appendix \ref{A:KM} and is of ${\cal O}(p^2)$.
   If we were to work at leading order, only, we would have to replace
\begin{displaymath}
\mathcal{K}_A\to{\mathbbm 1},\quad
\mathcal{M}^2_A\to{\mathcal{M}^2_A}^{\!(0)}=
\begin{pmatrix}
\stackrel{\circ}{M^2_8}&\stackrel{\circ}{M^2_{81}}\\
\stackrel{\circ}{M^2_{81}}&\stackrel{\circ}{M^2_1}\!+M^2_0
\end{pmatrix},\quad
\mathcal{C}_A\to 0.
\end{displaymath}
    The elements of the (leading-order) mass matrix ${\mathcal{M}^2_A}^{\!(0)}$ read
\begin{align}
\label{lomasses1}
\stackrel{\circ}{M^2_8}&=\frac{2}{3} B\left(\hat m+2m_s\right) = \frac{1}{3} \Big(4\!\stackrel{\circ}{M^2_K}-\stackrel{\circ}{M^2_{\pi}} \Big),\\
\label{lomasses2}
\stackrel{\circ}{M^2_1}&=\frac{2}{3} B\left(2\hat m+m_s\right) = \frac{1}{3} \Big(2\!\stackrel{\circ}{M^2_K}+\stackrel{\circ}{M^2_{\pi}} \Big),\\
\label{lomasses3}
M^2_0&=6\frac{\tau}{F^2},\\
\label{lomasses4}
\stackrel{\circ}{M^2_{81}}&=-\frac{2\sqrt{2}}{3}B\left(m_s - \hat m\right)=-\frac{2\sqrt{2}}{3}\Big(\!\stackrel{\circ}{M^2_K}-\stackrel{\circ}{M^2_{\pi}}\Big),
\end{align}
where $\stackrel{\circ}{M^2_K}\,=\! B\left(\hat m+m_s\right)$ and $\stackrel{\circ}{M^2_{\pi}}\,=\!2 B \hat m$ are the leading-order kaon and pion masses squared,
respectively, and $M^2_0$ denotes the $\text{U}(1)_A$ anomaly contribution to the $\eta_1$ mass squared.
   The mixing already shows up at leading order, because the mass matrix $\mathcal{M}^2_A$ is non-diagonal at
that order.
   The ``kinetic'' matrix $\mathcal{K}_A$ receives NLO and NNLO corrections.
   Finally, the last term in Eq.~(\ref{eq:quadL}), containing higher derivatives of $\eta_{A}$,
originates from the $C_{12}$ term of the $\mathcal{O}(\delta^2)$ Lagrangian in Eq.~(\ref{L2c}).

   Our first step is to perform a field redefinition to get rid of the higher-derivative structure in
Eq.~(\ref{eq:quadL}) \cite{Scherer:1994wi,Fearing:1999fw},
\begin{equation}
\label{eq:etaaetab}
\eta_A=\left({\mathbbm 1}+\frac{1}{2}{\cal C}_A\Box\right)\eta_B.
\end{equation}
   The entries of ${\cal C}_A$ are of ${\cal O}(p^2)$ and the d'Alembertian operator
counts as ${\cal O}(p^2)$.
   The field transformation is constructed such that, after inserting Eq.~(\ref{eq:etaaetab}) into
Eq.~(\ref{eq:quadL}), the last term is canceled by a term originating from the first term in Eq.~(\ref{eq:quadL}).
   Moreover, we obtain additional terms originating from the ``mass term'' of Eq.~(\ref{eq:quadL})
which now contribute to the new kinetic matrix.
   Finally, we neglect any term generated by the field transformation which is beyond the accuracy of a NNLO calculation.
   Using the relation $\phi\Box\phi=\partial_\mu(\phi\partial^\mu\phi)-\partial_\mu\phi\partial^\mu\phi$
for the components of $\eta_B$, and neglecting total-derivative terms, the Lagrangian after the first
field redefinition is of the form
\begin{align}
\label{Lmixing2}
\mathcal{L}_B=\frac{1}{2}\partial_{\mu}\eta^T_B\mathcal{K}_B\partial^{\mu}\eta_B-\frac{1}{2}\eta^T_B\mathcal{M}^2_B\eta_B,
\end{align}
where
\begin{align}
\label{Kappamatrix}
\mathcal{K}_B&= \mathcal{K}_A
+\frac{1}{2}\begin{pmatrix} 2c_8\!\stackrel{\circ}{M^2_8}+\,2c_{81}\!\stackrel{\circ}{M^2_{81}}
&(c_1+c_8)\!\stackrel{\circ}{M^2_{81}}+\,c_{81}(\stackrel{\circ}{M^2_1}\!+M^2_0\,+\!\stackrel{\circ}{M^2_8})\\
(c_1+c_8)\!\stackrel{\circ}{M^2_{81}}+\,c_{81}(\stackrel{\circ}{M^2_1}\!+M^2_0\,+\!\stackrel{\circ}{M^2_8})
&2c_1(\stackrel{\circ}{M^2_1}+\,M^2_0)+2c_{81}\!\stackrel{\circ}{M^2_{81}}
\end{pmatrix}\nonumber\\
&=\begin{pmatrix} 1+\delta^{(1)}_8+\delta^{(2)}_8
&\delta^{(1)}_{81}+\delta^{(2)}_{81}\\
\delta^{(1)}_{81}+\delta^{(2)}_{81}
&1+\delta^{(1)}_1+\delta^{(2)}_1
\end{pmatrix},
\end{align}
   where $\delta^{(i)}_j$ denotes corrections of $\mathcal{O}(\delta^i)$.
   The entries of the mass matrix $\mathcal{M}^2_B=\mathcal{M}^2_A$ are given by
\begin{align}
\label{masses28}
M^2_8&=\,\stackrel{\circ}{M^2_8}\!+\,\Delta {M^2_8}^{(1)}\!+\Delta {M^2_8}^{(2)},\\
\label{masses21}
M^2_1&=M^2_0+\!\stackrel{\circ}{M^2_1}+\,\Delta {M^2_1}^{(1)}\!+\Delta {M^2_1}^{(2)},\\
\label{masses281}
M^2_{81}&=\,\stackrel{\circ}{M^2_{81}}\!+\Delta {M^2_{81}}^{\!\!(1)}\!+\Delta {M^2_{81}}^{\!\!(2)},
\end{align}
where $\Delta{M^2_{j}}^{(i)}$ denotes corrections of $\mathcal{O}(\delta^i)$.

   The next step consists of diagonalizing the kinetic matrix $\mathcal{K}_B$ in Eq.~(\ref{Kappamatrix}) up to and including $\mathcal{O}(\delta^2)$
through the field redefinition
\begin{equation}
\label{fieldrdf}
\eta_B=\sqrt{Z}\eta_C,
\end{equation}
such that
\begin{equation}
\label{conditionZ}
\sqrt{Z}^{\,T}\mathcal{K}_B \sqrt{Z}={\mathbbm 1}.
\end{equation}
   Writing ${\cal K}_B$ as
\begin{displaymath}
{\cal K}_B={\mathbbm 1}+K^{(1)}+K^{(2)},
\end{displaymath}
and making an ansatz for the symmetric matrix $\sqrt{Z}$ of the form
\begin{displaymath}
\sqrt{Z}={\mathbbm 1}+\sqrt{Z}^{\,(1)}+\sqrt{Z}^{\,(2)},
\end{displaymath}
we obtain from Eq.~(\ref{conditionZ}) the conditions
\begin{displaymath}
2 \sqrt{Z}^{\,(1)}+K^{(1)}=0 \quad\Rightarrow\quad \sqrt{Z}^{\,(1)}=-\frac{1}{2}K^{(1)},
\end{displaymath}
and
\begin{displaymath}
2 \sqrt{Z}^{\,(2)}+K^{(2)}+{\sqrt{Z}^{\,(1)}}^2+\sqrt{Z}^{\,(1)}K^{(1)}+K^{(1)}\sqrt{Z}^{\,(1)}=0\quad
\Rightarrow\quad
\sqrt{Z}^{\,(2)}=-\frac{1}{2}K^{(2)}+\frac{3}{8}{K^{(1)}}^2.
\end{displaymath}
   The matrix $\sqrt{Z}$ is, therefore, given by
\begin{align}
\sqrt{Z}
&=\begin{pmatrix}
1-\frac{1}{2}\delta^{(1)}_8+\frac{3}{8} {\delta^{(1)}_8}^2+\frac{3}{8} {\delta^{(1)}_{81}}^2-\frac{1}{2} \delta^{(2)}_8
&-\frac{1}{2}\delta^{(1)}_{81}+ \frac{3}{8} \delta^{(1)}_1\delta^{(1)}_{81}
+\frac{3}{8}\delta^{(1)}_8 \delta^{(1)}_{81}-\frac{1}{2} \delta^{(2)}_{81}\\
-\frac{1}{2}\delta^{(1)}_{81}+\frac{3}{8}\delta^{(1)}_1\delta^{(1)}_{81}+\frac{3}{8}\delta^{(1)}_8 \delta^{(1)}_{81}
-\frac{1}{2}\delta^{(2)}_{81}
&1-\frac{1}{2}\delta^{(1)}_1+\frac{3}{8}{\delta^{(1)}_1}^2+\frac{3}{8}{\delta^{(1)}_{81}}^2-\frac{1}{2}\delta^{(2)}_1
\end{pmatrix}.
\end{align}
   In terms of $\eta_C$, the Lagrangian reads
\begin{equation}
\label{Lmixing3}
\mathcal{L}_C=\frac{1}{2}\partial_{\mu}\eta^T_C\partial^{\mu}\eta_C-\frac{1}{2}\eta^T_C\mathcal{M}^2_C\eta_C,
\end{equation}
with the mass matrix given by
\begin{align}
\mathcal{M}_C^2=\sqrt{Z}^{\,T}\mathcal{M}_B^2 \sqrt{Z}\equiv
\begin{pmatrix}
\hat{M}^2_8& \hat{M}^2_{81}\\
\hat{M}^2_{81}&\hat{M}^2_1
\end{pmatrix}.
\end{align}
   Up to and including second order in the corrections $\delta^{(i)}_j$ and ${\Delta M^2_j}^{(i)}$,
the entries of the matrix $\mathcal{M}_C^2$ read
\begin{align}
\hat{M}^2_8
&=\,\stackrel{\circ}{M^2_8}\left(1-\delta^{(1)}_8+{\delta^{(1)}_8}^2+\frac{3}{4}{\delta^{(1)}_{81}}^2-\delta^{(2)}_8\right)
+{\Delta M_8^2}^{(1)}\left(1-\delta^{(1)}_8\right)+{{\Delta M}_8^2}^{(2)}\nonumber \\
&\quad+\!\stackrel{\circ}{M_{81}^2}\left(-\delta^{(1)}_{81} +\frac{3}{4}\delta^{(1)}_1 \delta^{(1)}_{81}+\frac{5}{4}\delta^{(1)}_8 \delta^{(1)}_{81}-\delta^{(2)}_{81}\right)
+{\Delta M_{81}^2}^{(1)}\left(-\delta^{(1)}_{81}\right) + \frac{1}{4}\left (M_0^2+\stackrel{\circ}{M_{1}^2}\right){\delta^{(1)}_{81}}^2,\\
\hat{M}^2_1
&=\left(M_0^2+\stackrel{\circ}{M^2_1}\right)\left(1-\delta^{(1)}_1+{\delta^{(1)}_1}^2+\frac{3}{4}{\delta^{(1)}_{81}}^2-\delta^{(2)}_1\right)
+{\Delta M_1^2}^{(1)}\left(1-\delta^{(1)}_1\right)+{{\Delta M}_1^2}^{(2)}\nonumber\\
&\quad+\!\stackrel{\circ}{M_{81}^2}\left(-\delta^{(1)}_{81}+\frac{3}{4}\delta^{(1)}_8\delta^{(1)}_{81}+\frac{5}{4}\delta^{(1)}_1\delta^{(1)}_{81}-\delta^{(2)}_{81}\right)
+{\Delta M_{81}^2}^{(1)}\left(-\delta^{(1)}_{81}\right)+\frac{1}{4}\stackrel{\circ}{M_{8}^2}{\delta^{(1)}_{81}}^2,\\
\hat{M}^2_{81}&=\,\stackrel{\circ}{M^2_{81}}\left(1-\frac{1}{2}\delta^{(1)}_1-\frac{1}{2}\delta^{(1)}_8
+\frac{3}{8}{\delta^{(1)}_1}^2+\frac{1}{4}\delta^{(1)}_1\delta^{(1)}_8+\frac{3}{8}{\delta^{(1)}_8}^2
+{\delta^{(1)}_{81}}^2-\frac{1}{2}\delta^{(2)}_1-\frac{1}{2}\delta^{(2)}_8\right)\nonumber\\
&\quad+{\Delta M_{81}^2}^{(1)}\left(1-\frac{1}{2}\delta^{(1)}_{1}-\frac{1}{2}\delta^{(1)}_{8}\right)+{\Delta M_{81}^2}^{(2)}\nonumber\\
&\quad+\!\stackrel{\circ}{M^2_{8}}\left(-\frac{1}{2}\delta^{(1)}_{81}+\frac{3}{8}\delta^{(1)}_1 \delta^{(1)}_{81}
+\frac{5}{8}\delta^{(1)}_8\delta^{(1)}_{81}-\delta^{(2)}_{81}\right)+{\Delta M_{8}^2}^{(1)}\left(-\frac{1}{2}\delta^{(1)}_{81}\right)\nonumber \\
&\quad+\left(M_0^2+ \stackrel{\circ}{M^2_{1}}\right)\left(-\frac{1}{2}\delta^{(1)}_{81}+\frac{3}{8}\delta^{(1)}_8\delta^{(1)}_{81}
+\frac{5}{8}\delta^{(1)}_1\delta^{(1)}_{81}-\delta^{(2)}_{81}\right) + {\Delta M_{1}^2}^{(1)}\left(-\frac{1}{2}\delta^{(1)}_{81}\right).
\end{align}
   Finally, to obtain the physical mass eigenstates, we diagonalize the matrix $\mathcal{M}_C^2$ by means of
an orthogonal transformation,
\begin{align}
\eta_D&=R\eta_C,\\
R&\equiv\begin{pmatrix}
\text{cos}\:\theta^{[2]} & -\text{sin}\:\theta^{[2]}\\
\text{sin}\:\theta^{[2]} & \text{cos}\:\theta^{[2]}
\end{pmatrix},
\end{align}
such that
\begin{equation}
\label{eq:mdiag}
R\mathcal{M}_C^2R^T=\mathcal{M}_D^2=\begin{pmatrix}M_\eta^2&0\\0&M_{\eta'}^2\end{pmatrix}.
\end{equation}
   The superscript $[2]$ refers to corrections up to and including second
order in the $\delta$ expansion.
   Introducing the nomenclature $\eta_P$ for the physical fields and ${\cal M}^2_P$ for the diagonal
mass matrix,
\begin{displaymath}
\eta_P=\eta_{D}=\begin{pmatrix}\eta\\ \eta'\end{pmatrix},
\quad {\cal M}_P^2=\begin{pmatrix}M_\eta^2&0\\0&M_{\eta'}^2\end{pmatrix},
\end{displaymath}
   the Lagrangian is now of the ``free-field'' type,
\begin{displaymath}
{\cal L}={\cal L}_D=\frac{1}{2}\partial_\mu \eta_P^T\partial^\mu \eta_P-\frac{1}{2}\eta_P^T{\cal M}_P^2 \eta_P
=\frac{1}{2}\partial_\mu\eta\partial^\mu\eta-\frac{1}{2}M^2_\eta \eta^2
+\frac{1}{2}\partial_\mu\eta'\partial^\mu\eta'-\frac{1}{2}M^2_{\eta'}{\eta'}^2.
\end{displaymath}
   Equation (\ref{eq:mdiag}) yields three relations,
\begin{align}
\label{eqs1}
\hat{M}^2_8&= M_{\eta}^2 \cos^2 \theta^{[2]} + M_{\eta'}^2 \sin^2 \theta^{[2]},\\
\label{eqs2}
\hat{M}^2_1&= M_{\eta}^2 \sin^2 \theta^{[2]}+ M_{\eta'}^2 \cos^2 \theta^{[2]},\\
\label{eqs3}
\hat{M}^2_{81}& =\left(M_{\eta'}^2 - M_{\eta}^2\right) \sin \theta^{[2]} \cos \theta^{[2]},
\end{align}
which {\it define} the mixing angle $\theta^{[2]}$ calculated up to and including $\mathcal{O}(\delta^2)$.
   First, from Eq.~(\ref{eqs3}) we infer
\begin{equation}
\label{angle}
\text{sin}\:2\theta^{[2]}=\frac{2\hat{M}^2_{81}}{M^2_{\eta'}-M^2_{\eta}}.
\end{equation}
   Adding Eqs.~(\ref{eqs1}) and (\ref{eqs2}), we obtain
\begin{equation}
\label{sum_squared_masses}
M^2_{\eta'}+M^2_{\eta}=\hat{M}^2_8+\hat{M}^2_1.
\end{equation}
   In the end, we subtract Eq.~(\ref{eqs2}) from Eq.~(\ref{eqs1}), take the square of the result,
add the square of 2 $\times$ Eq.~(\ref{eqs3}), and take the square root of the result to obtain
\begin{equation}
\label{difference_squared_masses}
M^2_{\eta'}-M^2_{\eta}=\sqrt{\left(\hat{M}^2_8-\hat{M}^2_1\right)^2+4\left(\hat{M}^2_{81}\right)^2}.
\end{equation}
   This equation implies that Eq.~(\ref{angle}) can also be written as
\begin{equation}
\label{angle2}
\text{sin}\:2\theta^{[2]}=\frac{2\hat{M}^2_{81}}{\sqrt{\left(\hat{M}^2_8-\hat{M}^2_1\right)^2+4\left(\hat{M}^2_{81}\right)^2}}.
\end{equation}

   The transformation from the octet fields $\eta_{A}$ to the physical fields $\eta_D$ can be summarized
as
\begin{equation}
\eta_A=T\eta_D=\left(1+\frac{1}{2}{\cal C}_A\Box\right)\sqrt{Z}R^T\eta_D,
\end{equation}
where the transformation matrix $T$ is given by
\begin{equation}
T=
\begin{pmatrix}
-A \sin \theta^{[2]} + B_8 \cos \theta^{[2]} & A  \cos \theta^{[2]} +B_8 \sin \theta^{[2]}\\
A\cos \theta^{[2]}-B_1\sin \theta^{[2]}  & A\sin \theta^{[2]} + B_1 \cos \theta^{[2]}
\end{pmatrix},
\label{TransfT}
\end{equation}
with
\begin{align}
A&=-\delta^{(1)}_{81}\left(\frac{1}{2}-\frac{3}{8}\delta^{(1)}_1-\frac{3}{8}\delta^{(1)}_{8}\right) -\frac{1}{2}\delta^{(2)}_{81}+\frac{c_{81}}{2} \Box,\\
B_i& = 1-\frac{1}{2}\delta_i^{(1)}+\frac{3}{8}{\delta_i^{(1)}}^2+\frac{3}{8}{\delta_{81}^{(1)}}^2-\frac{1}{2}\delta_i^{(2)}+\frac{c_i}{2} \Box.
\end{align}

\begin{figure}[htbp]
    \centering
        \includegraphics[width=0.8\textwidth]{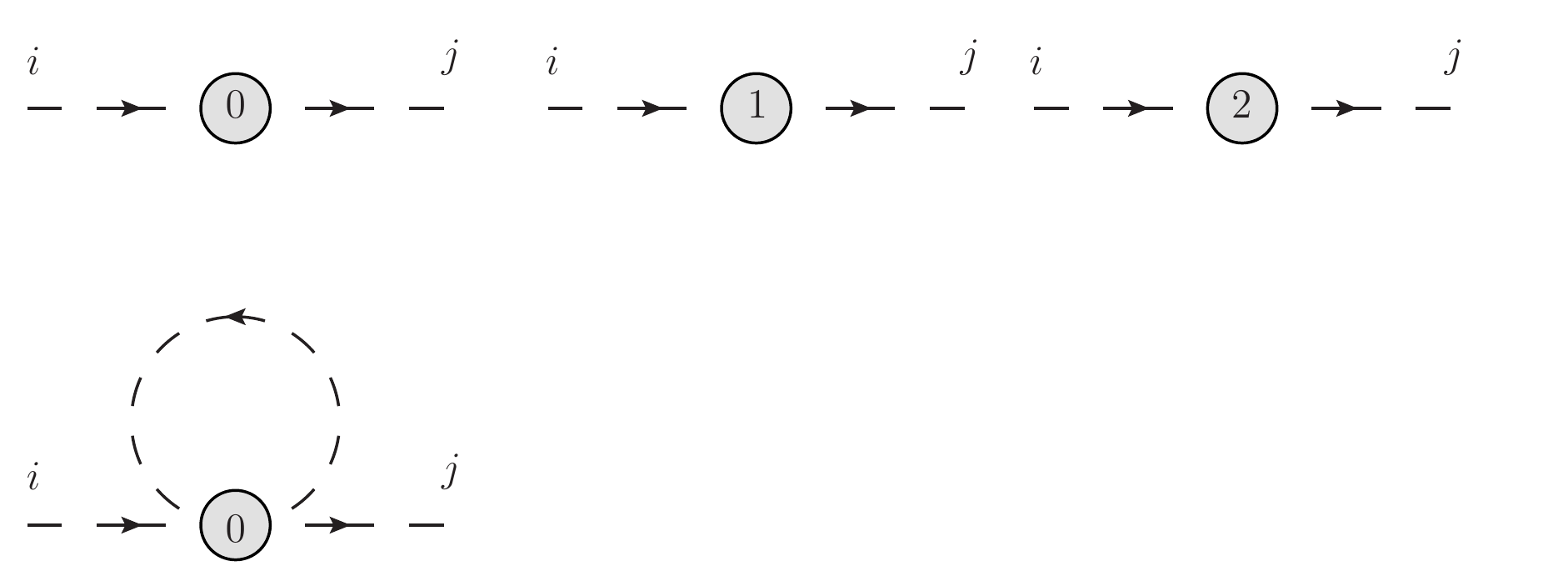}
        \caption{Self-energy diagrams up to and including $\mathcal{O}(\delta^2)$: Dashed lines refer to pseudoscalar mesons, and the numbers $k$ in the interaction blobs refer to vertices derived from the corresponding Lagrangians $\mathcal{L}^{(k)}$.}
    \label{fig:Mixingp}
\end{figure}

   Up to this point, the procedure for defining a mixing angle in terms of successive transformations is rather general.
   We now turn to a determination of the quantities $\delta_i^{(j)}$ as well as the $M_i^2$ terms within L$N_c$ChPT.
   To identify $\mathcal{K}_A$ and $\mathcal{M}^2_A$ at NNLO, we calculate the self-energy insertions
$-i\Sigma_{ij}(p^2)$, $(i,j=1,8)$ corresponding to the Feynman diagrams in Fig.~\ref{fig:Mixingp}.
   The Feynman rules are derived from the Lagrangians $\mathcal{L}^{(0)}$, $\mathcal{L}^{(1)}$, and $\mathcal{L}^{(2)}$
of Eqs.~(\ref{eq:lolagrangian}), (\ref{L1}), and (\ref{L2a})--(\ref{L2c}).
   The self energy calculated from the Lagrangian in Eq.~(\ref{eq:quadL}) takes the form\footnote{Since both the
singlet and the octet states are massless in the combined chiral and $N_c\to\infty$ limits, we consider the
lowest-order mass terms as part of the self-energy contributions.}
\begin{align}
\Sigma(p^2)=\left(\begin{matrix} \Sigma_{88}(p^2) & \Sigma_{81}(p^2) \\
                             \Sigma_{18}(p^2)  &\Sigma_{11}(p^2) \\
                                                       \end{matrix}\right),
\end{align}
where the $\Sigma_{ij}(p^2)$ are parametrized up to and including ${\cal O}(\delta^2)$ as
\begin{align}
\label{selfenergy88}
\Sigma_{88}(p^2)&=-\left(k_8+c_8p^2\right)p^2+M^2_8,\\
\label{selfenergy81}
\Sigma_{81}(p^2)&= \Sigma_{18}(p^2)=-\left(k_{81}+c_{81}p^2\right)p^2+M^2_{81},\\
\label{selfenergy11}
\Sigma_{11}(p^2)&=-\left(k_1+c_1p^2\right)p^2+M^2_1.
\end{align}
   We now obtain the elements of the kinematic matrix $\mathcal{K}_A$, the mass matrix $\mathcal{M}^2_A$, and the matrix $\mathcal{C}_A$
by comparing the results for the self energies calculated by means of the Feynman diagrams (Fig.~\ref{fig:Mixingp}) with the parametrization
given in Eqs.~(\ref{selfenergy88})--(\ref{selfenergy11}).

   The NLO contributions to the kinetic matrix read
\begin{align}
\label{delta1_8}
\delta^{(1)}_8&=\frac{8 \left(4 M_K^2-M_{\pi }^2\right)\text{\textit{$L$}}_5^{}}{3 F_{\pi }^2},\\
\label{delta1_1}
\delta^{(1)}_1&=\frac{8 \left(2 M_K^2+M_{\pi }^2\right) \text{\textit{$L$}}_5^{}}{3 F_{\pi }^2}+ \Lambda _1,\\
\label{delta1_81}
\delta^{(1)}_{81}&=-\frac{16 \sqrt{2} \left(M_K^2-M_{\pi }^2\right)\text{\textit{$L$}}_5^{}}{3 F_{\pi }^2},
\end{align}
where $M_{\pi}$, $M_K$, and $F_{\pi}$ denote the physical pion and kaon masses, and the physical pion-decay constant, respectively.
   The difference between using physical values instead of leading-order expressions in
Eqs.~(\ref{delta1_8})--(\ref{delta1_81}) is of NNLO and is compensated by an appropriate
 modification of the $\mathcal{O}(\delta^2)$ terms.
   The NNLO expressions for $M_{\pi}, M_K$, and $F_{\pi}$ are displayed in Appendix \ref{A:MF}.

   The entries of the mass matrix $\mathcal{M}^2_A$ are defined in Eqs.~(\ref{masses28})--(\ref{masses281})
in terms of leading-order, $\delta^1$, and $\delta^2$ pieces.
   The leading-order masses are given in Eqs.~(\ref{lomasses1})--(\ref{lomasses4}).
   In terms of the physical pion and kaon masses, and the physical pion-decay constant,
the first-order corrections read
\begin{align}
\label{deltaM28}
\Delta {M^2_8}^{(1)}&=\frac{16 \left(8 M_K^4-8 M_{\pi }^2 M_K^2+3 M_{\pi }^4\right)  \text{\textit{$L$}}_8^{}}{3 F_{\pi }^2},\\
\label{deltaM21}
\Delta {M^2_1}^{(1)}&=\frac{16 \left(4 M_K^4-4 M_{\pi }^2 M_K^2+3 M_{\pi}^4\right) \text{\textit{$L$}}_8^{}}{3 F_{\pi }^2}
+ \frac{2\Lambda _2}{3} \left(2 M_K^2+M_{\pi }^2\right),\\
\label{deltaM281}
\Delta {M^2_{81}}^{(1)}&=-\frac{64 \sqrt{2} \left(M_K^2-M_{\pi }^2\right)M_K^2 \text{\textit{$L$}}_8^{}}{3 F_{\pi}^2}- \frac{2\sqrt{2}\Lambda _2}{3}(M_K^2-M_{\pi }^2) .
\end{align}
   The corresponding NNLO expressions for the kinetic and mass matrix elements can be found in Appendix \ref{A:KM}.

\section{Decay constants}
\label{section_decay_constants}

   The decay constants of the $\eta$-$\eta'$ system are defined via the matrix element of the axial-vector-current
operator $A^a_\mu=\bar{q}\gamma_\mu\gamma_5\frac{\lambda^a}{2}q$,
\begin{align}
\bra{0}A^a_\mu(0)\ket{P(p)}=iF^a_P p_\mu,
\label{eq:f}
\end{align}
where $a=8,\ 0$ and $P=\eta,\ \eta'$.
   Since both mesons have octet and singlet components, Eq.~(\ref{eq:f}) defines four independent decay constants, $F^a_P$.
   We parametrize them according to the convention in \cite{Leutwyler:1997yr}
\begin{align}
\{F^a_P\}=\begin{pmatrix} F^8_\eta & F^0_\eta\\
                          F^8_{\eta'} & F^0_{\eta'}
                                                    \end{pmatrix}
=\begin{pmatrix}F_8\cos\theta_8 & -F_0\sin\theta_0 \\
                F_8\sin\theta_8 & F_0\cos\theta_0
    \end{pmatrix}.
\end{align}
   This parametrization is a popular way to define the $\eta$-$\eta'$ mixing within the so-called two-angle scheme
\cite{Feldmann:1998vh,Feldmann:1998sh,Benayoun:1999au,Escribano:2005qq,Escribano:2010wt,Escribano:2013kba,Escribano:2015nra,Escribano:2015yup}.
   The angles $\theta_8$ and $\theta_0$ and the constants $F_8$ and $F_0$ are given by
\begin{align}
\label{angles}
\tan\theta_8=\frac{F^8_{\eta'}}{F^8_{\eta}},\quad\quad
\tan\theta_0=-\frac{F^0_\eta}{F^0_{\eta'}},
\end{align}
\begin{align}
\label{F80}
F_8=\sqrt{\left(F^8_\eta\right)^2+\left(F^8_{\eta'}\right)^2},\quad\quad
F_0=\sqrt{\left(F^0_\eta\right)^2+\left(F^0_{\eta'}\right)^2}.
\end{align}

\begin{figure}[htbp]
    \centering
     \includegraphics[width=0.8\textwidth]{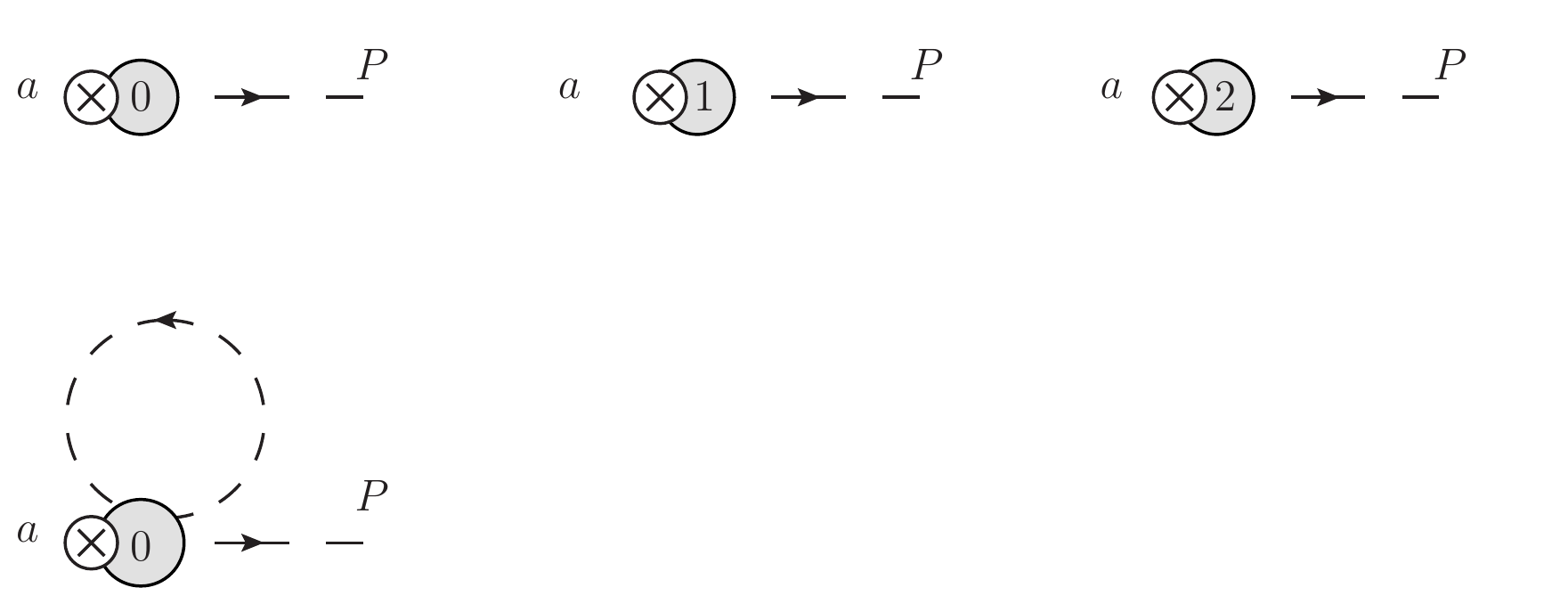}
    \caption{Feynman diagrams contributing to the calculation of the decay constants up to and including ${\cal O}(\delta^2)$. Dashed lines refer to pseudoscalar mesons, crossed dots refer to axial-vector sources, and the numbers $k$ in the interaction blobs refer to vertices derived from the Lagrangians $\mathcal{L}^{(k)}$ in Sec.~\ref{sec_lagrangian_power_counting}.}
    \label{fig:DecayConstp}
\end{figure}

To determine the decay constants $F^a_P$, we calculate the Feynman diagrams in Fig.~\ref{fig:DecayConstp}.
First, we calculate the coupling of the axial-vector current to the octet and singlet fields $\phi_b$, collected in the doublet $\eta_A$, at the one-loop level up to NNLO in the $\delta$ counting. The result, which should be interpreted as a Feynman rule, is represented by the ``matrix elements'' $\mathcal{F}_{ab}=\bra{0}A^a_\mu(0)\ket{b}$. In a next step, we transform the bare fields $\eta_A$ to the physical states using the transformation $T$ in Eq.~(\ref{TransfT}). The decay constants $F^a_P$ are then given by
\begin{align}
\left\{F^a_P\right\}^T=\begin{pmatrix} F^8_\eta & F^0_\eta\\
                          F^8_{\eta'} & F^0_{\eta'}
                                                    \end{pmatrix}^T=(\mathcal{F}\cdot T).
\end{align}
At leading order, the decay constants read
\begin{align}
F_\eta^8 &= F_{\eta'}^0 = F \cos \theta^{[0]},\\
-F_\eta^0 &= F_{\eta'}^8 = F \sin \theta^{[0]},
\end{align}
in terms of the leading-order mixing angle $\theta^{[0]}$ given in Eq.~(\ref{angle}). Equation (\ref{angles}) then yields $\theta_0 =\theta_8 = \theta^{[0]}$. The NLO decay constants are given by
\begin{align}\label{anglesNLO}
F_\eta^8/F &= \left(1+ \frac{1}{2}\delta_8^{(1)}\right)\cos \theta^{[1]} - \frac{1}{2}\delta_{81}^{(1)} \sin \theta^{[1]}, \\
F_\eta^0/F &= -\left(1+ \frac{1}{2}\delta_1^{(1)}\right)\sin \theta^{[1]} + \frac{1}{2}\delta_{81}^{(1)} \cos \theta^{[1]}, \\
F_{\eta'}^8/F &= \left(1+\frac{1}{2} \delta_8^{(1)}\right)\sin \theta^{[1]} + \frac{1}{2}\delta_{81}^{(1)} \cos \theta^{[1]}, \\
F_{\eta'}^0/F &= \left(1+ \frac{1}{2}\delta_1^{(1)}\right)\cos \theta^{[1]} + \frac{1}{2}\delta_{81}^{(1)}\sin \theta^{[1]},
\end{align}
now in terms of the NLO mixing angle $\theta^{[1]}$.
Using Eqs.~(\ref{angles}) and (\ref{F80}), one obtains
\begin{align}
 F_8 &= F\left(1+ \frac{\delta_8^{(1)}}{2}\right),\no\\
F_0 &= F\left(1+ \frac{\delta_1^{(1)}}{2}\right),
\label{fNLO}
\end{align}
and
\begin{align}
\theta_8&=\theta^{[1]} + \arctan\left(\frac{\delta_{81}^{(1)}}{2}\right),\no\\
\theta_0&=\theta^{[1]} - \arctan\left(\frac{\delta_{81}^{(1)}}{2}\right).
\label{theta80}
\end{align}
The results for the decay constants at NNLO are lengthy and are given in Appendix \ref{A:MF}.

\section{Numerical analysis}
\label{section_numerical_analysis}

In the following, we perform the numerical evaluation of the mixing angle, the masses of the pseudoscalar mesons, and their decay constants. We present the results in a systematic way, order by order.

\subsection{LO}
At leading order, the mixing angle is given by Eq.~(\ref{angle2})  which reduces to
\begin{align}
\sin 2\theta^{[0]}=\frac{-4 \sqrt{2}\left(M_K^2-M_{\pi }^2\right)}{\sqrt{12 M_0^2 \left(M_{\pi }^2-M_K^2\right)+36
   \left(M_K^2-M_{\pi }^2\right){}^2+9 M_0^4}}.
\label{LOangle}
\end{align}
This equation is well suited to study the two limits, the flavor-symmetric case, i.e., $M^2_\pi=M^2_K$, and the limit $N_c\to\infty$. In the flavor-symmetric limit, the mixing angle vanishes, $\theta^{[0]}=0$. On the other hand, in the L$N_c$ limit, the $\text{U}(1)_A$ contribution to  the $\eta'$ mass vanishes, i.e., $M^2_0=0$, and the mixing angle becomes independent of the pseudoscalar masses
\begin{align}
\sin 2\theta^{[0]}=-\frac{2 \sqrt{2}}{3},
\end{align}
which yields $\theta^{[0]}=-35.3\degree$.
We then turn to the physical case. Employing Eqs.~(\ref{sum_squared_masses}) and (\ref{difference_squared_masses}), we fix $M^2_0$ to the physical $M^2_{\eta'}$ mass
\begin{align}
M^2_0=\frac{3 \left(M_{\text{$\eta'$}}^2-M_{\pi }^2\right) \left(2 M_K^2-M_{\text{$\eta'$}}^2-M_{\pi }^2\right)}{4 M_K^2-3 M_{\text{$\eta'$}}^2-M_{\pi }^2},
\end{align}
and obtain
\begin{align}
\sin 2\theta^{[0]}=-\frac{4 \sqrt{2} \left(M_K^2-M_{\pi }^2\right) \left(-4 M_K^2+3 M_{\text{$\eta'$}}^2+M_{\pi }^2\right)}{3 \left[-8 M_K^2 \left(M_{\text{$\eta'$}}^2+M_{\pi
   }^2\right)+8 M_K^4+3 M_{\text{$\eta'$}}^4+2 M_{\pi }^2 M_{\text{$\eta'$}}^2+3
   M_{\pi }^4\right]}.
\end{align}
Evaluating these results for physical masses $M^2_\pi$, $M^2_K$, and $M^2_{\eta'}$ yields
\begin{align}
\theta^{[0]}=-19.6\degree\quad\text{and}\quad
M_0=0.820\ \text{GeV}.
\end{align}

\subsection{NLO}\label{sec:mixNLO}

At NLO, still only tree diagrams contribute, since loop contributions are relegated to NNLO. Beyond $F$, $B\hat{m}$, $Bm_s$, and $\tau$, the four NLO LECs $L_5$, $L_8$, $\Lambda_1$, $\Lambda_2$ appear and need to be fixed.
Since there are, at present, no values for all of the NLO
LECs in U(3) ChPT available in the literature,
we follow two different strategies to fix the coupling constants:

\begin{enumerate}
    \item We design a compact system of observables calculated within our framework of L$N_c$ChPT and determine the LECs by fixing them to the physical values of the observables. Our set of observables consists of $M^2_\pi$, $M^2_K$, $F_K/F_\pi$, $M^2_\eta$, $M^2_{\eta'}$. In addition, we need the quark mass ratio $m_s/\hat{m}$, which we take from Ref.~\cite{Aoki:2013ldr}. The experimental values for the masses and decay constants are taken from Ref.~\cite{Agashe:2014kda}, reading
    \begin{align}
    M_\pi&=0.135\ \text{GeV},\quad M_K=0.494\ \text{GeV},\quad M_\eta=0.548\ \text{GeV},\no\\
    M_{\eta'}&=0.958\ \text{GeV},\quad F_\pi=0.0922(1)\ \text{GeV},\quad F_K/F_\pi=1.198(6).
    \end{align}
\item We use phenomenological determinations of some constants obtained in SU(3) ChPT, for example, Tab.~1 from Ref.~\cite{Bijnens:2014lea}.
\end{enumerate}
We start with the first strategy and
begin by fixing $M^2_0$ to the physical $M^2_{\eta'}$ using the relation
\begin{align}
\left(2M^2_{\eta'}-\hat{M}^2_8-\hat{M}^2_1\right)^2=\left(\hat{M}^2_8-\hat{M}^2_1\right)^2+4\left(\hat{M}^{2}_{81}\right)^2,
\end{align}
which follows from Eqs.~(\ref{sum_squared_masses}) and (\ref{difference_squared_masses}). After expressing $M^2_0$ in terms of $M^2_{\eta'}$, the parameters $\Lambda_1$ and $\Lambda_2$ appear only in the QCD-scale-invariant combination $\tilde{\Lambda}=\Lambda_1-2\Lambda_2$ \cite{Kaiser:2000gs} in the expressions for our observables and the mixing angle. Using the ratio $m_s/\hat{m}=27.5$ from Ref.~\cite{Aoki:2013ldr}, the parameters $B\hat{m}$, $L_5$, $L_8$, $\tilde{\Lambda}$ can be unambiguously obtained from the NLO relations for the physical values of $M^2_\pi$, $M^2_K$, $F_K/F_\pi$, $M^2_\eta$, given in Appendix \ref{A:MF}. The results for the LECs are shown in Tab.~\ref{tab:LECNLO} labeled NLO I. Notice that at this order no EFT-scale dependence is introduced yet, so these LECs are scale independent. We also display errors for all calculated quantities. These errors are only due to the input errors.
We do not give estimates for the errors due to neglecting higher orders or particular assumptions of our models.
As input errors we consider the errors of $F_K/F_\pi$, $F_\pi$, $m_s/\hat{m}$ and later, when we make use of LECs determined in SU(3) ChPT \cite{Bijnens:2014lea}, we also take their errors into account.

Once the set of LECs is determined, we can evaluate the LO pseudoscalar masses, the $\eta$-$\eta'$ mixing angle, and the pseudoscalar decay constants. For the calculation of the parameters $\theta_8$, $\theta_0$, $F_8$, $F_0$, we use the simplified formula at NLO given in Eqs.~(\ref{fNLO}) and (\ref{theta80}). The quantities $M^2_0$ and $F_0$ depend on the QCD-renormalization scale \cite{Kaiser:2000gs}.
Therefore, we can only provide the QCD-scale-invariant quantities $M^2_0/(1+\Lambda_1)$ and $F_0/(1+\Lambda_1/2)$. We are not able to extract a value for $\Lambda_1$ from our observables, since physical observables do not depend on the QCD scale and we can only determine the invariant combination $\tilde{\Lambda}=\Lambda_1-2\Lambda_2$.
The expressions for $M^2_0/(1+\Lambda_1)$ and $F_0/(1+\Lambda_1/2)$ are expanded up to NLO yielding results which depend on $\Lambda_1$ only through $\tilde{\Lambda}$.
Table~\ref{tab:massNLO} shows the leading-order masses $\stackrel{\circ}{M^2_\pi}$, $\stackrel{\circ}{M^2_K}$, $M^2_0/(1+\Lambda_1)$, and $M^2_\eta$ for $\tilde{\Lambda}=0$. The mixing angle $\theta^{[1]}$, the angles $\theta_8$, $\theta_0$ and the constants $F_8$, $F_0/(1+\Lambda_1/2)$ are shown in Tab.~\ref{tab:angNLO}, again under the label NLO I.

The second scenario uses values for the LECs determined phenomenologically in the framework of SU(3) ChPT. Since our calculations are performed in U(3) ChPT, we apply the appropriate matching between the two EFTs \cite{Kaiser:2000gs,HerreraSiklody:1998cr} when we make use of SU(3) determinations. We set the matching scale of the two theories to be $\mu_0=M_0=0.85\ \text{GeV}$, which is basically the value of $M_{\eta'}$ in the chiral limit: $M^2_0=6\tau/(F^2(1+\Lambda_1))$. Since SU(3) ChPT contains one-loop corrections already at NLO, the LECs depend on the scale of the effective theory $\mu$. The SU(3) LECs are typically provided at $\mu_1=0.77$ GeV. To study the scale dependence of our results, we evaluate them at $\mu=0.77\ \text{GeV}$ and at $\mu=1\ \text{GeV}$, which is the scale of $M_{\eta'}$. Combining the matching at $\mu_0$ and the running from $\mu_1$ to $\mu$ results in \cite{Kaiser:2000gs,HerreraSiklody:1998cr}:
\begin{align}
&L^r_5(\mu)=L^{\text{SU}_3,r}_5(\mu_1)+\frac{3}{8}\frac{1}{16\pi^2}\ln\left(\frac{\mu_1}{\mu}\right),\no\\
&L^r_8(\mu)=L^{\text{SU}_3,r}_8(\mu_1)+\frac{5}{48}\frac{1}{16\pi^2}\ln\left(\frac{\mu_1}{\mu}\right)+\frac{1}{12}\frac{1}{16\pi^2}\ln\left(\frac{\mu_0}{\mu}\right),\no\\
&L^r_4(\mu)=L^{\text{SU}_3,r}_4(\mu_1)+\frac{1}{8}\frac{1}{16\pi^2}\ln\left(\frac{\mu_1}{\mu}\right),\no\\
&L^r_6(\mu)=L^{\text{SU}_3,r}_6(\mu_1)+\frac{11}{144}\frac{1}{16\pi^2}\ln\left(\frac{\mu_1}{\mu}\right)+\frac{1}{72}\frac{1}{16\pi^2}\left(\frac{1}{2}-\ln\left(\frac{\mu_0}{\mu}\right)\right),\no\\
&L^r_7(\mu)=L^{\text{SU}_3,r}_7+\frac{F^4(1+\Lambda_2)^2}{288\tau}\no,\\
&L^r_{18}(\mu)=L^{r}_{18}(\mu_2)-\frac{1}{4}\frac{1}{16\pi^2}\ln\left(\frac{\mu_2}{\mu}\right).
\label{eq:runningLECs}
\end{align}
The constant $L_{18}$ does not appear in SU(3) ChPT, but we include its running for completeness, since the running from the scale $\mu_2=1\ \text{GeV}$ will be needed later.

The LO quantities $\stackrel{\circ}{M^2_\pi}$, $\stackrel{\circ}{M^2_K}$, $F$ are expressed in terms of the physical quantities $M^2_\pi$, $M^2_K$, $F_\pi$, and, again, $M^2_0$ is determined from the relation to $M^2_{\eta'}$ at this order. The parameters $\theta_8$, $\theta_0$, $F_8$, $F_0$ are calculated using Eqs.~(\ref{fNLO}) and (\ref{theta80}). For the LECs $L_5$ and $L_8$ we use the values determined at $\mathcal{O}(p^4)$ in SU(3) ChPT, i.e., column ``$p^4$ fit'' in Tab.~1 in Ref.~\cite{Bijnens:2014lea}. The OZI-rule-violating parameter $\tilde{\Lambda}$ is fixed to $M^2_\eta$. The results are given in Tabs.~\ref{tab:LECNLO}--\ref{tab:angNLO} labeled NLO II. The dependence of $M^2_\eta$ on $\tilde{\Lambda}$ is shown in Fig.~\ref{fig:mEtax}.

\begin{table}[tbp]
    \centering
        \tabcolsep0.5em\renewcommand{\arraystretch}{1.2}\begin{tabular}{l c r@{$\,\pm\,$}l r@{$\,\pm\,$}l r@{$\,\pm\,$}l}
        \hline\hline
   &$\mu\ [\text{GeV}]$ & \multicolumn{2}{c}{$L_5\ [10^{-3}]$} &  \multicolumn{2}{c}{$L_8\ [10^{-3}]$} &  \multicolumn{2}{c}{$\tilde{\Lambda}$}\\ \hline
           \text{NLO I} & - & $1.86$&$0.06$ & $0.78$&$0.05$ & $-0.34$&$0.05$ \\
 \text{NLO II} & $0.77$ & $1.20$&$0.10$ & $0.55$&$0.20$ & $0.02$&$0.13$ \\
 \text{NLO II} & $1$ & $0.58$&$0.10$ & $0.24$&$0.20$ & $0.41$&$0.13$ \\
        \hline\hline
        \end{tabular}
    \caption{LECs at NLO.
        }
    \label{tab:LECNLO}
\end{table}

\begin{table}[tbp]
    \centering
       \tabcolsep0.5em\renewcommand{\arraystretch}{1.2}\begin{tabular}{l c r@{$\,\pm\,$}l r@{$\,\pm\,$}l r@{$\,\pm\,$}l r@{$\,\pm\,$}l}
        \hline\hline
        &$\mu\ [\text{GeV}]$ &  \multicolumn{2}{c}{$\stackrel{\circ}{M^2_\pi}$} &  \multicolumn{2}{c}{$\stackrel{\circ}{M^2_K}$} &  \multicolumn{2}{c}{$\frac{M^2_0}{(1+\Lambda_1)}$} &  \multicolumn{2}{c}{$M^2_\eta(\tilde{\Lambda}=0)$} \\ \hline
         \text{NLO I} & - & $0.018$&$0.000$ & $0.261$&$0.005$ & $0.902$&$0.013$ & $0.326$&$0.003$ \\
 \text{NLO II} & $0.77$ & $0.018$&$0.000$ & $0.249$&$0.023$ & $0.871$&$0.061$ & $0.299$&$0.010$ \\
 \text{NLO II} & $1$ & $0.018$&$0.000$ & $0.249$&$0.023$ & $0.871$&$0.061$ & $0.269$&$0.010$ \\
\hline\hline
        \end{tabular}
    \caption{Pseudoscalar masses at NLO in $\text{GeV}^2$.}
    \label{tab:massNLO}
\end{table}

\begin{table}[tbp]
    \centering
        \tabcolsep0.5em\renewcommand{\arraystretch}{1.2}\begin{tabular}{l c r@{$\,\pm\,$}l r@{$\,\pm\,$}l r@{$\,\pm\,$}l r@{$\,\pm\,$}l r@{$\,\pm\,$}l}
      \hline\hline
         & $\mu\ [\text{GeV}]$& \multicolumn{2}{c}{$\theta\ [^{\circ}]$} & \multicolumn{2}{c}{$\theta_8\ [^{\circ}]$}& \multicolumn{2}{c}{$\theta_0\ [^{\circ}]$}&\multicolumn{2}{c}{$F_8/F_\pi$}&\multicolumn{2}{c}{$\frac{F_0}{1+\Lambda_1/2}/F_\pi$}\\ \hline
 \text{NLO I} & - & $-11.1$&$0.6$ & $-21.7$&$0.7$ & $-0.5$&$0.7$ & $1.26$&$0.01$ & $1.13$&$0.00$ \\
 \text{NLO II} & $0.77$ & $-12.6$&$3.0$ & $-19.5$&$3.0$ & $-5.7$&$3.2$ & $1.17$&$0.01$ & $1.09$&$0.01$ \\
 \text{NLO II} & $1$ & $-12.6$&$3.0$ & $-15.9$&$3.0$ & $-9.3$&$3.2$ & $1.08$&$0.01$ & $1.04$&$0.01$ \\
 \hline\hline
        \end{tabular}
    \caption{Mixing angles and decay constants at NLO.}
    \label{tab:angNLO}
\end{table}

\subsection{NLO+Loops}

Before considering the full NNLO corrections, we first discuss the case where we just add the loop contributions to the NLO expressions. Since the loop corrections do not contain any unknown parameters, we can use exactly the same system of equations from the NLO I scenario in the previous section to obtain the desired LECs. We augment the system of linear equations with the one-loop corrections and extract the values of $B\hat{m}$, $L_5$, $L_8$, $\tilde{\Lambda}$. The results depend now on the scale of the effective theory and we choose to extract the LECs at $\mu=1\ \text{GeV}$. The parameters $\theta_8$, $\theta_0$, $F_8$, $F_0$ are obtained from Eqs.~(\ref{eq:f8nlo2})--(\ref{eq:theta0nlo2}) in Appendix \ref{A:MF}, now also including the one-loop corrections. The results are given in Tabs.~\ref{tab:LECNLOLp}--\ref{tab:angNLOLp}
labeled NLO+Lps I.

\begin{figure}[tbp]
    \centering
        \includegraphics[width=0.7\textwidth]{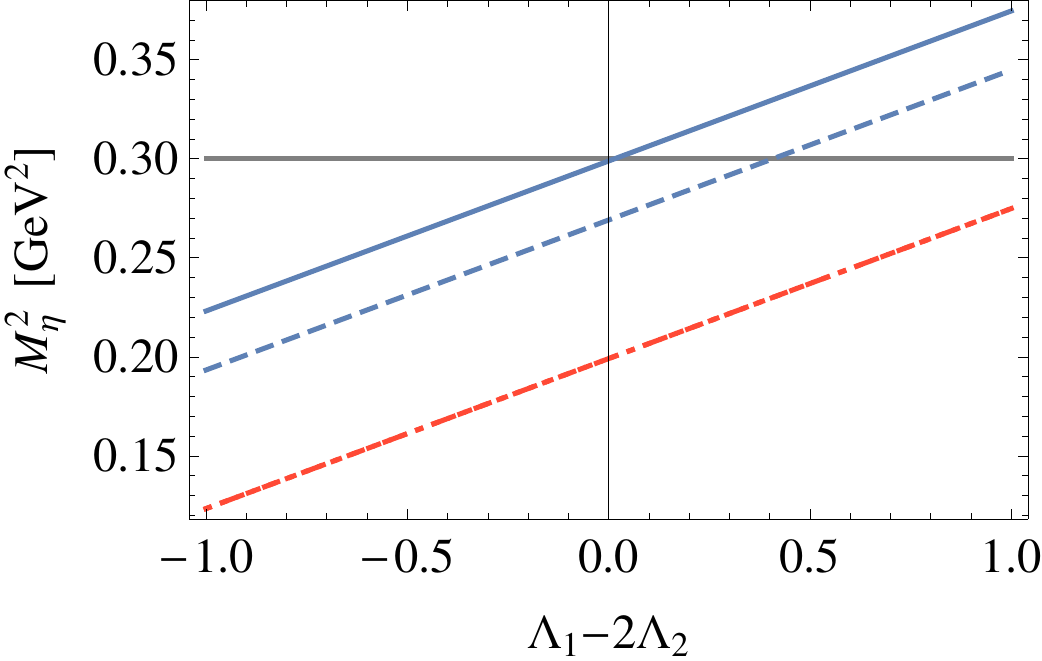}
    \caption{(Color online) $M^2_\eta$ as a function of $\tilde{\Lambda}=\Lambda_1-2\Lambda_2$. Solid (blue) line: NLO II at 0.77 GeV; dashed (blue) line: NLO II at 1 GeV. Dotdashed (red) line: NLO+Loop II at 0.77 GeV; dotted (red) line: NLO+Loop II at 1 GeV. The dotdashed and dotted lines coincide. Horizontal line: Physical value.}
    \label{fig:mEtax}
\end{figure}

We compare the results with the values obtained in SU(3) ChPT. For $L_5$ and $L_8$ we use the same values as in the NLO II case. To compensate the scale dependence of the loop contributions we include the scale-dependent parts of the LECs $L_4$, $L_6$, $L_7$, $L_{18}$ [see Eqs.~(\ref{eq:runningLECs})],
which would appear only at NNLO. These constants are included without the SU(3)-U(3) matching and we choose $L^r_4=L^r_6=L^r_7=L^r_{18}=0$ at $\mu_1=1\ \text{GeV}$. Eventually, we again use $M^2_\eta$ to extract $\tilde{\Lambda}$. Equations (\ref{eq:f8nlo2})--(\ref{eq:theta0nlo2}) in Appendix \ref{A:MF} provide then our values for $\theta_8$, $\theta_0$, $F_8$, $F_0$. The results can be found in Tabs.~\ref{tab:LECNLOLp}--\ref{tab:angNLOLp}, denoted by NLO+Lps II. Figure \ref{fig:mEtax} shows the dependence of $M^2_\eta$ on $\tilde{\Lambda}$ for the different scenarios discussed so far.
We notice that the dependence is quite strong.
After the inclusion of the loops and the scale-dependent parts of the $1/N_c$-suppressed $L_i$, $M^2_\eta$ is independent of the renormalization scale $\mu$ (compare solid and dashed red lines).

\begin{table}[tbp]
    \centering
        \tabcolsep0.5em\renewcommand{\arraystretch}{1.2}\begin{tabular}{l c r@{$\,\pm\,$}l r@{$\,\pm\,$}l r@{$\,\pm\,$}l}
        \hline\hline
        &$\mu\ [\text{GeV}]$ & \multicolumn{2}{c}{$L_5\ [10^{-3}]$} &  \multicolumn{2}{c}{$L_8\ [10^{-3}]$} &  \multicolumn{2}{c}{$\tilde{\Lambda}$}\\ \hline
\text{NLO+Lps I} & $0.77$ & $1.37$&$0.06$ & $0.85$&$0.05$ & $0.52$&$0.05$ \\
 \text{NLO+Lps I} & $1$ & $0.75$&$0.06$ & $0.55$&$0.05$ & $1.09$&$0.04$ \\
 \text{NLO+Lps II} & $0.77$ & $1.20$&$0.10$ & $0.55$&$0.20$ & $1.34$&$0.13$ \\
 \text{NLO+Lps II} & $1$ & $0.58$&$0.10$ & $0.24$&$0.20$ & $1.34$&$0.13$\\
        \hline\hline
        \end{tabular}
    \caption{LECs at NLO with loops added.
        }
    \label{tab:LECNLOLp}
\end{table}

\begin{table}[tbp]
    \centering
        \tabcolsep0.5em\renewcommand{\arraystretch}{1.2}\begin{tabular}{l c r@{$\,\pm\,$}l r@{$\,\pm\,$}l r@{$\,\pm\,$}l r@{$\,\pm\,$}l}
       \hline\hline
        &$\mu\ [\text{GeV}]$ &  \multicolumn{2}{c}{$\stackrel{\circ}{M^2_\pi}$} &  \multicolumn{2}{c}{$\stackrel{\circ}{M^2_K}$} &  \multicolumn{2}{c}{$\frac{M^2_0}{(1+\Lambda_1)}$} &  \multicolumn{2}{c}{$M^2_\eta(\tilde{\Lambda}=0)$} \\ \hline
       \text{NLO+Lps I} & 0.77 & $0.018$&$0.000$ & $0.263$&$0.005$ & $0.927$&$0.013$ & $0.261$&$0.003$ \\
 \text{NLO+Lps I} & 1 & $0.017$&$0.000$ & $0.240$&$0.005$ & $0.867$&$0.012$ & $0.218$&$0.003$ \\
 \text{NLO+Lps II} & 0.77 & $0.019$&$0.000$ & $0.287$&$0.023$ & $0.933$&$0.061$ & $0.199$&$0.010$ \\
 \text{NLO+Lps II} & 1 & $0.017$&$0.000$ & $0.265$&$0.023$ & $0.933$&$0.061$ & $0.199$&$0.010$ \\
\hline\hline
        \end{tabular}
    \caption{Pseudoscalar masses at NLO with loops added in $\text{GeV}^2$.}
    \label{tab:massNLOLp}
\end{table}

\begin{table}[tbp]
    \centering
        \tabcolsep0.5em\renewcommand{\arraystretch}{1.2}\begin{tabular}{l c r@{$\,\pm\,$}l r@{$\,\pm\,$}l r@{$\,\pm\,$}l r@{$\,\pm\,$}l r@{$\,\pm\,$}l}
        \hline\hline
         & $\mu\ [\text{GeV}]$& \multicolumn{2}{c}{$\theta\ [^{\circ}]$} & \multicolumn{2}{c}{$\theta_8\ [^{\circ}]$}& \multicolumn{2}{c}{$\theta_0\ [^{\circ}]$}&\multicolumn{2}{c}{$F_8/F_\pi$}&\multicolumn{2}{c}{$\frac{F_0}{1+\Lambda_1/2}/F_\pi$}\\\hline
 \text{NLO+Lps I} & 0.77 & $-10.2$&$0.6$ & $-18.0$&$0.7$ & $-2.4$&$0.7$ & $1.31$&$0.01$ & $0.97$&$0.00$ \\
 \text{NLO+Lps I} & 1 & $-13.4$&$0.6$ & $-17.7$&$0.7$ & $-9.1$&$0.7$ & $1.31$&$0.01$ & $0.87$&$0.00$ \\
 \text{NLO+Lps II} & 0.77 & $-10.2$&$2.9$ & $-13.5$&$2.9$ & $-6.8$&$3.1$ & $1.28$&$0.01$ & $0.86$&$0.01$ \\
 \text{NLO+Lps II} & 1 & $-10.2$&$2.9$ & $-13.5$&$2.9$ & $-6.8$&$3.1$ & $1.28$&$0.01$ & $0.86$&$0.01$ \\
  \hline\hline
        \end{tabular}
    \caption{Mixing angles and decay constants at NLO with loops added.}
    \label{tab:angNLOLp}
\end{table}

\subsection{NNLO}

\begin{figure}[tbp]
    \centering
        \includegraphics[width=0.7\textwidth]{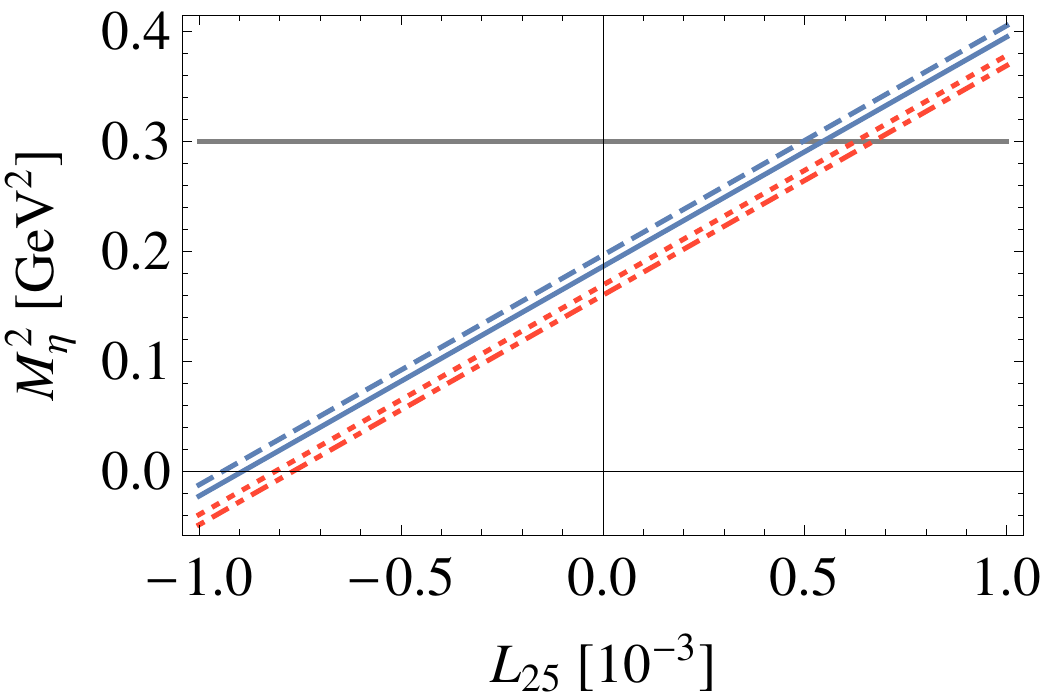}
    \caption{(Color online) $M^2_\eta$ as a function of $L_{25}$. Solid (blue) line: NNLO without $C_i$ at 0.77 GeV; dashed (blue) line: NNLO without $C_i$ at 1 GeV. Dotdashed (red) line: NNLO with $C_i$ at 0.77 GeV; dotted (red) line: NNLO with $C_i$ at 1 GeV. Horizontal line: Physical value.}
    \label{fig:mEtaL25}
\end{figure}

At NNLO, there are too many unknown LECs, which cannot be determined from our chosen set of observables. This means that it is not possible to consistently determine all LECs appearing at NNLO within our framework of L$N_c$ChPT. So we can only employ the second strategy and make use of phenomenological determinations of the LECs $L_i$ and $C_i$ in SU(3) ChPT.
We are then left with five completely unknown LECs, $\Lambda_1$, $\Lambda_2$, $L_{18}$, $L_{25}$, $v^{(2)}_2$, and the combination $L_{46}+L_{53}$, which are related to the singlet field.
First, we investigate the case with $C_i=0$. We match the $L_i$ from SU(3) to U(3), according to Eq.~(\ref{eq:runningLECs}), and take their values from the column ``$p^4$ fit'' in Tab.~1 in Ref.~\cite{Bijnens:2014lea}. Since a NNLO calculation in the $\delta$ counting includes contributions of the type NLO$\times$NLO, e.g., products of $L_i$, the results depend on the EFT scale $\mu$. We display results for two different scales, $\mu=0.77\ \text{GeV}$ and $\mu=1\ \text{GeV}$. We choose $\Lambda_1=\Lambda_2=L^r_{18}=v^{(2)}_2=L_{46}=L_{53}=0$ at $\mu_2=1\ \text{GeV}$, which, together with the U(3)-SU(3) matching, results in $L^r_7\approx0$ (at $\mu=1\ \text{GeV}$). We can then fix one OZI-rule-violating LEC, which we choose to be $L_{25}$, to the physical value of $M^2_\eta$. In this way, $L_{25}$ accounts for the contributions to $M^2_\eta$ of all other OZI-rule-violating LECs, which are put to zero.
 At NNLO including $C_{12}$ terms, the simplified expressions for $\theta_8$, $\theta_0$, $F_8$, $F_0$ in Eqs.~(\ref{fNLO}) and (\ref{theta80}) no longer hold. We therefore use the general formulae in Eqs.~(\ref{angles}) and (\ref{F80}) to calculate the parameters of the two-angle scheme in the NNLO scenarios.
The results are given in Tabs.~\ref{tab:LECNNLO}--\ref{tab:angNNLO} labeled NNLO w/o Ci. Figure \ref{fig:mEtaL25} shows $M^2_\eta$ as a function of $L_{25}$.

Finally, we include the contributions of the $C_i$. The $L_i$ are treated as before in terms of running and matching, but now we use the $\mathcal{O}(p^6)$ values from Ref.~\cite{Bijnens:2014lea}, i.e., column ``BE14'' in Tab.~3. For the $C_i$ we employ the values from Tab.~4 in Ref.~\cite{Bijnens:2014lea}. In order to obtain values for the $C_i$ in U(3) ChPT, we employ the tree-level-matching relations between SU(3) and U(3) ChPT, given by \cite{Kaiser:2007zz}
\begin{align}
C_{19}&=C^{\text{SU}_3}_{19}+\frac{1}{3}\frac{F^2}{48 M^4_0},\no\\
C_{31}&=C^{\text{SU}_3}_{31}+\frac{F^2}{48 M^4_0},
\end{align}
where we take $F=F_\pi$ and the LO value $M^2_0=0.673\ \text{GeV}^2$.
We do not consider the matching at the loop level, because this is a correction beyond the accuracy of our calculation. We also do not include the dependence of the $C_i$ on the EFT renormalization scale, since this would be introduced only by two-loop effects, which are again higher-order contributions beyond our accuracy. The SU(3) values of the $C_i$ are provided without errors. They are also not very well constrained in Ref.~\cite{Bijnens:2014lea} and might be only suited for the SU(3) observables studied in this reference. Therefore, we assume an error of 50\% on the SU(3) values and propagate it to our results.
The dependence of $M^2_\eta$ on $L_{25}$ is shown in Fig. \ref{fig:mEtaL25}, and eventually $L_{25}$ is fixed to the physical value of $M^2_\eta$. The results are given in Tabs.~\ref{tab:LECNNLO}--\ref{tab:angNNLO} labeled NNLO w/ Ci.

Another source for the $L_i$ and $C_i$ is provided in Ref.~\cite{Jiang:2015dba}, where the LECs are computed in a chiral quark model. Since the LECs are calculated in the L$N_c$ limit, loop effects are not included, and the LECs do not depend on the EFT renormalization scale. Thus, to obtain values for the LECs in U(3) ChPT, we consider only the tree-level SU(3)-U(3) matching relations (for $L_7$, $C_{19}$, $C_{31}$). Further, we do not take the running of the LECs with the EFT scale into account.
The one-loop contributions are evaluated at $\mu=0.77\ \text{GeV}$ and $\mu=1\ \text{GeV}$. The OZI-rule-violating couplings are treated as in the other NNLO scenarios described above.
The results are provided in Tabs.~\ref{tab:LECNNLO}--\ref{tab:angNNLO} labeled NNLO w/ Ci J, where the errors are obtained from the errors of the $L_i$ and $C_i$ given in Ref.~\cite{Jiang:2015dba}.

Figure \ref{fig:mEtaL25} shows a strong dependence of $M^2_\eta$ on $L_{25}$. The renormalization-scale dependence is now much smaller than in the NLO cases. The small residual scale dependence stems from products of $L_5$ and $L_8$, whose scale dependence would be compensated by products of one-loop terms in the full two-loop calculation.
The inclusion of the one-loop corrections decreases the value of $M^2_\eta(L_{25}=0)$ by about $30\%$. This would rather match the expected order of magnitude of a NLO correction.
Taking the $C_i$ into account, further decreases $M^2_\eta(L_{25}=0)$.
According to the $\delta$ counting, we would expect the value for $L_{25}$ to be of the same order of magnitude as $L_5$ and $L_8$, since the operator structure is similar, with an additional $1/N_c$ suppression leading to $|L_{25}|\sim\frac{1}{3}\cdot 10^{-3}$. The fit to the physical $M^2_\eta$ results in values for $L_{25}$ which match this expectation pretty well.

\begin{table}[tbp]
    \centering
        \tabcolsep0.5em\renewcommand{\arraystretch}{1.2}\begin{tabular}{l c r@{$\,\pm\,$}l r@{$\,\pm\,$}l r@{$\,\pm\,$}l}
      \hline\hline
        &$\mu\ [\text{GeV}]$ & \multicolumn{2}{c}{$L_5\ [10^{-3}]$} &  \multicolumn{2}{c}{$L_8\ [10^{-3}]$} &  \multicolumn{2}{c}{$L_{25}\ [10^{-3}]$}\\ \hline
 \text{NNLO w/o Ci} & 0.77 & $1.20$&$0.10$ & $0.55$&$0.20$ &  $0.55$&$0.08$ \\
 \text{NNLO w/o Ci} & 1 & $0.58$&$0.10$ & $0.24$&$0.20$ &  $0.50$&$0.08$ \\
 \text{NNLO w/ Ci} & 0.77 & $1.01$&$0.06$ & $0.52$&$0.10$ &  $0.67$&$0.13$ \\
 \text{NNLO w/ Ci} & 1 & $0.39$&$0.06$ & $0.21$&$0.10$ &  $0.63$&$0.13$ \\
 \text{NNLO w/ Ci J} & 0.77 & $1.26$&$0.06$ & $0.84$&$0.05$ & $0.70$&$0.07$ \\
 \text{NNLO w/ Ci J} & 1 & $1.26$&$0.06$ & $0.84$&$0.05$ &  $0.77$&$0.07$ \\
        \hline\hline
        \end{tabular}
    \caption{LECs at NNLO.
        }
    \label{tab:LECNNLO}
\end{table}

\begin{table}[tbp]
    \centering
        \tabcolsep0.5em\renewcommand{\arraystretch}{1.2}\begin{tabular}{l c r@{$\,\pm\,$}l r@{$\,\pm\,$}l r@{$\,\pm\,$}l r@{$\,\pm\,$}l}
        \hline\hline
        &$\mu\ [\text{GeV}]$ &  \multicolumn{2}{c}{$\stackrel{\circ}{M^2_\pi}$} &  \multicolumn{2}{c}{$\stackrel{\circ}{M^2_K}$} &  \multicolumn{2}{c}{$\frac{M^2_0}{(1+\Lambda_1)}$} &  \multicolumn{2}{c}{$M^2_\eta(L_{25}=0)$} \\ \hline
  \text{NNLO w/o Ci} & 0.77 & $0.018$&$0.007$ & $0.277$&$0.101$ & $0.840$&$0.154$ & $0.186$&$0.016$ \\
 \text{NNLO w/o Ci} & 1 & $0.016$&$0.007$ & $0.257$&$0.102$ & $0.841$&$0.158$ & $0.197$&$0.017$ \\
 \text{NNLO w/ Ci} & 0.77 & $0.018$&$0.001$ & $0.267$&$0.040$ & $0.521$&$0.170$ & $0.160$&$0.028$ \\
 \text{NNLO w/ Ci} & 1 & $0.017$&$0.001$ & $0.246$&$0.041$ & $0.518$&$0.171$ & $0.169$&$0.028$ \\
 \text{NNLO w/ Ci J} & 0.77 & $0.018$&$0.000$ & $0.232$&$0.024$ & $0.729$&$0.088$ & $0.153$&$0.014$ \\
 \text{NNLO w/ Ci J} & 1 & $0.017$&$0.000$ & $0.210$&$0.024$ & $0.670$&$0.088$ & $0.140$&$0.014$ \\
\hline\hline
        \end{tabular}
    \caption{Pseudoscalar masses at NNLO in $\text{GeV}^2$.}
    \label{tab:massNNLO}
\end{table}

\begin{table}[tbp]
    \centering
        \tabcolsep0.5em\renewcommand{\arraystretch}{1.2}\begin{tabular}{l c r@{$\,\pm\,$}l r@{$\,\pm\,$}l r@{$\,\pm\,$}l r@{$\,\pm\,$}l r@{$\,\pm\,$}l}
        \hline\hline
         & $\mu\ [\text{GeV}]$& \multicolumn{2}{c}{$\theta\ [^{\circ}]$} & \multicolumn{2}{c}{$\theta_8\ [^{\circ}]$}& \multicolumn{2}{c}{$\theta_0\ [^{\circ}]$}&\multicolumn{2}{c}{$F_8/F_\pi$}&\multicolumn{2}{c}{$\frac{F_0}{1+\Lambda_1/2}/F_\pi$}\\ \hline
 \text{NNLO w/o Ci} & 0.77 & $-9.6$&$6.0$ & $-11.7$&$5.8$ & $-6.6$&$6.4$ & $1.27$&$0.02$ & $0.85$&$0.01$ \\
 \text{NNLO w/o Ci} & 1 & $-10.1$&$6.3$ & $-12.6$&$6.1$ & $-6.3$&$6.5$ & $1.28$&$0.02$ & $0.86$&$0.01$ \\
 \text{NNLO w/ Ci} & 0.77 & $-33.8$&$18.8$ & $-31.8$&$18.5$ & $-32.4$&$21.1$ & $1.17$&$0.07$ & $0.82$&$0.01$ \\
 \text{NNLO w/ Ci} & 1 & $-35.2$&$21.5$ & $-33.7$&$21.5$ & $-33.3$&$24.2$ & $1.18$&$0.08$ & $0.83$&$0.01$ \\
 \text{NNLO w/ Ci J} & 0.77 & $-16.8$&$4.9$ & $-16.0$&$4.4$ & $-11.7$&$6.1$ & $1.16$&$0.04$ & $0.90$&$0.02$ \\
 \text{NNLO w/ Ci J} & 1 & $-20.2$&$5.4$ & $-19.4$&$4.9$ & $-15.3$&$6.7$ & $1.24$&$0.04$ & $0.84$&$0.02$ \\
    \hline\hline
        \end{tabular}
    \caption{Mixing angles and decay constants at NNLO.}
    \label{tab:angNNLO}
\end{table}

\subsection{Discussion of the results}

In the following, we discuss the summaries of our results in Tabs.~\ref{tab:massSum}--\ref{tab:decSum}. A summary of the LECs used in the different scenarios is provided in Tabs.~\ref{tabinputLECs1}--\ref{tabinputLECs3} in Appendix \ref{app:Par}.
We start with the results for the masses summarized in Tab.~\ref{tab:massSum}.
The values for the squared pion mass at LO are very close to the physical squared pion mass with deviations of ca.~$10\%$.
The LO squared kaon masses are larger than the physical value, up to about $25\%$, except for the NNLO w/ Ci J scenario. The positive NLO and NNLO corrections are in accordance with the findings in Ref.~\cite{Bijnens:2014lea}.
The LO squared pion and kaon masses, $2\hat{m}B$ and $(\hat{m}+m_s)B$, respectively, show a renormalization-scale dependence, which is caused by the renormalization of the parameter $B$ in U(3) ChPT.
The squared singlet mass in the chiral limit, $M^2_0/(1+\Lambda_1)$, increases by about $30\%$ in most of the higher-order scenarios compared to the LO value, except for the NNLO w/ Ci case, where we can see a strong influence of the $C_i$ contributions. However, a direct comparison to the LO value, $M^2_0$, remains difficult, since we do not know the value of $\Lambda_1$.
The column $M^2_\eta(x=0)$ shows the value of $M^2_\eta$ if the OZI-rule-violating parameter $\tilde{\Lambda}$ or $L_{25}$, which is fixed to the physical $M^2_\eta$, is switched off. Especially in the NNLO scenarios, the resulting values are only ca.~$50\%$ of the physical $M^2_\eta$. Therefore, we conclude that employing the LECs determined in SU(3) ChPT is not sufficient in a L$N_c$ChPT calculation and OZI-rule-violating couplings need to be included to adequately describe $M^2_\eta$. The same conclusion applies to the NNLO w/ Ci J case.
The contributions of the OZI-rule-violating parameters $\tilde{\Lambda}$ and $L_{25}$ are very important. One should also keep in mind that we only retained $L_{25}$ and omitted all other OZI-rule-violating LECs in the NNLO cases.

\begin{figure}
    \centering
        \includegraphics[width=0.85\textwidth]{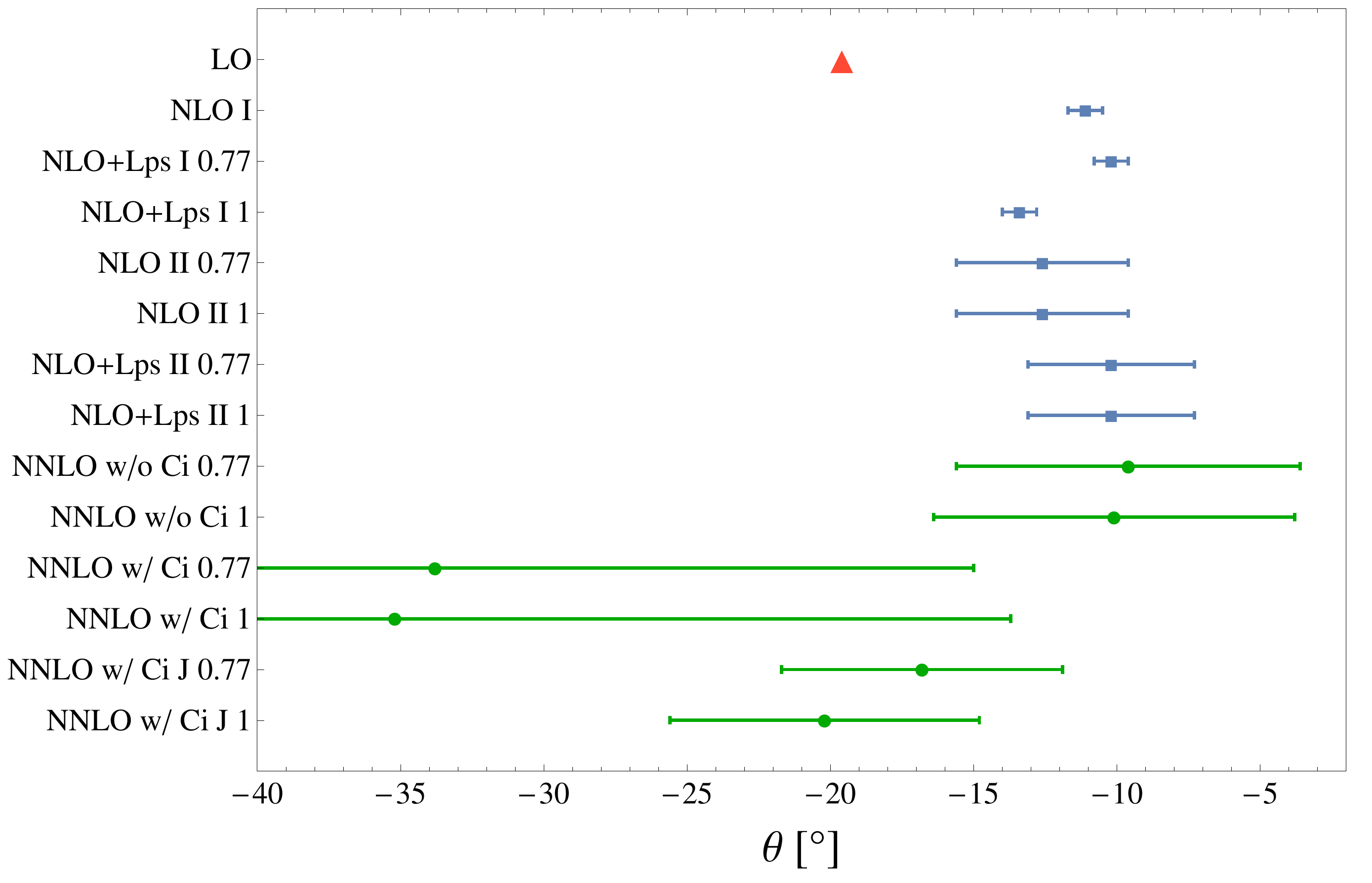}
    \caption{Results for the mixing angle $\theta$ obtained within the different scenarios in this work.}
    \label{fig:thetap}
\end{figure}

\begin{figure}
    \centering
        \includegraphics[width=1.00\textwidth]{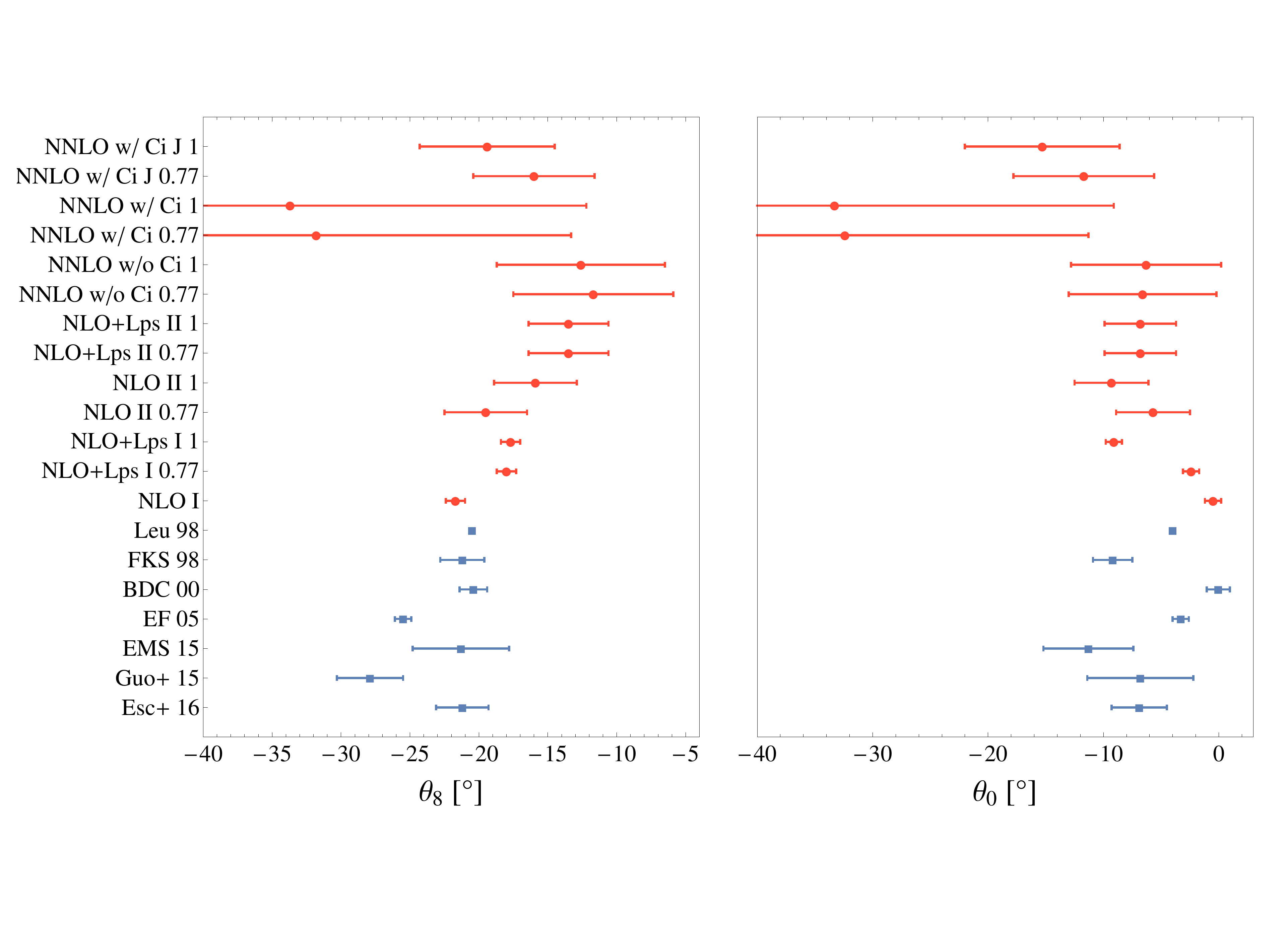}
    \caption{Results for $\theta_8$ and $\theta_0$ obtained within the different scenarios in this work and compared to phenomenological determinations from Leu 98 \cite{Leutwyler:1997yr}, FKS 98 \cite{Feldmann:1998vh}, BDC 00 \cite{Benayoun:1999au}, EF 05 \cite{Escribano:2005qq}, EMS 15 \cite{Escribano:2015nra}, Guo+ 15 \cite{Guo:2015xva}, Esc+ 16 \cite{Escribano:2015yup}.}
    \label{fig:theta80}
\end{figure}

\begin{figure}
    \centering
        \includegraphics[width=1.00\textwidth]{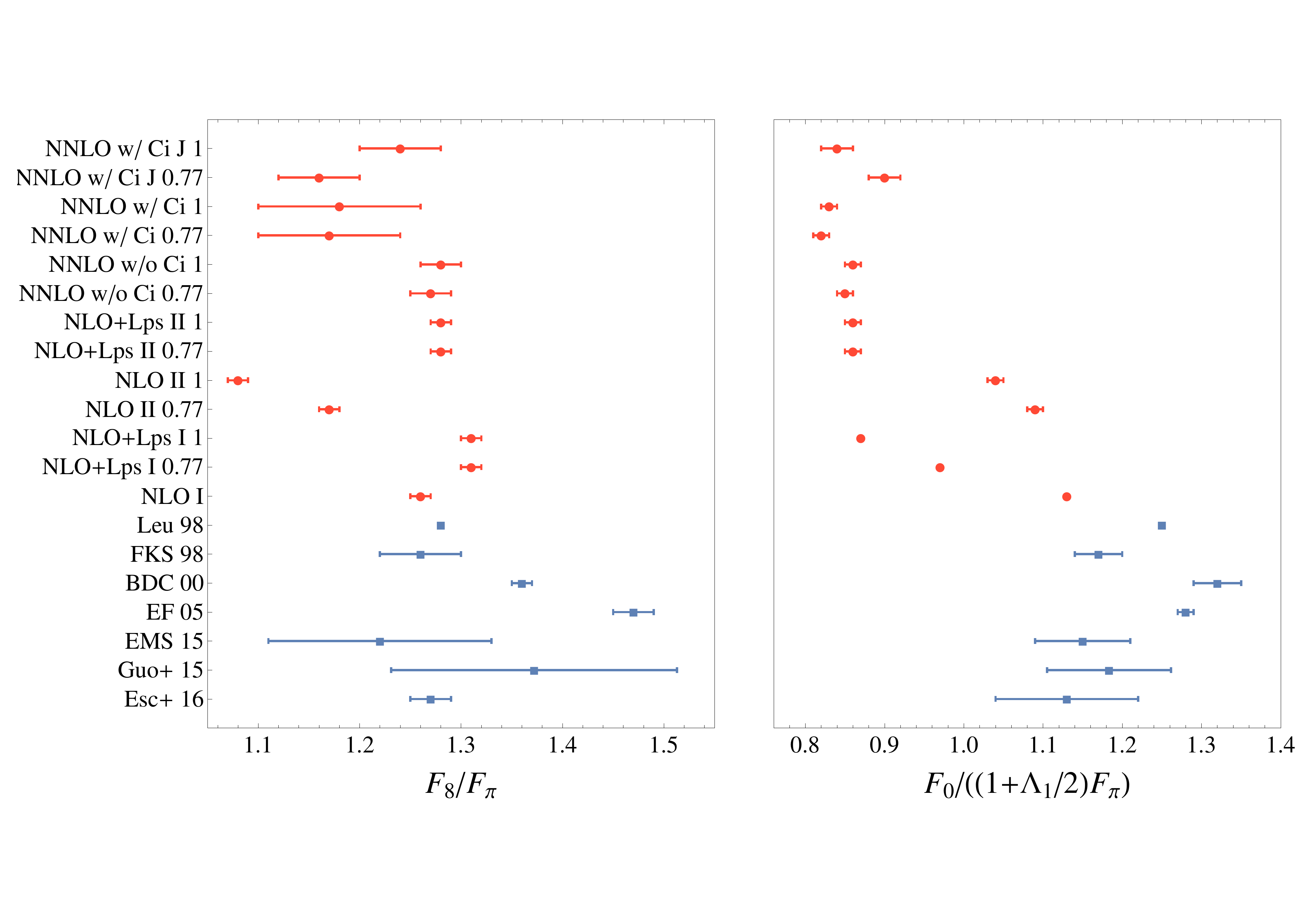}
    \caption{Results for $F_8$ and $F_0/(1+\Lambda_1/2)$ obtained within the different scenarios in this work and compared to phenomenological determinations from Leu 98 \cite{Leutwyler:1997yr}, FKS 98 \cite{Feldmann:1998vh}, BDC 00 \cite{Benayoun:1999au}, EF 05 \cite{Escribano:2005qq}, EMS 15 \cite{Escribano:2015nra}, Guo+ 15 \cite{Guo:2015xva}, Esc+ 16 \cite{Escribano:2015yup}.}
    \label{fig:f80}
\end{figure}

A summary of the results for the mixing angle $\theta$ is shown in Fig.~\ref{fig:thetap}. In comparison to the LO value $\theta=-19.6\degree$, in the cases without $C_i$, $\theta$ gets shifted to values between $-9\degree$ and $-14\degree$. The results of the NNLO w/ Ci J scenario are close to the LO value. Including the $C_i$ obtained in SU(3) ChPT (NNLO w/ Ci) leads to a drastic change of $\theta$, where the large errors are mainly caused by the assumed 50\% errors of the input $C_i$. The mixing angle seems to be very sensitive on the values of the $C_i$, although they are supposed to give only small contributions since they are NNLO corrections.
We display the results for the angles $\theta_{8}$, $\theta_0$ and the constants $F_8$, $F_0$ in Figs.~\ref{fig:theta80} and \ref{fig:f80}, respectively. They are compared to other phenomenological determinations. Reference \cite{Leutwyler:1997yr} determined the mixing parameters at NLO in L$N_c$ChPT using additional input from the two-photon decays of $\eta$ and $\eta'$. References \cite{Feldmann:1998vh,Benayoun:1999au,Escribano:2005qq,Escribano:2015nra,Escribano:2015yup} employed the two-angle scheme to extract the mixing parameters phenomenologically from decays involving $\eta$ and $\eta'$, mostly the two-photon decays but other processes, e.g., $\eta^{(')}V\gamma$ with vector mesons $V$, were used as well \cite{Benayoun:1999au}.
Note, however, these other determinations were performed only in a NLO framework and under certain assumptions, e.g., neglecting OZI-rule-violating couplings \cite{Feldmann:1998vh}. A study of the $\eta$-$\eta'$ mixing at NNLO in L$N_c$ChPT was performed in Ref.~\cite{Guo:2015xva}, and the mixing parameters were obtained from a fit to data from Lattice QCD and input from the two-photon decays. However, the authors of Ref.~\cite{Guo:2015xva} were also not able to determine all LECs at NNLO and therefore they put some of the LECs to zero.
For $\theta_0$, in the cases without $C_i$, we find values between $-10\degree$ and $0\degree$, which agree approximately with the other calculations. For $\theta_8$, the values in these cases range from $-22\degree$ to $-11\degree$, and their absolute values are slightly smaller than those obtained from phenomenology at NLO. Also the NNLO w/ Ci J values for $\theta_8$ and $\theta_0$ tend to agree with the other scenarios and determinations. Again the NNLO w/ Ci scenario is an exception, with values for $\theta_8$ and $\theta_0$ around $-33\degree$. These large negative values are related to the similar values for $\theta$ in this case and strongly depend on the $C_i$.
Our values for $F_8$ agree with most of the other calculations. Note that $F_8$ depends only on LECs which appear in SU(3) ChPT as well and $F_8$ is not affected by neglecting unknown OZI-rule-violating LECs. The errors of $F_8$ and $F_0/(1+\Lambda_1/2)$ due to the errors of the input parameters are very small, and the variation of our values in the different scenarios could serve as a better estimate of our systematic errors. For $F_0/(1+\Lambda_1/2)$ we find smaller values than the other works. The constant $F_0$ depends on the OZI-rule-violating couplings $\Lambda_1$, $L_{18}$, $L_{46}+L_{53}$. In our NNLO scenarios, however, all of them are set to zero, since they cannot be determined independently from the observables we study. Allowing values for $\Lambda_1$ and $L_{18}$ which are different from zero, e.g.~$\Lambda_1\approx 0.3$ and $L_{18}\approx 0.3\cdot 10^{-3}$, shifts $F_0$ to higher values in the range of the determinations of the other works.
The values for $F$ are mostly smaller than the physical value. This is consistent with the findings in Ref.~\cite{Bijnens:2014lea}.

The NLO I case is the most consistent scenario, since it is a full calculation up to NLO in L$N_c$ChPT and does not rely on input from other theories with different degrees of freedom or a different power-counting scheme. However, our aim was a calculation of the mixing at the one-loop level up to NNLO in the $\delta$ counting. Among these scenarios, the most complete one is NNLO w/ Ci. Note that even in this case we could not fix all parameters and set five OZI-rule-violating LECs equal to zero.

\begin{table}[tbp]
  \hspace{-0.75cm}
        \tabcolsep0.5em\renewcommand{\arraystretch}{1.2}\begin{tabular}{l c r@{$\,\pm\,$}l r@{$\,\pm\,$}l r@{$\,\pm\,$}l r@{$\,\pm\,$}l}
        \hline\hline
        &$\mu\ [\text{GeV}]$ &  \multicolumn{2}{c}{$\stackrel{\circ}{M^2_\pi}$} &  \multicolumn{2}{c}{$\stackrel{\circ}{M^2_K}$} &  \multicolumn{2}{c}{$\frac{M^2_0}{(1+\Lambda_1)}$} &  \multicolumn{2}{c}{$M^2_\eta(x=0)$} \\ \hline
                \text{LO} & - & $0.018$&$0$ & $0.244$&$0$ & $0.673$&$0$ & $0.244$&$0$ \\
 \text{NLO I} & - & $0.018$&$0.000$ & $0.261$&$0.005$ & $0.902$&$0.013$ & $0.326$&$0.003$ \\
 \text{NLO+Lps I} & 0.77 & $0.018$&$0.000$ & $0.263$&$0.005$ & $0.927$&$0.013$ & $0.261$&$0.003$ \\
 \text{NLO+Lps I} & 1 & $0.017$&$0.000$ & $0.240$&$0.005$ & $0.867$&$0.012$ & $0.218$&$0.003$ \\
 \text{NLO II} & 0.77 & $0.018$&$0.000$ & $0.249$&$0.023$ & $0.871$&$0.061$ & $0.299$&$0.010$ \\
 \text{NLO II} & 1 & $0.018$&$0.000$ & $0.249$&$0.023$ & $0.871$&$0.061$ & $0.269$&$0.010$ \\
 \text{NLO+Lps II} & 0.77 & $0.019$&$0.000$ & $0.287$&$0.023$ & $0.933$&$0.061$ & $0.199$&$0.010$ \\
 \text{NLO+Lps II} & 1 & $0.017$&$0.000$ & $0.265$&$0.023$ & $0.933$&$0.061$ & $0.199$&$0.010$ \\
  \text{NNLO w/o Ci} & 0.77 & $0.018$&$0.007$ & $0.277$&$0.101$ & $0.840$&$0.154$ & $0.186$&$0.016$ \\
 \text{NNLO w/o Ci} & 1 & $0.016$&$0.007$ & $0.257$&$0.102$ & $0.841$&$0.158$ & $0.197$&$0.017$ \\
 \text{NNLO w/ Ci} & 0.77 & $0.018$&$0.001$ & $0.267$&$0.040$ & $0.521$&$0.170$ & $0.160$&$0.028$ \\
 \text{NNLO w/ Ci} & 1 & $0.017$&$0.001$ & $0.246$&$0.041$ & $0.518$&$0.171$ & $0.169$&$0.028$ \\
 \text{NNLO w/ Ci J} & 0.77 & $0.018$&$0.000$ & $0.232$&$0.024$ & $0.729$&$0.088$ & $0.153$&$0.014$ \\
 \text{NNLO w/ Ci J} & 1 & $0.017$&$0.000$ & $0.210$&$0.024$ & $0.670$&$0.088$ & $0.140$&$0.014$ \\
\hline\hline
        \end{tabular}
    \caption{Summary of the results for the pseudoscalar masses in $\text{GeV}^2$. The parameter $x$ denotes $\tilde{\Lambda}$ or $L_{25}$.}
    \label{tab:massSum}
\end{table}

\begin{table}[tbp]
    \centering
        \tabcolsep0.5em\renewcommand{\arraystretch}{1.2}\begin{tabular}{l c r@{$\,\pm\,$}l r@{$\,\pm\,$}l r@{$\,\pm\,$}l}
      \hline\hline
       & $\mu\ [\text{GeV}]$& \multicolumn{2}{c}{$\theta\ [^{\circ}]$} & \multicolumn{2}{c}{$\theta_8\ [^{\circ}]$}& \multicolumn{2}{c}{$\theta_0\ [^{\circ}]$}\\ \hline
\text{LO} & - & $-19.6$&$0$ & $-19.6$&$0$ & $-19.6$&$0$ \\
 \text{NLO I} & - & $-11.1$&$0.6$ & $-21.7$&$0.7$ & $-0.5$&$0.7$ \\
 \text{NLO+Lps I} & 0.77 & $-10.2$&$0.6$ & $-18.0$&$0.7$ & $-2.4$&$0.7$ \\
 \text{NLO+Lps I} & 1 & $-13.4$&$0.6$ & $-17.7$&$0.7$ & $-9.1$&$0.7$ \\
 \text{NLO II} & 0.77 & $-12.6$&$3.0$ & $-19.5$&$3.0$ & $-5.7$&$3.2$ \\
 \text{NLO II} & 1 & $-12.6$&$3.0$ & $-15.9$&$3.0$ & $-9.3$&$3.2$ \\
 \text{NLO+Lps II} & 0.77 & $-10.2$&$2.9$ & $-13.5$&$2.9$ & $-6.8$&$3.1$ \\
 \text{NLO+Lps II} & 1 & $-10.2$&$2.9$ & $-13.5$&$2.9$ & $-6.8$&$3.1$ \\
\text{NNLO w/o Ci} & 0.77 & $-9.6$&$6.0$ & $-11.7$&$5.8$ & $-6.6$&$6.4$ \\
 \text{NNLO w/o Ci} & 1 & $-10.1$&$6.3$ & $-12.6$&$6.1$ & $-6.3$&$6.5$ \\
 \text{NNLO w/ Ci} & 0.77 & $-33.8$&$18.8$ & $-31.8$&$18.5$ & $-32.4$&$21.1$ \\
 \text{NNLO w/ Ci} & 1 & $-35.2$&$21.5$ & $-33.7$&$21.5$ & $-33.3$&$24.2$ \\
 \text{NNLO w/ Ci J} & 0.77 & $-16.8$&$4.9$ & $-16.0$&$4.4$ & $-11.7$&$6.1$ \\
 \text{NNLO w/ Ci J} & 1 & $-20.2$&$5.4$ & $-19.4$&$4.9$ & $-15.3$&$6.7$ \\
   \hline\hline
        \end{tabular}
    \caption{Summary of the results for the mixing angles.}
    \label{tab:angSum}
\end{table}

\begin{table}[tbp]
    \centering
        \tabcolsep0.5em\renewcommand{\arraystretch}{1.2}\begin{tabular}{l c  r@{$\,\pm\,$}l r@{$\,\pm\,$}l r@{$\,\pm\,$}l}
       \hline\hline
         & $\mu\ [\text{GeV}]$& \multicolumn{2}{c}{$F_8/F_\pi$}&\multicolumn{2}{c}{$\frac{F_0}{1+\Lambda_1/2}/F_\pi$}&\multicolumn{2}{c}{$F\ [\text{MeV}]$}\\ \hline
 \text{LO} & - & $1$&$0$ & $1$&$0$ & $92.2$&$0.1$ \\
\text{NLO I} & - & $1.26$&$0.01$ & $1.13$&$0.00$ & $90.73$&$0.11$ \\
 \text{NLO+Lps I} & 0.77 & $1.31$&$0.01$ & $0.97$&$0.00$ & $79.31$&$0.12$ \\
 \text{NLO+Lps I} & 1 & $1.31$&$0.01$ & $0.87$&$0.00$ & $74.77$&$0.12$ \\
 \text{NLO II} & 0.77 & $1.17$&$0.01$ & $1.09$&$0.01$ & $91.25$&$0.13$ \\
 \text{NLO II} & 1 & $1.08$&$0.01$ & $1.04$&$0.01$ & $91.74$&$0.13$ \\
 \text{NLO+Lps II} & 0.77 & $1.28$&$0.01$ & $0.86$&$0.01$ & $74.91$&$0.14$ \\
 \text{NLO+Lps II} & 1 & $1.28$&$0.01$ & $0.86$&$0.01$ & $74.91$&$0.14$ \\
\text{NNLO w/o Ci} & 0.77 & $1.27$&$0.02$ & $0.85$&$0.01$ & $79.46$&$6.59$ \\
 \text{NNLO w/o Ci} & 1 & $1.28$&$0.02$ & $0.86$&$0.01$ & $79.45$&$6.59$ \\
 \text{NNLO w/ Ci} & 0.77 & $1.17$&$0.07$ & $0.82$&$0.01$ & $73.02$&$0.13$ \\
 \text{NNLO w/ Ci} & 1 & $1.18$&$0.08$ & $0.83$&$0.01$ & $73.02$&$0.13$ \\
 \text{NNLO w/ Ci J} & 0.77 & $1.16$&$0.04$ & $0.90$&$0.02$ & $79.44$&$0.12$ \\
 \text{NNLO w/ Ci J} & 1 & $1.24$&$0.04$ & $0.84$&$0.02$ & $74.40$&$0.13$ \\
   \hline\hline
        \end{tabular}
    \caption{Summary of the results for the decay constants.}
    \label{tab:decSum}
\end{table}

\section{Summary and outlook}
\label{section_summary}

We have derived an expression for the $\eta$-$\eta'$ mixing in the framework of L$N_c$ChPT up to NNLO, including higher-derivative, kinetic, and mass terms. Furthermore, we have calculated the axial-vector-current decay constants of the $\eta$-$\eta'$ system at NNLO and determined the mixing parameters $F_8$, $F_0$, $\theta_8$, $\theta_0$ of the two-angle scheme.

The numerical evaluation of the results has been performed successively at LO, NLO, and NNLO. At NLO, we have determined all LECs by fixing them to the physical values of the pseudoscalar masses $M^2_\pi$, $M^2_K$, $M^2_\eta$, $M^2_{\eta'}$, the decay constants $F_\pi$, $F_K$, and the quark mass ratio $m_s/m$. We have compared our results with the values for the LECs obtained in SU(3) ChPT \cite{Bijnens:2014lea}.
Due to the large number of LECs at NNLO, we have not been able to determine all of them through our aforementioned input quantities. Therefore, we have made use of the values obtained in SU(3) ChPT and have applied the matching relations between SU(3) and U(3) ChPT. One OZI-rule-violating parameter, $L_{25}$, has been fixed to the physical value of $M^2_\eta$. The impact of OZI-rule-violating parameters on our observables is rather large and they cannot be neglected.
In addition to using input from SU(3) ChPT, we also investigated the case where we employed LECs which were computed in a chiral quark model \cite{Jiang:2015dba}.
We have compared our results for the parameters of the two-angle scheme with other phenomenological determinations of those quantities.

The mixing angle $\theta$ and the angles $\theta_8$, $\theta_0$ of the two-angle scheme depend strongly on the values of the NNLO correction given by $C_i$ terms. This leads to results which deviate very much from the determinations at LO, NLO, or NNLO without $C_i$ terms. From this observation, we conclude that the mixing angles are particularly sensitive to the expansion scheme and it remains unclear to which extent the convergence is under control.

At NNLO, it has not been possible to determine all LECs from the available experimental data. In the future, Lattice QCD may provide further information on these LECs, since it will make it possible to study the quark-mass dependence of the pseudoscalar masses and decay constants.

Our NNLO expressions for the $\eta$-$\eta'$ mixing can now be used to study anomalous decays, e.g. $\eta^{(')}\to\gamma\gamma$ and $\eta^{(')}\to\pi^+\pi^-\gamma$, consistently at the one-loop level. A further step would be the inclusion of vector mesons as explicit degrees of freedom and the investigation of $PV\gamma$ processes, where $P$ refers to pseudoscalar mesons and $V$ to vector mesons.

\begin{acknowledgements}

 Supported by the Deutsche Forschungsgemeinschaft DFG through the Collaborative Research Center ``The Low-Energy Frontier of the Standard Model" (SFB 1044).
 P.M. wants to thank G.~Ecker and J.~J.~Sanz-Cillero for useful discussions.

\end{acknowledgements}

\begin{appendix}

\section{Building blocks and transformation behavior}
\label{appendix_building_blocks}
   The effective dynamical degrees of freedom are contained in the U(3) matrix
\begin{displaymath}
U=\exp\left(i\sum_{a=0}^8 \frac{\phi_a\lambda_a}{F}\right)=e^{\frac{i}{3}\psi}\hat{U},
\end{displaymath}
where
\begin{displaymath}
\text{det}(\hat{U})=1,\quad \text{det}(U)=e^{i\psi},\quad \psi=-i\ln(\text{det}(U)).
\end{displaymath}
   The external fields $s$, $p$, $l_\mu$, and $r_\mu$ are Hermitian, color-neutral $3\times 3$ matrices
coupling to the corresponding quark bilinears, and $\theta$ is a real field
coupling to the winding number density \cite{Gasser:1984gg}.
   The traceless components of $r_\mu$ and $l_\mu$ are defined as
\begin{align*}
r_\mu&=\hat{r}_\mu+\frac{1}{3}\langle r_\mu\rangle,\quad \langle \hat{r}_\mu\rangle=0,\\
l_\mu&=\hat{l}_\mu+\frac{1}{3}\langle l_\mu\rangle,\quad \langle \hat{l}_\mu\rangle=0.
\end{align*}
   We parametrize the group elements $(V_L,V_R)\in\mbox{U}(3)_L\times\mbox{U}(3)_R$ in terms
of
\begin{align*}
V_R&=\exp\left(-i\sum_{a=0}^8\theta_{Ra}\frac{\lambda_a}{2}\right)=e^{-\frac{i}{3}\theta_R}\hat{V}_R,\\
\text{det}\left(\hat{V}_R\right)&=1,\quad \theta_R=i\ln\left(\text{det}\left(V_R\right)\right),\\
V_L&=\exp\left(-i\sum_{a=0}^8\theta_{La}\frac{\lambda_a}{2}\right)=e^{-\frac{i}{3}\theta_L}\hat{V}_L,\\
\text{det}\left(\hat{V}_L\right)&=1,\quad \theta_L=i\ln\left(\text{det}\left(V_L\right)\right).
\end{align*}
   We define $v_\mu=\frac{1}{2}(r_\mu+l_\mu)$, $a_\mu=\frac{1}{2}(r_\mu-l_\mu)$, and $\chi=2B(s+ip)$.
   Under the group $G=\text{U}(3)_L\times\text{U}(3)_R$, the transformation properties of the dynamical degrees
of freedom and of the external fields read
\begin{align}
\label{transformations_properties}
U&\mapsto V_RUV_L^\dagger,\nonumber\\
\psi&\mapsto\psi-i\ln(\text{det}(V_R))+i\ln(\text{det}(V_L))=\psi
-(\theta_R-\theta_L),\nonumber\\
r_\mu&\mapsto V_Rr_\mu V_R^\dagger+iV_R\partial_\mu V_R^\dagger,\nonumber\\
\hat{r}_\mu&\mapsto\hat{V}_R\hat{r}_\mu\hat{V}_R^\dagger+i\hat{V}_R\partial_\mu\hat{V}_R^\dagger,\nonumber\\
\langle r_\mu\rangle&\mapsto\langle r_\mu\rangle-\partial_\mu\theta_R,\nonumber\\
l_\mu&\mapsto V_Ll_\mu V_L^\dagger+iV_L\partial_\mu V_L^\dagger,\nonumber\\
\hat{l}_\mu&\mapsto\hat{V}_L\hat{l}_\mu\hat{V}_L^\dagger+i\hat{V}_L\partial_\mu\hat{V}_L^\dagger,\nonumber\\
\langle l_\mu\rangle&\mapsto\langle l_\mu\rangle-\partial_\mu\theta_L,\nonumber\\
\langle a_\mu\rangle&\mapsto\langle a_\mu\rangle-\frac{1}{2}(\partial_\mu\theta_R-\partial_\mu\theta_L),\nonumber\\
\chi&\mapsto V_R\chi V_L^\dagger,\nonumber\\
\theta&\mapsto \theta+(\theta_R-\theta_L).
\end{align}
   We define covariant derivatives according to the transformation behavior of the object to
which they are applied:
\begin{align}
\label{appendix_covariant_derivatives}
D_\mu U&=\partial_\mu U-ir_\mu U+iU l_\mu\mapsto V_RD_\mu U V_L^\dagger,\nonumber\\
D_\mu U^\dagger&=\partial_\mu U^\dagger+iU^\dagger r_\mu-il_\mu U^\dagger
\mapsto V_L D_\mu U^\dagger V_R^\dagger,\nonumber\\
D_\mu \hat{U}&=\partial_\mu\hat{U}-i\hat{r}_\mu\hat{U}+i\hat{U}\hat{l}_\mu,\nonumber\\
D_\mu \psi&=\partial_\mu \psi-2\langle a_\mu\rangle\mapsto D_\mu\psi,\nonumber\\
D_\mu U&=e^{\frac{i}{3}\psi}\left(D_\mu\hat{U}+\frac{i}{3}D_\mu\Psi\hat{U}\right),\nonumber\\
D_\mu \theta&=\partial_\mu \theta+2\langle a_\mu\rangle\mapsto D_\mu \theta.
\end{align}
   Finally, the parity transformation behavior reads
\begin{align*}
U(t,\vec x)&\mapsto U^\dagger(t,-\vec x),\\
\psi(t,\vec x)&\mapsto -\psi(t,-\vec x),\\
\theta(t,\vec x)&\mapsto -\theta(t,-\vec x),\\
D_\mu U(t,\vec x)&\mapsto D^\mu U^\dagger(t,-\vec x).
\end{align*}

\section{Masses and decay constants}\label{A:MF}

The pion decay constant $F$ in the chiral limit is given by
\begin{align}\label{FoverFpi}
F&=F_{\pi }\left[1-\frac{4 M_{\pi }^2 \text{\textit{$L$}}_5^{}}{F_{\pi }^2}\right.\no\\
&\quad -\frac{1}{F_{\pi }^4}\left(4 \left(2 M_{\pi }^4 \left(3
   \left(\text{\textit{$L$}}_5^{}\right){}^2-8
   \text{\textit{$L$}}_8^{} \text{\textit{$L$}}_5^{}+\left(C_{14}+C_{17}\right) F_{\pi }^2\right)+F_{\pi }^2
   \left(2 M_K^2+M_{\pi }^2\right)
   \text{\textit{$L$}}_4^{}\right)\right)\no\\
    &\quad\left.-\frac{A_0\left(M_K^2\right)+2 A_0\left(M_{\pi }^2\right)}{32 \pi
   ^2 F_{\pi }^2}\right]
\end{align}
in terms of the physical decay constant $F_\pi$ and the physical pion and kaon masses, $M_{\pi}$ and $M_K$, respectively.
The expression for the LO pion mass $\stackrel{\circ}{M^2_\pi}$ reads
\begin{align}\label{Mpi0}
\stackrel{\circ}{M^2_\pi}&=2B\hat{m}\no\\
&=M_{\pi }^2\left[1+\frac{8 M_{\pi }^2 \left(\text{\textit{$L$}}_5^{}-2
   \text{\textit{$L$}}_8^{}\right)}{F_{\pi }^2}\right.\no\\
    &\quad+\frac{1}{F_{\pi }^4}\left(8 \left(2 M_{\pi }^4 \left(8
   \left(\text{\textit{$L$}}_5^{}-2
   \text{\textit{$L$}}_8^{}\right){}^2+\left(2
   C_{12}+C_{14}+C_{17}-3 C_{19}-2 C_{31}\right) F_{\pi
   }^2\right)\right.\right.\no\\
    &\quad\left.\left.+2 F_{\pi }^2 M_K^2 \left(\text{\textit{$L$}}_4^{}-2
   \text{\textit{$L$}}_6^{}\right)+F_{\pi }^2 M_{\pi }^2
   \left(\text{\textit{$L$}}_4^{}-2
   \text{\textit{$L$}}_6^{}\right)\right)\right)\no\\
    &\quad+\frac{1}{192 F_{\pi
   }^2}\left(\left(2 \sqrt{2} \sin (2 \theta^{[0]} )+\cos (2 \theta^{[0]} )-3\right)
   A_0\left(M_{\eta }^2\right)\right.\no\\
    &\quad\left.\left.-\left(2 \sqrt{2} \sin (2 \theta^{[0]}
   )+\cos (2 \theta^{[0]} )+3\right) A_0\left(M_{\text{$\eta'
   $}}^2\right)+6 A_0\left(M_{\pi }^2\right)\right)\right],
\end{align}
and the LO kaon mass $\stackrel{\circ}{M^2_K}$ is given by
\begin{align}\label{Mk0}
\stackrel{\circ}{M^2_K}&=B(\hat{m}+ms)\no\\
&=M_K^2\left[1+\frac{8 M_K^2 \left(\text{\textit{$L$}}_5^{}-2
   \text{\textit{$L$}}_8^{}\right)}{F_{\pi }^2}\right.\no\\
    &\quad+\frac{1}{F_{\pi }^4}\left(8 \left(4 M_K^4 \left(2 \left(\text{\textit{$L$}}_5^{}-4
   \text{\textit{$L$}}_8^{}\right)
   \left(\text{\textit{$L$}}_5^{}-2
   \text{\textit{$L$}}_8^{}\right)+\left(C_{12}+C_{14}-3
   C_{19}-C_{31}\right) F_{\pi }^2\right)\right.\right.\no\\
    &\quad+2 M_K^2 \left(F_{\pi }^2
   \left(\text{\textit{$L$}}_4^{}-2 \text{\textit{$L$}}_6^{}+2
   \left(-C_{14}+C_{17}+3 C_{19}\right) M_{\pi }^2\right)+4 M_{\pi
   }^2 \text{\textit{$L$}}_5^{} \left(\text{\textit{$L$}}_5^{}-2
   \text{\textit{$L$}}_8^{}\right)\right)\no\\
    &\quad\left.\left.-F_{\pi }^2 M_{\pi }^2
   \left(2 \left(\text{\textit{$L$}}_6^{}+\left(-C_{14}+C_{17}+3
   C_{19}\right) M_{\pi
   }^2\right)-\text{\textit{$L$}}_4^{}\right)\right)\right)\no\\
    &\quad+\frac{1}{192F_{\pi
   }^2 M_K^2}\left(\sin ^2(\theta^{[0]} ) \left(\left(3 M_{\text{$\eta'$}}^2+M_{\pi
   }^2\right) A_0\left(M_{\text{$\eta'$}}^2\right)-4 M_K^2
   A_0\left(M_{\eta }^2\right)\right)\right.\no\\
    &\quad+\sqrt{2} \left(2
   M_K^2-M_{\pi }^2\right) \sin (2 \theta^{[0]} )
   \left(A_0\left(M_{\text{$\eta'$}}^2\right)-A_0\left(M_{\eta
   }^2\right)\right)\no\\
    &\quad\left.\left.+\cos ^2(\theta^{[0]} ) \left(\left(3 M_{\eta
   }^2+M_{\pi }^2\right) A_0\left(M_{\eta }^2\right)-4 M_K^2
   A_0\left(M_{\text{$\eta'$}}^2\right)\right)\right)\right].
\end{align}
In loop contributions, we always use the LO mixing angle
\begin{align}
\theta^{[0]}=-\arctan\left(\frac{2 \sqrt{2} \left(M_K^2-M_{\pi
   }^2\right)}{3 \left(\frac{1}{3} \left(M_{\pi }^2-4
   M_K^2\right)+M_{\text{$\eta'$}}^2\right)}\right),
\end{align}
which yields $\theta^{[0]}=-19.6\degree$.
The ratio of the physical kaon and pion decay constants is given by
\begin{align}\label{FkoverFpi}
&F_K/F_\pi\no\\
&=1+\frac{4 \left(M_K^2-M_{\pi }^2\right)
   \text{\textit{$L$}}_5^{}}{F_{\pi }^2}\no\\
    &\quad+\frac{1}{F_{\pi }^4}\left(8 \left(\left(3 M_K^4+2 M_{\pi }^2 M_K^2-3 M_{\pi }^4\right)
   \left(\text{\textit{$L$}}_5^{}\right){}^2+8 \left(M_{\pi
   }^4-M_K^4\right) \text{\textit{$L$}}_8^{}
   \text{\textit{$L$}}_5^{}\right.\right.\no\\
    &\quad\left.\left.+2 F_{\pi }^2 \left(M_K^2-M_{\pi
   }^2\right) \left(C_{14}
   M_K^2+C_{17} M_{\pi
   }^2\right)\right)\right)\no\\
    &\quad+\frac{1}{128 \pi ^2 F_{\pi
   }^2}\left(2 A_0\left(M_K^2\right)+3 \cos ^2(\theta^{[0]} ) A_0\left(M_{\eta
   }^2\right)+3 \sin ^2(\theta^{[0]} ) A_0\left(M_{\text{$\eta'
   $}}^2\right)\right.\no\\
    &\quad\left.-5 A_0\left(M_{\pi }^2\right)\right)
\end{align}

The NNLO expressions for the decay constants of the $\eta$-$\eta'$ system are given by
\begin{align}
F^8_\eta&=F_{\pi } \cos (\theta^{[2]})+\frac{1}{3 F_{\pi }}\left(8 \left(M_K^2-M_{\pi }^2\right) \text{\textit{$L$}}_5^{} \left(\sqrt{2} \sin (\theta^{[2]})+2 \cos (\theta^{[2]})\right)\right)\no\\
&\quad+\frac{\cos(\theta^{[2]}) }{48 \pi ^2 F_{\pi }^3}\left[256 \pi ^2 \left(\left(4 M_K^4+8 M_{\pi }^2 M_K^2-9 M_{\pi }^4\right) \left(\text{\textit{$L$}}_5^{}\right){}^2+16 \left(M_{\pi }^4-M_K^4\right)
   \text{\textit{$L$}}_8^{} \text{\textit{$L$}}_5^{}\right.\right.\no\\
    &\quad\left.\left.+4 \left(C_{14}+C_{17}\right) F_{\pi }^2 M_K^2 \left(M_K^2-M_{\pi }^2\right)\right)+3 F_{\pi }^2
   \left(A_0\left(M_K^2\right)-A_0\left(M_{\pi }^2\right)\right)\right]\no\\
    &\quad+\frac{\sin(\theta^{[2]})}{3 F_{\pi }^3}\left[2 \sqrt{2} \left(M_K^2-M_{\pi }^2\right) \left(F_{\pi }^2 \left(-\Lambda _1 \text{\textit{$L$}}_5^{}+16 \left(C_{14}+C_{17}\right) M_K^2+6 L_{18}\right)\right.\right.\no\\
    &\quad\left.\left.+16
   \text{\textit{$L$}}_5^{} \left(\left(M_K^2+3 M_{\pi }^2\right) \text{\textit{$L$}}_5^{}-4 \left(M_K^2+M_{\pi }^2\right)
   \text{\textit{$L$}}_8^{}\right)\right)\right]\no\\
    &\quad+\frac{C_{12}}{3
   F_{\pi } \left(4 M_K^2-3 M_{\text{$\eta'$}}^2-M_{\pi }^2\right)}\left[16 \left(M_K^2-M_{\pi }^2\right) \left(-2 \left(M_K^2-M_{\pi }^2\right) M_{\text{$\eta'$}}^2\right.\right.\no\\
    &\quad\times\left(\sqrt{2} \sin (\text{$\theta^{[2]} $})-4 \cos (\text{$\theta^{[2]}
   $})\right)+3 \sqrt{2} M_{\pi }^2 \left(M_{\pi }^2-2 M_K^2\right) \sin (\text{$\theta^{[2]} $})\no\\
    &\quad\left.\left.+3 \sqrt{2} M_{\text{$\eta'$}}^4 \sin (\text{$\theta^{[2]} $})\right)\right],
\end{align}
\begin{align}
F^8_{\eta'}&=F_{\pi } \sin (\theta^{[2]})-\frac{1}{3 F_{\pi }}\left(8 \left(M_K^2-M_{\pi }^2\right) \text{\textit{$L$}}_5^{}
   \left(\sqrt{2} \cos (\theta^{[2]})-2 \sin (\theta^{[2]})\right)\right)\no\\
&\quad-\frac{\cos(\theta^{[2]})}{3 F_{\pi }^3}\left[2 \sqrt{2} \left(M_K^2-M_{\pi }^2\right) \left(F_{\pi }^2 \left(-\Lambda _1 \text{\textit{$L$}}_5^{}+16 \left(C_{14}+C_{17}\right) M_K^2+6 L_{18}\right)\right.\right.\no\\
&\quad\left.\left.+16
   \text{\textit{$L$}}_5^{} \left(\left(M_K^2+3 M_{\pi }^2\right) \text{\textit{$L$}}_5^{}-4 \left(M_K^2+M_{\pi }^2\right)
   \text{\textit{$L$}}_8^{}\right)\right)\right]\no\\
&\quad+\frac{\sin(\theta^{[2]})}{48 \pi ^2 F_{\pi }^3}\left[256 \pi ^2 \left(\left(4 M_K^4+8 M_{\pi }^2 M_K^2-9 M_{\pi }^4\right) \left(\text{\textit{$L$}}_5^{}\right){}^2+16 \left(M_{\pi }^4-M_K^4\right)
   \text{\textit{$L$}}_8^{} \text{\textit{$L$}}_5^{}\right.\right.\no\\
    &\quad\left.\left.+4 \left(C_{14}+C_{17}\right) F_{\pi }^2 M_K^2 \left(M_K^2-M_{\pi }^2\right)\right)+3 F_{\pi }^2
   \left(A_0\left(M_K^2\right)-A_0\left(M_{\pi }^2\right)\right)\right]\no\\
    &\quad+\frac{C_{12}}{3 F_{\pi }}\left[16 \left(\frac{1}{4 M_K^2-3 M_{\text{$\eta'$}}^2-M_{\pi }^2}\left[\sqrt{2} \left(M_K^2-M_{\pi }^2\right) \cos (\text{$\theta^{[2]} $})\no\right.\right.\right.\\
    &\quad\left.\left(-2 M_K^2 \left(7 M_{\text{$\eta'$}}^2+5 M_{\pi }^2\right)+16 M_K^4+3
   M_{\text{$\eta'$}}^4+2 M_{\pi }^2 M_{\text{$\eta'$}}^2+3 M_{\pi }^4\right)\right]\no\\
    &\quad\left.\left.-\sin (\text{$\theta^{[2]} $}) \left(-4
   M_K^2 \left(M_{\text{$\eta'$}}^2+2 M_{\pi }^2\right)+8 M_K^4+M_{\pi }^2 \left(M_{\text{$\eta'$}}^2+3 M_{\pi }^2\right)\right)\right)\right],
\end{align}
\begin{align}
F^0_\eta&=-\frac{1}{6 F_{\pi }}\left[16 \left(M_K^2-M_{\pi }^2\right) \text{\textit{$L$}}_5^{} \left(\sin (\text{$\theta^{[2]} $})+\sqrt{2} \cos (\text{$\theta^{[2]} $})\right)+3 F_{\pi }^2 \left(\Lambda
   _1+2\right) \sin (\text{$\theta^{[2]} $})\right]\no\\
&\quad-\frac{\cos(\theta^{[2]})}{3 F_{\pi }^3}\left[2 \sqrt{2} \left(M_K^2-M_{\pi }^2\right) \left(F_{\pi }^2 \left(-\Lambda _1 \text{\textit{$L$}}_5^{}+16 \left(C_{14}+C_{17}\right) M_K^2\right.\right.\right.\no\\
&\quad\left.\left.\left.+6 \left(L_{18}+2
   L_{46}+2 L_{53}\right)\right)+16 \text{\textit{$L$}}_5^{} \left(\left(M_K^2+3 M_{\pi }^2\right) \text{\textit{$L$}}_5^{}-4 \left(M_K^2+M_{\pi }^2\right)
   \text{\textit{$L$}}_8^{}\right)\right)\right]\no\\
    &\quad +\frac{\sin(\theta^{[2]})}{96 \pi ^2 F_{\pi }^3}\left[2 \left(2 \pi ^2 \left(32 \left(4 \text{\textit{$L$}}_5^{} \left(8 \left(M_K^4-M_{\pi }^4\right) \text{\textit{$L$}}_8^{}+\left(-2 M_K^4-4 M_{\pi }^2 M_K^2+3
   M_{\pi }^4\right) \text{\textit{$L$}}_5^{}\right)\right.\right.\right.\right.\no\\
    &\quad+F_{\pi }^2 \left(M_{\pi }^2 \left(8 \left(C_{14}+C_{17}\right) M_K^2-3
   \left(L_{18}+L_{46}+L_{53}\right)\right)-8 \left(C_{14}+C_{17}\right) M_K^4\right.\no\\
    &\quad\left.\left.\left.\left.-6 \left(L_{18}+L_{46}+L_{53}\right) M_K^2\right)\right)+32 F_{\pi }^2 \Lambda _1
   \left(M_K^2+2 M_{\pi }^2\right) \text{\textit{$L$}}_5^{}+3 F_{\pi }^4 \Lambda _1^2\right)\right.\no\\
    &\quad\left.\left.+3 F_{\pi }^2 A_0\left(M_{\pi }^2\right)\right)+3 F_{\pi }^2
   A_0\left(M_K^2\right)\right]\no\\
    &\quad+\frac{C_{12}}{3 F_{\pi } \left(4
   M_K^2-3 M_{\text{$\eta'$}}^2-M_{\pi }^2\right)}\left[16 \left(\sin (\text{$\theta^{[2]} $}) \left(4 M_K^2-3 M_{\text{$\eta'$}}^2-M_{\pi }^2\right)\right.\right.\no\\
    &\quad\times\left(2 M_K^2 \left(M_{\text{$\eta'$}}^2-3 M_{\pi
   }^2\right)+M_{\pi }^2 \left(M_{\text{$\eta'$}}^2+3 M_{\pi }^2\right)\right)+\sqrt{2} \left(M_K^2-M_{\pi }^2\right) \cos (\text{$\theta^{[2]} $}) \no\\
    &\quad\left.\left.\times\left(2 M_K^2
   \left(M_{\text{$\eta'$}}^2+3 M_{\pi }^2\right)-2 M_{\pi }^2 M_{\text{$\eta'$}}^2-3 M_{\text{$\eta'$}}^4-3 M_{\pi }^4\right)\right)\right],
\end{align}
\begin{align}
F^0_{\eta'}&=\frac{1}{6 F_{\pi }}\left[16 \left(M_K^2-M_{\pi }^2\right) \text{\textit{$L$}}_5^{} \left(\cos (\text{$\theta^{[2]} $})-\sqrt{2} \sin (\text{$\theta^{[2]} $})\right)+3 F_{\pi }^2 \left(\Lambda
   _1+2\right) \cos (\text{$\theta^{[2]} $})\right]\no\\
&\quad\frac{\cos(\theta^{[2]})}{96 \pi ^2 F_{\pi
   }^3}\left[2 \left(2 \pi ^2 \left(32 \left(4 \text{\textit{$L$}}_5^{} \left(\left(2 M_K^4+4 M_{\pi }^2 M_K^2-3 M_{\pi }^4\right) \text{\textit{$L$}}_5^{}+8 \left(M_{\pi
   }^4-M_K^4\right) \text{\textit{$L$}}_8^{}\right)\right.\right.\right.\right.\no\\
    &\quad+F_{\pi }^2 \left(M_{\pi }^2 \left(3 \left(L_{18}+L_{46}+L_{53}\right)-8 \left(C_{14}+C_{17}\right)
   M_K^2\right)+8 \left(C_{14}+C_{17}\right) M_K^4\right.\no\\
    &\quad\left.\left.\left.\left.+6 \left(L_{18}+L_{46}+L_{53}\right) M_K^2\right)\right)-32 F_{\pi }^2 \Lambda _1 \left(M_K^2+2 M_{\pi }^2\right)
   \text{\textit{$L$}}_5^{}-3 F_{\pi }^4 \Lambda _1^2\right)\right.\no\\
    &\quad\left.\left.-3 F_{\pi }^2 A_0\left(M_{\pi }^2\right)\right)-3 F_{\pi }^2 A_0\left(M_K^2\right)\right]\no\\
    &\quad-\frac{\sin(\theta^{[2]})}{3 F_{\pi }^3}\left[2 \sqrt{2} \left(M_K^2-M_{\pi }^2\right) \left(F_{\pi }^2 \left(-\Lambda _1 \text{\textit{$L$}}_5^{}+16 \left(C_{14}+C_{17}\right) M_K^2\right.\right.\right.\no\\
    &\quad\left.\left.\left.+6 \left(L_{18}+2
   L_{46}+2 L_{53}\right)\right)+16 \text{\textit{$L$}}_5^{} \left(\left(M_K^2+3 M_{\pi }^2\right) \text{\textit{$L$}}_5^{}-4 \left(M_K^2+M_{\pi }^2\right)
   \text{\textit{$L$}}_8^{}\right)\right)\right]\no\\
    &\quad+\frac{C_{12}}{3 F_{\pi } \left(4 M_K^2-3 M_{\text{$\eta'$}}^2-M_{\pi }^2\right)}\left[16 \left(M_K^2-M_{\pi }^2\right) \left(\sqrt{2} \sin (\text{$\theta^{[2]} $})\right.\right.\no\\
    &\quad\times\left(-2 M_K^2 \left(7 M_{\text{$\eta'$}}^2+5 M_{\pi }^2\right)+16 M_K^4+3
   M_{\text{$\eta'$}}^4+2 M_{\pi }^2 M_{\text{$\eta'$}}^2+3 M_{\pi }^4\right)\no\\
    &\quad\left.\left.-8 \left(M_K^2-M_{\pi }^2\right) \cos (\text{$\theta^{[2]} $}) \left(2
   M_K^2-M_{\text{$\eta'$}}^2\right)\right)\right],
\end{align}
in terms of the physical masses $M^2_\pi$, $M^2_K$, $M^2_{\eta'}$ and the physical pion decay constant $F_\pi$. The mixing angle $\theta^{[2]}$ is the NNLO mixing angle given in Eq.~(\ref{angle2}) in Sec.~\ref{section_mixing_angle}.
In the case where the loop contributions are added to the NLO results, the parameters of the two-angle scheme can be simplified to read
\begin{align}
F_8&=F_\pi+\frac{1}{48 \pi ^2 F_{\pi }}\left[256 \pi ^2 \left(M_K^2-M_{\pi }^2\right) \text{\textit{$L$}}_5^{}+3 A_0\left(M_K^2\right)-3 A_0\left(M_{\pi }^2\right)\right],\label{eq:f8nlo2}\\
F_0&=F_\pi+\frac{1}{96 \pi ^2 F_{\pi }}\left[16 \pi ^2 \left(16 M_K^2 \left(\text{\textit{$L$}}_5^{}+3 L_{18}\right)+8 M_{\pi }^2 \left(3 L_{18}-2 \text{\textit{$L$}}_5^{}\right)+3 F_{\pi }^2 \Lambda
   _1\right)\right.\no\\
    &\quad\left.-3 A_0\left(M_K^2\right)-6 A_0\left(M_{\pi }^2\right)\right],\\
\theta_8&=\theta^{[2]}+\arctan\left(-\frac{4 \sqrt{2} \left(M_K^2-M_{\pi }^2\right) \left(2 \text{\textit{$L$}}_5^{}+3 L_{18}\right)}{3
   F_{\pi }^2}\right),\\
\theta_0&=\theta^{[2]}-\arctan\left(-\frac{4 \sqrt{2} \left(M_K^2-M_{\pi }^2\right) \left(2 \text{\textit{$L$}}_5^{}+3 L_{18}\right)}{3
   F_{\pi }^2}\right).\label{eq:theta0nlo2}
\end{align}

\section{Kinetic matrix and mass matrix at NNLO}
\label{A:KM}

In the following, the NNLO expressions for the matrix $\mathcal{C}_A$ defined in Eq.~(\ref{eq:quadL}) in Sec.~\ref{section_mixing_angle}, the kinetic matrix $\mathcal{K}_B$ and the mass matrix $\mathcal{M}_B$ defined in Eq.~(\ref{Lmixing2}) are provided.
The components of $\mathcal{C}_A$ are given by
\begin{align}
\label{matrixCAc8}
c_8&=\frac{32 C_{12} \left(4 M_K^2-M_{\pi }^2\right)}{3 F_{\pi }^2},\\
\label{matrixCAc1}
c_1&=\frac{32 C_{12} \left(2 M_K^2+M_{\pi }^2\right)}{3 F_{\pi }^2},\\
\label{matrixCAc81}c_{81}&=\frac{64 \sqrt{2} C_{12} \left(M_{\pi }^2-M_K^2\right)}{3 F_{\pi}^2}.
\end{align}
At NNLO, both tree and loop corrections occur. The second-order tree contributions to the kinetic matrix read
\begin{align}
\delta^{(2,\text{tr})}_8&=\frac{1}{3 F_{\pi }^4}\left[8 \left(2 \left(8 \left(2 M_K^4+2 M_{\pi }^2 M_K^2-M_{\pi
   }^4\right) \left(\text{\textit{$L$}}_5^{}\right){}^2+8
   \left(M_{\pi }^4-4 M_K^4\right) \text{\textit{$L$}}_8^{}
   \text{\textit{$L$}}_5^{}\right.\right.\right.\no\\
    &\quad\left.+\left(C_{14}+C_{17}\right) F_{\pi }^2
   \left(8 M_K^4-8 M_{\pi }^2 M_K^2+3 M_{\pi }^4\right)\right)\no\\
    &\quad\left.\left.+3
   F_{\pi }^2 \left(2 M_K^2+M_{\pi }^2\right)
   \text{\textit{$L$}}_4^{}\right)+32 C_{12} F_{\pi }^2 \left(8 M_K^4-8 M_{\pi }^2 M_K^2+3 M_{\pi }^4\right)\right],\\
\delta^{(2,\text{tr})}_1&=\frac{1}{3 F_{\pi }^4}\left[8 \left(3 F_{\pi }^2 \left(2 M_K^2+M_{\pi }^2\right)
   \text{\textit{$L$}}_4^{}+16 \left(M_K^4+M_{\pi }^2 M_K^2+M_{\pi
   }^4\right) \left(\text{\textit{$L$}}_5^{}\right){}^2\right.\right.\no\\
    &\quad-16 \left(2
   M_K^4+M_{\pi }^4\right) \text{\textit{$L$}}_8^{}
   \text{\textit{$L$}}_5^{}\no\\
    &\quad\left.\left.\left.+2 \left(C_{14}+C_{17}\right) \left(4
   M_K^4-4 M_{\pi }^2 M_K^2+3 M_{\pi }^4\right)+3 L_{18} \left(2
   M_K^2+M_{\pi }^2\right)\right)\right)\right.\no\\
    &\quad\left. +32 C_{12} F_{\pi }^2 \left(4 M_K^4-4 M_{\pi }^2 M_K^2+M_0^2 \left(2
   M_K^2+M_{\pi }^2\right)+3 M_{\pi }^4\right)\right],
    \end{align}
\begin{align}
\delta^{(2,\text{tr})}_{81}&=-\frac{1}{3 F_{\pi }^4}\left[8 \sqrt{2} \left(M_K^2-M_{\pi }^2\right) \left(16
   \text{\textit{$L$}}_5^{} \left(\left(M_K^2+2 M_{\pi }^2\right)
   \text{\textit{$L$}}_5^{}-2 \left(M_K^2+M_{\pi }^2\right)
   \text{\textit{$L$}}_8^{}\right)\right.\right.\no\\
    &\quad\left.\left.+F_{\pi }^2 \left(8
   \left(C_{14}+C_{17}\right) M_K^2+3
   L_{18}\right)\right)\right.\no\\
    &\quad\left.+32 \sqrt{2} C_{12} F_{\pi }^2 \left(4 M_K^2+M_0^2\right) \left(M_K^2-M_{\pi
   }^2\right)\right],
\end{align}
and the loop contributions read
\begin{align}
\delta^{(2,\text{lo})}_8=\frac{A_0\left(M_K^2\right)}{16\pi^2F_{\pi }^2},\ \ \
    \delta^{(2,\text{lo})}_1=0,\ \ \
        \delta^{(2,\text{lo})}_{81}=0.
\end{align}
The second-order tree contributions to the mass matrix are
\begin{align}
&\Delta {M^2_8}^{(2,\text{tr})}\no\\
&=\frac{1}{3
   F_{\pi }^4}\left[16 \left(16 M_K^6 \left(8 \left(\text{\textit{$L$}}_5^{}-2
   \text{\textit{$L$}}_8^{}\right)
   \text{\textit{$L$}}_8^{}+\left(3 C_{19}+2 C_{31}\right) F_{\pi
   }^2\right)\right.\right.\no\\
    &\quad+8 M_{\pi }^2 M_K^4 \left(16
   \left(\text{\textit{$L$}}_8^{}\right){}^2-3 \left(3 C_{19}+2
   C_{31}\right) F_{\pi }^2\right)+4 M_{\pi }^4 M_K^2 \left(32
   \text{\textit{$L$}}_8^{}
   \left(\text{\textit{$L$}}_8^{}-\text{\textit{$L$}}_5^{}\right)\right.\no\\
    &\quad\left.+3 \left(3 C_{19}+2 C_{31}\right) F_{\pi }^2\right)+M_{\pi }^6
   \left(24 \left(3 \text{\textit{$L$}}_5^{}-4
   \text{\textit{$L$}}_8^{}\right)
   \text{\textit{$L$}}_8^{}-\left(3 C_{19}+2 C_{31}\right) F_{\pi
   }^2\right)\no\\
    &\quad\left.\left.+8 F_{\pi }^2 \left(M_K^2-M_{\pi }^2\right){}^2
   \text{\textit{$L$}}_7^{}+F_{\pi }^2 \left(8 M_K^4+2 M_{\pi }^2
   M_K^2-M_{\pi }^4\right) \text{\textit{$L$}}_6^{}\right)\right],
    \end{align}
\begin{align}
&\Delta {M^2_1}^{(2,\text{tr})}\no\\
&=\frac{1}{3
   F_{\pi }^4}\left[16 \left(F_{\pi }^2 \Lambda _2 \left(2 M_K^4+M_{\pi
   }^4\right) \left(\text{\textit{$L$}}_5^{}-2
   \text{\textit{$L$}}_8^{}\right)+F_{\pi }^2 \left(2 M_K^2+M_{\pi
   }^2\right){}^2 \text{\textit{$L$}}_6^{}\right.\right.\no\\
    &\quad+F_{\pi }^2 \left(2
   M_K^2+M_{\pi }^2\right){}^2 \text{\textit{$L$}}_7^{}-128 M_K^6
   \left(\text{\textit{$L$}}_8^{}\right){}^2+64 M_K^6
   \text{\textit{$L$}}_5^{} \text{\textit{$L$}}_8^{}+64 M_{\pi }^2
   M_K^4 \left(\text{\textit{$L$}}_8^{}\right){}^2\no\\
    &\quad+64 M_{\pi }^4
   M_K^2 \left(\text{\textit{$L$}}_8^{}\right){}^2-64 M_{\pi }^4
   M_K^2 \text{\textit{$L$}}_5^{} \text{\textit{$L$}}_8^{}-96
   M_{\pi }^6 \left(\text{\textit{$L$}}_8^{}\right){}^2+72 M_{\pi
   }^6 \text{\textit{$L$}}_5^{} \text{\textit{$L$}}_8^{}\no\\
    &\quad+24 C_{19}
   F_{\pi }^2 M_K^6+16 C_{31} F_{\pi }^2 M_K^6-36 C_{19} F_{\pi
   }^2 M_{\pi }^2 M_K^4-24 C_{31} F_{\pi }^2 M_{\pi }^2 M_K^4\no\\
    &\quad+18
   C_{19} F_{\pi }^2 M_{\pi }^4 M_K^2+12 C_{31} F_{\pi }^2 M_{\pi
   }^4 M_K^2+3 C_{19} F_{\pi }^2 M_{\pi }^6+2 C_{31} F_{\pi }^2
   M_{\pi }^6\no\\
    &\quad\left.\left.-12 F_{\pi }^2 L_{25} M_K^4+12 F_{\pi }^2 L_{25}
   M_{\pi }^2 M_K^2-9 F_{\pi }^2 L_{25} M_{\pi }^4\right)\right]\no\\
    &\quad+6(2M^2_K+M^2_\pi)v^{(2)}_2,
    \end{align}
\begin{align}
&\Delta {M^2_{81}}^{(2,\text{tr})}\no\\
&=-\frac{1}{3 F_{\pi }^4}\left[16 \sqrt{2} \left(M_K^2-M_{\pi }^2\right) \left(2 \left(4
   M_K^4 \left(8 \left(\text{\textit{$L$}}_5^{}-2
   \text{\textit{$L$}}_8^{}\right)
   \text{\textit{$L$}}_8^{}+\left(3 C_{19}+2 C_{31}\right) F_{\pi
   }^2\right)\right.\right.\right.\no\\
    &\quad+M_{\pi }^2 \left(F_{\pi }^2
   \left(\text{\textit{$L$}}_6^{}+\text{\textit{$L$}}_7^{}-2
   \left(3 C_{19}+2 C_{31}\right) M_K^2\right)+32 M_K^2
   \left(\text{\textit{$L$}}_5^{}-\text{\textit{$L$}}_8^{}\right)
   \text{\textit{$L$}}_8^{}\right)\no\\
    &\quad\left.+F_{\pi }^2 M_K^2 \left(2
   \left(\text{\textit{$L$}}_6^{}+\text{\textit{$L$}}_7^{}\right)-
   3 L_{25}\right)+\left(3 C_{19}+2 C_{31}\right) F_{\pi }^2
   M_{\pi }^4\right)\no\\
    &\quad\left.\left.+F_{\pi }^2 \Lambda _2 \left(M_K^2+M_{\pi
   }^2\right) \left(\text{\textit{$L$}}_5^{}-2
   \text{\textit{$L$}}_8^{}\right)\right)\right],
\end{align}
and the loop corrections are given by
\begin{align}
\Delta {M^2_8}^{(2,\text{lo})}&=\frac{1}{576 F_{\pi }^2}\left(2 \sqrt{2} \left(8 M_K^2-5 M_{\pi }^2\right) \sin (2 \theta^{[0]}
   ) \left(A_0\left(M_{\eta }^2\right)-A_0\left(M_{\text{$\eta'
   $}}^2\right)\right)\right.\no\\
    &\quad+\left(8 M_K^2-5 M_{\pi }^2\right) \cos (2
   \theta^{[0]} ) \left(A_0\left(M_{\eta
   }^2\right)-A_0\left(M_{\text{$\eta'$}}^2\right)\right)\no\\
    &\quad+3
   \left(8 M_K^2-3 M_{\pi }^2\right) \left(A_0\left(M_{\eta
   }^2\right)+A_0\left(M_{\text{$\eta'$}}^2\right)\right)\no\\
    &\quad\left.+6
   M_{\pi }^2 \left(3 A_0\left(M_{\pi }^2\right)-2
   A_0\left(M_K^2\right)\right)\right),
    \end{align}
\begin{align}
\Delta {M^2_1}^{(2,\text{lo})}&=\frac{1}{144F_{\pi }^2}\left(2 \sqrt{2} \left(M_K^2-M_{\pi }^2\right) \sin (2 \theta^{[0]} )
   \left(A_0\left(M_{\eta }^2\right)-A_0\left(M_{\text{$\eta'
   $}}^2\right)\right)\right.\no\\
    &\quad+\left(M_K^2-M_{\pi }^2\right) \cos (2
   \theta^{[0]} ) \left(A_0\left(M_{\eta
   }^2\right)-A_0\left(M_{\text{$\eta'$}}^2\right)\right)\no\\
    &\quad\left.+3 M_K^2
   \left(4 A_0\left(M_K^2\right)+A_0\left(M_{\eta
   }^2\right)+A_0\left(M_{\text{$\eta'$}}^2\right)\right)+9
   M_{\pi }^2 A_0\left(M_{\pi }^2\right)\right),
    \end{align}
\begin{align}
\Delta {M^2_{81}}^{(2,\text{lo})}&=\frac{1}{576 F_{\pi }^2}\left(4 \left(4 M_K^2-M_{\pi }^2\right) \sin (2 \theta^{[0]} )
   \left(A_0\left(M_{\text{$\eta'$}}^2\right)-A_0\left(M_{\eta
   }^2\right)\right)\right.\no\\
    &\quad+\sqrt{2} \left(4 M_K^2-M_{\pi }^2\right) \cos
   (2 \theta^{[0]} ) \left(A_0\left(M_{\text{$\eta'
   $}}^2\right)-A_0\left(M_{\eta }^2\right)\right)\no\\
    &\quad-3 \sqrt{2}
   \left(\left(4 M_K^2-3 M_{\pi }^2\right) \left(A_0\left(M_{\eta
   }^2\right)+A_0\left(M_{\text{$\eta'$}}^2\right)\right)\right.\no\\
    &\quad\left.\left.+\left(8
   M_K^2-4 M_{\pi }^2\right) A_0\left(M_K^2\right)-6 M_{\pi }^2
   A_0\left(M_{\pi }^2\right)\right)\right).
\end{align}

\section{Input parameters}
\label{app:Par}

\begin{table}[htbp]
    \centering
        \tabcolsep0.5em\renewcommand{\arraystretch}{1.2}\begin{tabular}{l c r@{$\,\pm\,$}l r@{$\,\pm\,$}l r@{$\,\pm\,$}l r@{$\,\pm\,$}l}
        \hline\hline
         & $\mu\ [\text{GeV}]$& \multicolumn{2}{c}{$L_5\ [10^{-3}]$} & \multicolumn{2}{c}{$L_8\ [10^{-3}]$} & \multicolumn{2}{c}{$\tilde{\Lambda}$} & \multicolumn{2}{c}{$L_{25}\ [10^{-3}]$}\\ \hline
        \text{NLO I} & - & $1.86$&$0.06$ & $0.78$&$0.05$ & $-0.34$&$0.05$ & $0$&$0$ \\
 \text{NLO+Lps I} & 0.77 & $1.37$&$0.06$ & $0.85$&$0.05$ & $0.52$&$0.05$ & $0$&$0$ \\
 \text{NLO+Lps I} & 1 & $0.75$&$0.06$ & $0.55$&$0.05$ & $1.09$&$0.04$ & $0$&$0$ \\
 \text{NLO II} & 0.77 & $1.20$&$0.10$ & $0.55$&$0.20$ & $0.02$&$0.13$ & $0$&$0$ \\
 \text{NLO II} & 1 & $0.58$&$0.10$ & $0.24$&$0.20$ & $0.41$&$0.13$ & $0$&$0$ \\
 \text{NLO+Lps II} & 0.77 & $1.20$&$0.10$ & $0.55$&$0.20$ & $1.34$&$0.13$ & $0$&$0$ \\
 \text{NLO+Lps II} & 1 & $0.58$&$0.10$ & $0.24$&$0.20$ & $1.34$&$0.13$ & $0$&$0$ \\
  \text{NNLO w/o Ci} & 0.77 & $1.20$&$0.10$ & $0.55$&$0.20$ & $0$&$0$ & $0.55$&$0.08$ \\
 \text{NNLO w/o Ci} & 1 & $0.58$&$0.10$ & $0.24$&$0.20$ & $0$&$0$ & $0.50$&$0.08$ \\
 \text{NNLO w/ Ci} & 0.77 & $1.01$&$0.06$ & $0.52$&$0.10$ & $0$&$0$ & $0.67$&$0.13$ \\
 \text{NNLO w/ Ci} & 1 & $0.39$&$0.06$ & $0.21$&$0.10$ & $0$&$0$ & $0.63$&$0.13$ \\
 \text{NNLO w/ Ci J} & 0.77 & $1.26$&$0.06$ & $0.84$&$0.05$ & $0$&$0$ & $0.70$&$0.07$ \\
 \text{NNLO w/ Ci J} & 1 & $1.26$&$0.06$ & $0.84$&$0.05$ & $0$&$0$ & $0.77$&$0.07$ \\
    \hline\hline
        \end{tabular}
    \caption{Summary of the results for the LECs determined in the numerical analysis of the $\eta$-$\eta'$ mixing in Sec.~\ref{section_numerical_analysis}.}
        \label{tabinputLECs1}
\end{table}

\begin{table}[htbp]
    \centering
        \tabcolsep0.5em\renewcommand{\arraystretch}{1.2}\begin{tabular}{l c r@{$\,\pm\,$}l r@{$\,\pm\,$}l r@{$\,\pm\,$}l r@{$\,\pm\,$}l}
        \hline\hline
         & $\mu\ [\text{GeV}]$& \multicolumn{2}{c}{$L_4\ [10^{-3}]$} & \multicolumn{2}{c}{$L_6\ [10^{-3}]$} & \multicolumn{2}{c}{$L_7\ [10^{-3}]$} & \multicolumn{2}{c}{$L_{18}\ [10^{-3}]$}\\ \hline
                 \text{NLO I} & - & $0$&$0$ & $0$&$0$ & $0$&$0$ & $0$&$0$ \\
 \text{NLO+Lps I} & 0.77 & $0$&$0$ & $0$&$0$ & $0$&$0$ & $0$&$0$ \\
 \text{NLO+Lps I} & 1 & $0$&$0$ & $0$&$0$ & $0$&$0$ & $0$&$0$ \\
 \text{NLO II} & 0.77 & $0$&$0$ & $0$&$0$ & $0$&$0$ & $0$&$0$ \\
 \text{NLO II} & 1 & $0$&$0$ & $0$&$0$ & $0$&$0$ & $0$&$0$ \\
 \text{NLO+Lps II} & 0.77 & $0.21$&$0$ & $0.10$&$0$ & $0$&$0$ & $-0.41$&$0$ \\
 \text{NLO+Lps II} & 1 & $0$&$0$ & $0$&$0$ & $0$&$0$ & $0$&$0$ \\
  \text{NNLO w/o Ci} & 0.77 & $0$&$0.30$ & $0.04$&$0.40$ & $0$&$0.20$ & $-0.41$&$0$ \\
 \text{NNLO w/o Ci} & 1 & $-0.21$&$0.30$ & $-0.07$&$0.40$ & $0$&$0.20$ & $0$&$0$ \\
 \text{NNLO w/ Ci} & 0.77 & $0.30$&$0$ & $0.18$&$0.05$ & $0$&$0.09$ & $-0.41$&$0$ \\
 \text{NNLO w/ Ci} & 1 & $0.09$&$0$ & $0.07$&$0.05$ & $0$&$0.09$ & $0$&$0$ \\
 \text{NNLO w/ Ci J} & 0.77 & $0$&$0$ & $0$&$0$ & $0$&$0.05$ & $0$&$0$ \\
 \text{NNLO w/ Ci J} & 1 & $0$&$0$ & $0$&$0$ & $0$&$0.05$ & $0$&$0$ \\
\hline\hline
        \end{tabular}
    \caption{Input LECs used in Sec.~\ref{section_numerical_analysis}.}
        \label{tabinputLECs2}
\end{table}

\begin{table}[htbp]
        \tabcolsep0.5em\renewcommand{\arraystretch}{1.2}\begin{tabular}{l c r@{$\,\pm\,$}l r@{$\,\pm\,$}l r@{$\,\pm\,$}l r@{$\,\pm\,$}l r@{$\,\pm\,$}l}
        \hline\hline
         & $\mu\ [\text{GeV}]$& \multicolumn{2}{c}{$C_{12}\ [10^{-3}]$}& \multicolumn{2}{c}{$C_{14}\ [10^{-3}]$}& \multicolumn{2}{c}{$C_{17}\ [10^{-3}]$}& \multicolumn{2}{c}{$C_{19}\ [10^{-3}]$}& \multicolumn{2}{c}{$C_{31}\ [10^{-3}]$} \\ \hline
                 \text{NLO I} & - & $0$&$0$ & $0$&$0$ & $0$&$0$ & $0$&$0$ & $0$&$0$ \\
 \text{NLO+Lps I} & 0.77 & $0$&$0$ & $0$&$0$ & $0$&$0$ & $0$&$0$ & $0$&$0$ \\
 \text{NLO+Lps I} & 1 & $0$&$0$ & $0$&$0$ & $0$&$0$ & $0$&$0$ & $0$&$0$ \\
 \text{NLO II} & 0.77 & $0$&$0$ & $0$&$0$ & $0$&$0$ & $0$&$0$ & $0$&$0$ \\
 \text{NLO II} & 1 & $0$&$0$ & $0$&$0$ & $0$&$0$ & $0$&$0$ & $0$&$0$ \\
 \text{NLO+Lps II} & 0.77 & $0$&$0$ & $0$&$0$ & $0$&$0$ & $0$&$0$ & $0$&$0$ \\
 \text{NLO+Lps II} & 1 & $0$&$0$ & $0$&$0$ & $0$&$0$ & $0$&$0$ & $0$&$0$ \\
 \text{NNLO w/o Ci} & 0.77 & $0$&$0$ & $0$&$0$ & $0$&$0$ & $0$&$0$ & $0$&$0$ \\
 \text{NNLO w/o Ci} & 1 & $0$&$0$ & $0$&$0$ & $0$&$0$ & $0$&$0$ & $0$&$0$ \\
\text{NNLO w/ Ci} & 0.77 & $-0.33$&$0.16$ & $-0.12$&$0.06$ & $-0.12$&$0.06$ & $-0.34$&$0.24$ & $0.63$&$0.12$ \\
 \text{NNLO w/ Ci} & 1 & $-0.33$&$0.16$ & $-0.12$&$0.06$ & $-0.12$&$0.06$ & $-0.34$&$0.24$ & $0.63$&$0.12$ \\
 \text{NNLO w/ Ci J} & 0.77 & $-0.34$&$0.01$ & $-0.87$&$0.21$ & $0.17$&$0.04$ & $-0.14$&$0.13$ & $-0.07$&$0.13$ \\
 \text{NNLO w/ Ci J} & 1 & $-0.34$&$0.01$ & $-0.87$&$0.21$ & $0.17$&$0.04$ & $-0.14$&$0.13$ & $-0.07$&$0.13$ \\
    \hline\hline
        \end{tabular}
    \caption{Input LECs used in Sec.~\ref{section_numerical_analysis} in $\text{GeV}^{-2}$.}
        \label{tabinputLECs3}
\end{table}

\end{appendix}


\begin{thebibliography}{99}


\bibitem{Agashe:2014kda}
  K.~A.~Olive {\it et al.} [Particle Data Group Collaboration],
  Chin.\ Phys.\ C {\bf 38}, 090001 (2014).


\bibitem{Adlarson:2012bi}
  P.~Adlarson {\it et al.},
  arXiv:1204.5509 [nucl-ex].

\bibitem{Amaryan:2013eja}
  K.~Kampf {\it et al.},
  arXiv:1308.2575 [hep-ph].

\bibitem{Adlarson:2014hka}
  P.~Adlarson {\it et al.},
  arXiv:1412.5451 [nucl-ex].


\bibitem{Adler:1969gk}
  S.~L.~Adler,
  Phys.\ Rev.\  {\bf 177}, 2426 (1969).

\bibitem{Bell:1969ts}
  J.~S.~Bell and R.~Jackiw,
  Nuovo Cim.\ A {\bf 60},  47 (1969).

\bibitem{Adler:1969er}
  S.~L.~Adler and W.~A.~Bardeen,
  Phys.\ Rev.\  {\bf 182}, 1517 (1969).


\bibitem{Scherer:2012zzd}
  S.~Scherer and M.~R.~Schindler,
  Lect.\ Notes Phys.\  {\bf 830}, 1 (2012).


\bibitem{Goldstone:1962es}
  J.~Goldstone, A.~Salam, and S.~Weinberg,
  Phys.\ Rev.\  {\bf 127}, 965 (1962).


\bibitem{'tHooft:1976up}
  G.~'t Hooft,
  Phys.\ Rev.\ Lett.\  {\bf 37}, 8 (1976).

\bibitem{Witten:1979vv}
  E.~Witten,
  Nucl.\ Phys.\ B {\bf 156}, 269 (1979).

\bibitem{Veneziano:1979ec}
  G.~Veneziano,
  Nucl.\ Phys.\ B {\bf 159}, 213 (1979).

\bibitem{'tHooft:1973jz}
  G.~'t Hooft,
  Nucl.\ Phys.\ B {\bf 72}, 461 (1974).

\bibitem{Witten:1979kh}
  E.~Witten,
  Nucl.\ Phys.\ B {\bf 160}, 57 (1979).

\bibitem{Bhaduri:1988gc}
  R.~K.~Bhaduri, {\it Models of the Nucleon: From Quarks to Soliton}
  (Addison-Wesley, Redwood City, Calif., 1988), Sec.~5.6.

\bibitem{Manohar:1998xv}
  A.~V.~Manohar, {\it Large N QCD}, in {\it
  Probing the Standard Model of Particle Interactions}. Proceedings, Summer School in Theoretical Physics,
  NATO Advanced Study Institute, 68th session, Les Houches, France, July 28 - September 5, 1997, Pt. 1,
  2, edited by R.~Gupta, A.~Morel, E.~de Rafael, and F.~David (Elsevier, Amsterdam, 1999),
  hep-ph/9802419.

\bibitem{DiVecchia:1980yfw}
  P.~Di Vecchia and G.~Veneziano,
  Nucl.\ Phys.\ B {\bf 171}, 253 (1980).

\bibitem{Coleman:1980mx}
  S.~R.~Coleman and E.~Witten,
  Phys.\ Rev.\ Lett.\  {\bf 45}, 100 (1980).


\bibitem{Leutwyler:2013wna}
  H.~Leutwyler,
  Mod.\ Phys.\ Lett.\ A {\bf 28}, 1360014 (2013).


\bibitem{Amsler}
  C.~Amsler, T.~DeGrand, and B.~Krusche, {\it Quark Model}, in
   Ref.~\cite{Agashe:2014kda}, p 259.


  \bibitem{Isgur:1976qg}
  N.~Isgur,
  Phys.\ Rev.\ D {\bf 13}, 122 (1976).

\bibitem{Fritzsch:1976qc}
  H.~Fritzsch and J.~D.~Jackson,
  Phys.\ Lett.\ B {\bf 66}, 365 (1977).


\bibitem{Gasser:1984gg}
  J.~Gasser and H.~Leutwyler,
  Nucl.\ Phys.\ B {\bf 250}, 465 (1985).

\bibitem{Donoghue:1986wv}
  J.~F.~Donoghue, B.~R.~Holstein, and Y.~C.~R.~Lin,
  Phys.\ Rev.\ Lett.\ {\bf 55}, 2766 (1985)
  [Phys.\ Rev.\ Lett.\  {\bf 61}, 1527 (1988)].

\bibitem{Gilman:1987ax}
  F.~J.~Gilman and R.~Kauffman,
  Phys.\ Rev.\ D {\bf 36}, 2761 (1987)
  [Phys.\ Rev.\ D {\bf 37}, 3348 (1988)].

\bibitem{Schechter:1992iz}
  J.~Schechter, A.~Subbaraman, and H.~Weigel,
  Phys.\ Rev.\ D {\bf 48}, 339 (1993).

\bibitem{Bramon:1997va}
  A.~Bramon, R.~Escribano, and M.~D.~Scadron,
  Eur.\ Phys.\ J.\ C {\bf 7}, 271 (1999).


 \bibitem{Moussallam:1994xp}
  B.~Moussallam,
  Phys.\ Rev.\ D {\bf 51}, 4939 (1995).

\bibitem{Leutwyler:1996sa}
  H.~Leutwyler,
  Phys.\ Lett.\ B {\bf 374}, 163 (1996).


\bibitem{HerreraSiklody:1996pm}
  P.~Herrera-Siklody, J.~I.~Latorre, P.~Pascual, and J.~Taron,
  Nucl.\ Phys.\ B {\bf 497}, 345 (1997).

\bibitem{Kaiser:2000gs}
  R.~Kaiser and H.~Leutwyler,
  Eur.\ Phys.\ J.\ C {\bf 17}, 623 (2000).

\bibitem{Leutwyler:1997yr}
  H.~Leutwyler,
  Nucl.\ Phys.\ Proc.\ Suppl.\  {\bf 64}, 223 (1998).

\bibitem{Kaiser:1998ds}
  R.~Kaiser and H.~Leutwyler,
  in {\it Nonperturbative methods in quantum field theory} (World Scientific, Singapore, 1998) [hep-ph/9806336].

\bibitem{HerreraSiklody:1998cr}
  P.~Herrera-Siklody,
  Phys.\ Lett.\ B {\bf 442}, 359 (1998).


\bibitem{Borasoy:2004ua}
  B.~Borasoy,
  Eur.\ Phys.\ J.\ C {\bf 34}, 317 (2004).

\bibitem{Guo:2015xva}
  X.~K.~Guo, Z.~H.~Guo, J.~A.~Oller, and J.~J.~Sanz-Cillero,
  JHEP {\bf 1506} 175, (2015).


\bibitem{Georgi:1993jn}
  H.~Georgi,
  Phys.\ Rev.\ D {\bf 49}, 1666 (1994).

\bibitem{Peris:1993np}
  S.~Peris,
  Phys.\ Lett.\ B {\bf 324}, 442 (1994).


\bibitem{Bijnens:2014lea}
  J.~Bijnens and G.~Ecker,
  Ann.\ Rev.\ Nucl.\ Part.\ Sci.\  {\bf 64}, 149 (2014).


\bibitem{Feldmann:1998vh}
  T.~Feldmann, P.~Kroll, and B.~Stech,
  Phys.\ Rev.\ D {\bf 58}, 114006 (1998).

\bibitem{Feldmann:1998sh}
  T.~Feldmann, P.~Kroll, and B.~Stech,
  Phys.\ Lett.\ B {\bf 449}, 339 (1999).

\bibitem{Benayoun:1999au}
  M.~Benayoun, L.~DelBuono, and H.~B.~O'Connell,
  Eur.\ Phys.\ J.\ C {\bf 17}, 593 (2000).

\bibitem{Escribano:2005qq}
  R.~Escribano and J.~M.~Frere,
  JHEP {\bf 0506}, 029 (2005).

\bibitem{Escribano:2010wt}
  R.~Escribano, P.~Masjuan, and J.~J.~Sanz-Cillero,
  JHEP {\bf 1105}, 094 (2011).

\bibitem{Escribano:2013kba}
  R.~Escribano, P.~Masjuan, and P.~Sanchez-Puertas,
  Phys.\ Rev.\ D {\bf 89}, no. 3, 034014 (2014).

\bibitem{Escribano:2015nra}
  R.~Escribano, P.~Masjuan, and P.~Sanchez-Puertas,
  Eur.\ Phys.\ J.\ C {\bf 75}, no. 9, 414 (2015).

\bibitem{Escribano:2015yup}
  R.~Escribano, S.~Gonzàlez-Solís, P.~Masjuan, and P.~Sanchez-Puertas,
  Phys.\ Rev.\ D {\bf 94}, no. 5, 054033 (2016).


\bibitem{Ball:1995zv}
  P.~Ball, J.~M.~Frere, and M.~Tytgat,
  Phys.\ Lett.\ B {\bf 365}, 367 (1996).

\bibitem{Thomas:2007uy}
  C.~E.~Thomas,
  JHEP {\bf 0710},  026 (2007).

\bibitem{Escribano:2007cd}
  R.~Escribano and J.~Nadal,
  JHEP {\bf 0705},  006 (2007).

\bibitem{Fearing:1994ga}
  H.~W.~Fearing and S.~Scherer,
  Phys.\ Rev.\ D {\bf 53}, 315 (1996).

\bibitem{Scherer:2002tk}
  S.~Scherer,
  Adv.\ Nucl.\ Phys.\  {\bf 27}, 277 (2003).

\bibitem{Bijnens:1999sh}
  J.~Bijnens, G.~Colangelo, and G.~Ecker,
  JHEP {\bf 9902}, 020 (1999).

\bibitem{Ebertshauser:2001nj}
  T.~Ebertsh\"auser, H.~W.~Fearing, and S.~Scherer,
  Phys.\ Rev.\ D {\bf 65}, 054033 (2002).

\bibitem{Bijnens:2001bb}
  J.~Bijnens, L.~Girlanda, and P.~Talavera,
  Eur.\ Phys.\ J.\ C {\bf 23}, 539 (2002).

\bibitem{Jiang:2014via}
  S.~Z.~Jiang, F.~J.~Ge, and Q.~Wang,
  Phys.\ Rev.\ D {\bf 89},  074048 (2014).

\bibitem{Scherer:1994wi}
  S.~Scherer and H.~W.~Fearing,
  Phys.\ Rev.\ D {\bf 52}, 6445 (1995).


\bibitem{Fearing:1999fw}
  H.~W.~Fearing and S.~Scherer,
  Phys.\ Rev.\ C {\bf 62}, 034003 (2000).



\bibitem{Aoki:2013ldr}
  S.~Aoki, Y.~Aoki, C.~Bernard, T.~Blum, G.~Colangelo, M.~Della Morte, S.~D\"urr, and A.~X.~El Khadra {\it et al.},
  Eur.\ Phys.\ J.\ C {\bf 74}, no. 9, 2890 (2014).

\bibitem{Kaiser:2007zz}
  R.~Kaiser,
  Nucl.\ Phys.\ Proc.\ Suppl.\  {\bf 174}, 97 (2007).

\bibitem{Jiang:2015dba}
  S.~Z.~Jiang, Z.~L.~Wei, Q.~S.~Chen, and Q.~Wang,
  Phys.\ Rev.\ D {\bf 92}, 025014 (2015).


\end{thebibliography}
\end{document}